\documentclass[iop, apj]{emulateapj}
\usepackage{graphics}      % standard graphics specifications
\usepackage{amsmath}
\usepackage{amssymb}
\usepackage{morefloats}
\usepackage{natbib}
\usepackage{CJK}
\usepackage{soul}

\makeatletter

\newcommand{\Rmnum}[1]{\expandafter\@slowromancap\romannumeral #1@}
\makeatother

\def\Msol{M_\odot}
\def\Rsol{R_\odot}

\shorttitle{Global MHD Simulation of CV Disks}
\shortauthors{Ju et al.}

\begin{document}

%%%%%%%%%%%%%%%%%%%%%%%%%%%%%%%%%%%%%%%%%%%%%%%%%%%%%%%%%%%%%%%%%%%%%%%%%%%%%%%%%%%%%

\title{Global MHD Simulations of Accretion Disks in Cataclysmic Variables (CVs): {\Rmnum 1} The Importance of Spiral Shocks}

\author{Wenhua Ju\altaffilmark{1}, James M. Stone\altaffilmark{1}, Zhaohuan Zhu\altaffilmark{1}}
\altaffiltext{1}{Dept. of Astrophysical Sciences, Princeton University, Princeton, NJ 08544, USA}

%%%%%%%%%%%%%%%%%%%%%%%%%%%%%%%% Abstract %%%%%%%%%%%%%%%%%%%%%%%%%%%%%%%%%%
\begin{abstract}
We present results from the first global 3D MHD simulations of accretion disks in Cataclysmic Variable (CV) systems in order to investigate the relative importance of angular momentum transport via turbulence driven by the magnetorotational instability (MRI) compared to that driven by spiral shock waves. Remarkably, we find that even with vigorous MRI turbulence, spiral shocks are an important component to the overall angular momentum budget, at least when temperatures in the disk are high (so that Mach numbers are low).  In order to understand the excitation, propagation, and damping of spiral density waves in our simulations more carefully, we perform a series of 2D global hydrodynamical simulations with various equation of states and both with and without mass inflow via the Lagrangian point (L1).  
Compared with previous similar studies, we find the following new results.
%1) The dependency of disk response to tidal forcing on disk properties is qualitatively consistent with but much steeper than linear theory \citep{1994MNRAS.268...13S}: spiral waves are significantly stronger if the disk size is larger or the Mach number is lower.
1) Linear wave dispersion relation fits the pitch angles of spiral density waves very well.
2) We demonstrate explicitly that mass accretion is driven by the deposition of negative angular momentum carried by the waves when they dissipate in shocks.
3) Using Reynolds stress scaled by gas pressure to represent the effective angular momentum transport rate $\alpha_{eff}$ is not accurate when mass accretion is driven by non-axisymmetric shocks. 
4) Using the mass accretion rate measured in our simulations to directly measure $\alpha$ defined in standard thin-disk theory, we find $0.02 \lesssim \alpha_{eff} \lesssim 0.05$ for CV disks, consistent with observed values in quiescent states of dwarf novae (DNe). In this regime the disk may be too cool and neutral for the MRI to operate and spiral shocks are a possible accretion mechanism.
However, we caution that our simulations use unrealistically low Mach numbers in this regime, and therefore future models with more realistic thermodynamics and non-ideal MHD are warranted.
\end{abstract}


%%%%%%%%%%%%%%%%%%%%%%%%%%%%%%%%%%%%%%%%%%%%%%%%%%%%%%%%%%%%%%%%%
%%%%%%%%%%%%%%%%%%%%%% Introduction %%%%%%%%%%%%%%%%%%%%%%%%%%%%%%%%%
%%%%%%%%%%%%%%%%%%%%%%%%%%%%%%%%%%%%%%%%%%%%%%%%%%%%%%%%%%%%%%%%%
\section{Introduction}

%% [Ideal laboratory] 
Accretion disks in cataclysmic variables (CVs) are ideal laboratories for studying accretion physics for several reasons. Firstly, a large body of observational data exists to constrain theoretical models \citep[see reviews by][]{warner95, lin-papa96, Hellier2001}. For example, the observed episodic outbursts of dwarf novae (DNe) provide tight constraints on the Shakura-Sunyaev $\alpha$ viscosity parameter. The disk instability model (DIM) indicates that $\alpha \sim 0.1-0.3$ during outburst state \citep{cannizzo93, king07, lasota01,kotko12}, and $\alpha \sim 0.01$ during quiescence state \citep{cannizzo93}.
%By interpreting the observed outburst-related timescales (e.g. the duration of outbursts, the decay time after outburst peak and burst recurrence timescales) in light curves of DNe ,
In addition, techniques such as eclipse mapping and Doppler tomography unveil the size, radial structure and kinematics of the disk \citep{kaitchuck94}. Secondly, length and time scales in CV disks are much smaller than in other known accretion disk systems (e.g. AGN disks, protoplanetary disks). For example, the radial range from the surface of the white dwarf to the inner Lagrangian point is only two orders of magnitude, to be compared with four orders of magnitude in protoplanetary disks and even larger in AGN disks. This allows numerical simulations to cover the entire range of both spatial and temporal scale in the flow. In addition, since the period of DNe outbursts is typically several days, such short timescales allow observations to follow many cycles. Lastly, the mass supply for the disk, e.g. gas streaming through the inner Lagrangian (L1) point at a rate of $\sim 10^{-10} - 10^{-9} \Msol / yr$ \citep{Hellier2001}, is better understood than in other accretion disk systems.

%% [Raise the question of accretion] 
An important ingredient to understanding the observed episodic outbursts in DNe is identifying the mechanisms of angular momentum transport in CV disks. There are two candidates. The first is the spiral shock model \citep{Lynden-Bell1974}, in which the density waves are excited by the non-axisymmetric gravitational potential and transport negative angular momentum which is deposited into the gas when the waves steepen into shocks. Two-armed spiral structures in CV disks have been identified in lots of CV observations with techniques of eclipse mapping and Doppler tomography and comparison with hydrodynamical models \citep{2005A&A...444..201B, 2006Ap&SS.304..315P,  2006A&A...448.1061K, 2007jena.confR..19G, 2008ARep...52..815K, 2010AIPC.1273..346Y, 2011MNRAS.410..963N, 2011AIPC.1356...17B, 2012A&A...538A..94K}. The second mechanism is magnetorotational instability (MRI) in which shear amplification of weak magnetic field induces turbulent transport \citep{balbus-hawley1998}. \citet{Gammie-Menou1998} proposed that the physical origin of episodic accretion in DNe disks may be that MRI turbulence is suppressed when the gas is predominantly neutral at low temperature during the quiescent state. This implies that during quiescence, spiral shocks may play a major role in angular momentum transport, while during outbursts MRI and spiral shocks both contribute. 

Controversy exists on whether MRI is sufficient to drive mass accretion during DNe outbursts. \citet{King-Pringle-livio2007} have suggested that even in the DNe outburst phase, MRI alone cannot explain observed accretion rates since some local shearing box MHD simulations with zero net vertical magnetic field ($B_z$) give $\alpha \le 0.02$ \citep*[although see][]{2015arXiv151201106S}. \citet{Kotko-lasota2012} reiterated this discrepancy using several methods to confirm the validity of $\alpha \sim 0.1 - 0.2$ from DNe outburst observations. \citet{2014ApJ...787....1H} conducted local, vertically stratified, radiation MHD shearing box simulations with zero net vertical magnetic field, and proposed one possible solution to the discrepancy: an enhanced stress to pressure ratio $\alpha$ is produced by strong thermal convection triggered due to large opacity near temperature of $>10^4$ K which significantly increases magnetic stress. However, to date, there are still no global MHD models to investigate the interplay of spiral shocks and MRI turbulence in angular momentum transport during DNe outbursts.

%The mechanism of angular momentum transport in accretion disks has remained a longstanding problem. The standard $\alpha$ viscosity (Shakura \& Sunyaev, 1973) is used to describe the efficiency of accretion, where $\alpha$ is the turbulent stress scaled by gas pressure. However, the origin of viscosity has been a mystery. In the context of CV disks, 

%% [This work, introduce spiral waves] 
Motivated by these questions, we have begun a series of global MHD simulations of unstratified CV disks to explore the angular momentum transport process. Remarkably, we find that spiral shocks can coexist with MRI turbulence and play a more important role than MRI in CV disks in certain specific parameter settings (see \S \ref{sec:result-mhd}). Reynolds stress is larger than Maxwell stress in some of our MHD models. Therefore, in order to understand the relative importance of spiral shocks and MRI in angular momentum transport in CV disks, it is clear we first need to thoroughly understand spiral shocks on their own.

%% [Theoretical background for spiral waves]
There is a rich literature that explores the theory of spiral waves in accretion disks of binary systems \citep[see reviews by][and references therein]{1981ARA&A..19..277S, 1995ARA&A..33..505P, 2001LNP...573...69B}. The basic picture is as follows. In binary systems, the gravitational force from the companion star provides perturbations to the gas orbits. In CV disks, or the inner region of protoplanetary disks, these perturbations are largest at inner Lindblad resonances of the companion star \citep{1991ApJ...381..259L}, and propagate inward in the form of waves \citep{1980ApJ...241..425G, 1984ApJ...285..818P,1994ApJ...421..651A}. The wave fronts form a ``spiral arm" pattern. These wave patterns usually rotate slower than the background Keplerian disk, so in principle they carry negative angular momentum \citep{1978ApJ...222..850G}. When the pattern speed exceeds the local gas sound speed, the waves become non-linear and steepen into shocks \citep{1987A&A...184..173S}. Non-linear shock dissipation locally deposits the negative angular momentum carried by the waves into the disk, and therefore the disk loses angular momentum \citep{1989ApJ...342.1075G, 2016arXiv160103009R}. Angular momentum transport is driven by the net negative torque exerted on the disk by the companion star, in other words, angular momentum is transported from the disk to the companion star \citep{1977MNRAS.181..441P, 1979MNRAS.188..191L, 1988ApJ...333..895P, 1994MNRAS.268...13S, 1995ARA&A..33..505P, 2001ApJ...552..793G, 2002ApJ...569..997R}. The rate of angular momentum transport driven by spiral shocks is found very sensitive to shock amplitude \citep{2016arXiv160103009R} and disk temperature \citep{1987A&A...184..173S, 2016arXiv160201721H}. Except for resonant forcing of a companion object, there are other possible mechanisms that can excite spiral density waves which may enhance angular momentum transport in the disk, such as MRI turbulence \citep{2009MNRAS.397...52H, 2009MNRAS.397...64H, 2012MNRAS.419.1085H, 2014ApJ...784..121S}, convection \citep{2011MNRAS.417..634M}, external inflow from a surrounding envelope \citep{2015ApJ...805...15B, 2015A&A...582L...9L, 2016arXiv160201721H}, and self-gravity of the disk \citep{2014ApJ...795...61B, 2015ApJ...812L..32D}.

%% [Previous modeling of spiral wave] 
%Currently, most models of CV disks focus either on reduced 1D models that follow the evolution of the vertically- and azimuthally-averaged disk structure in order to understand outbursts and variability, or on detailed, time-dependent hydrodynamical models in both 2D and 3D. For 1D models, besides the famous disk instability models (DIM) which successfully reproduced the observed outburst recurrence times and outburst durations \citep{cannizzo93, 2012ApJ...747..117C}, there is recent work by \citet{2014MNRAS.441..681P} proposed that a limit cycle of disk stability could be achieved by assuming $\alpha$ is correlated with magnetic Prandtl number instead of a constant over the disk. 
There is also a rich literature of 2D and 3D hydrodynamical simulations that explore spiral shock formation in accretion disks in close binaries in more detail \citep[see review by][and references therein]{2000Ap&SS.274..259M}. Early work by \citet{1979MNRAS.186..799L} and \citet{1986MNRAS.219...75S} and later numerical modeling for IP Peg \citep{1998MNRAS.295L..11G, 1998MNRAS.297L..81A} all observed spiral arms. \citet{1994MNRAS.268...13S} studied the tidal excitation of spirals in CV disks by understanding non-linear numerical results with linear tidal response calculation, and pointed out that the tidal torques exerted on the disk negatively correlates with Mach number ($\mathcal{M}$). \citet{2000NewA....5...53B} did 2D grid-based simulations of inviscid isothermal CV disks and measured an effective $\alpha$ of order $0.1$ in the outer edge of the disk which is much smaller at smaller radius. \citet{2000MNRAS.316..906M} did 2D and 3D grid-based simulations of CV disks with the simplified flux vector splitting (SFS) finite volume method, and found that the opening angle of spiral waves is correlated with ratio of specific heats ($\gamma$) which is consistent with the analytical work by \citet{1987A&A...184..173S}. \citet{2001PThPh.106..729F} did similar 3D simulations and investigated the interaction between the L1 stream and the disk in more detail. A series of 2D and 3D SPH simulations of accretion disks in binary systems were conducted to investigate the occurrence of spiral structures depending on the SPH particle concentration, the gas compressibility and physical viscosity \citep{2002aprm.conf..343L, 2002PASJ...54..781B, 2003A&A...403..593L, 2005ApJ...632..499L, 2007AIPC..924..907L, 2008ASPC..385..115L}. Evidence of elliptical instabilities driven by the mechanism proposed by \citet{1991ApJ...381..259L} were found in some of the hydro models \citep{2007A&AT...26...47B, 2008A&A...487..671K, 2011EAS....44..121B}, which possibly is responsible for superhumps observed during DNe outbursts. Spiral structures are also widely discussed in other astrophysical environment, such as accretion disks in supermassive black hole binaries \citep{2006ApJ...651..767M, 2007MNRAS.379..956D, 2010AIPC.1240..181M, 2012A&A...545A.127R}, protoplanetary disks \citep{2015ApJ...809L...5D, 2015ApJ...813...88Z}. However, regarding CV disks, very little numerical work has been done in quantifying the whole processes of tidal excitation, propagation and dissipation of spiral waves, and the associated angular momentum transport.
%In this paper, we do a series of 2D simulations of CV disks with various equations of states (EOS) and Mach numbers, with and without an accretion stream from the donor star via the first Lagrangian (L1) point. We track the angular momentum and torques exterted on the disk, and try to understand the tidal excitation and dissipation of spiral shocks, as well as angular momentum transport. 

%% [This work: hydro models] 
In this paper, we show that even when MHD turbulence driven by the MRI is present, spiral waves can be an important mechanism of angular momentum transport at least when the disk is very hot (or the Mach number of the flow is low). We present a comprehensive set of 2D hydrodynamical models that explore the excitation, propagation and dissipation of spiral waves, to try to answer questions such as what is the transport efficiency, whether it is consistent with observations of DNe in quiescence, and how the behavior of spiral shocks vary with different physical conditions. Although our 2D hydrodynamical simulations are similar to previous work , we present a number of new results that emerge from a comprehensive analysis of spiral waves in CV disks. For the first time, we show: 
\begin{enumerate}
\item The dependency of disk response to tidal forcing on disk properties is qualitatively consistent with but much steeper than linear theory \citep{1994MNRAS.268...13S}: spiral waves are significantly stronger if the disk size is larger or the Mach number is lower (see \S \ref{sec:waveexication}).

\item The pitch angles of spiral waves follow the linear wave dispersion relation with $m=2$, resulting in a unique relation between the pitch angles and local Mach numbers of the disk. The self-similar shock model by \citet{1987A&A...184..173S} is not a good fit to the spiral pattern, instead linear wave theory works well (see \S \ref{sec:wavepropagation}).

\item By explicitly analyzing the angular momentum budget of the disk, we conclude that the local shock dissipation leads to angular momentum loss of the disk gas and mass accretion. The effective $\alpha_{eff}$ is $\sim 0.02 - 0.05$ for CV disk models with mass accretion flow via the L1 point, similar to observed values of DNe in quiescence (see  \S \ref{sec:shockdissipation}).

\item The popular method of using Reynolds stress scaled by gas pressure to represent the effective angular momentum efficiency $\alpha_{eff}$ is not suitable for accretion disks in non-axisymmetric potentials since $\alpha_{eff}$ calculated in this way is not sufficient to account for the mass accretion rate according to the standard thin disk theory which assumes a kinetic viscosity. Tidal torques and effects from the inner boundary conditions are also important (see \S \ref{sec:effectivealpha}).
\end{enumerate}

Our paper is arranged as follows. We introduce our numerical methods, diagnostics and analytical analysis of angular momentum budget in \S \ref{sec:method}. We present the results of our MHD model in \S \ref{sec:result-mhd}. Our study of spiral waves with hydrodynamical models are shown in \S \ref{sec:result-noinflow} where we discuss the wave excitation in \S \ref{sec:waveexication}, the wave propagation and spiral patterns in \S \ref{sec:wavepropagation}, the angular momentum transport by spiral shocks in \S \ref{sec:shockdissipation}, the energy budget in \S \ref{sec:energybudget}, and a convergence study in \S \ref{sec:convergence}. Finally,  further discussions and major conclusions are presented in \S \ref{sec:discussion} and \S \ref{sec:conclusion}.

%%%%%%%%%%%%%%%%%%%%%%%%%%%%%%%%%%%%%%%%%%%%%%%%%%%%%%%%%%%%%%%%%
%%%%%%%%%%%%%%%% Method   %%%%%%%%%%%%%%%%%%%%%%%%%%%%%%%
%%%%%%%%%%%%%%%%%%%%%%%%%%%%%%%%%%%%%%%%%%%%%%%%%%%%%%%%%%%%%%%%%
\section{Method and Diagnostics}
\label{sec:method}

%%%%%%%%%%%%%%%%%%%
\subsection{Equations Solved}
\label{equation-solved}

We solve the equations of compressible ideal MHD in cylindrical coordinates using {\it Athena++}, a recent extension of the grid-based Godunov code package {\it Athena} \citep{2008ApJS..178..137S}. The equations for {\it inviscid} gas dynamics with an isothermal or adiabatic EOS are:
\begin{subequations}
\begin{align}
\frac{\partial \rho}{\partial t} + \nabla \cdot (\rho {\bf v}) &= 0,  \label{eq:massconservation} \\
\frac{\partial (\rho {\bf v})}{\partial t} + \nabla \cdot (\rho {\bf vv} - {\bf BB} + P^*{\bf I}) &= -\rho \nabla \Phi_{tot}, \\
P = \rho c_s^2, & \label{eq:momentumcons}\\
\qquad \text{for isothermal gas, or} & \nonumber \\
\frac{\partial E}{\partial t} + \nabla \cdot [ (E+P^*) {\bf v} - {\bf B (B \cdot v)} ] &= -\rho \nabla \Phi_{tot} \cdot {\bf v},  \label{eq:energycons}\\
\qquad \text{for adiabatic gas}  \nonumber \\
\frac{\partial B}{\partial t} - \nabla \times ({\bf v \times B}) &= 0,
\end{align}
\end{subequations}
where $\rho$ is the gas density (replaced with $\Sigma$ in 2D hydro simulations); $\bf B$ is the magnetic field vector; $P^* = P + |{\bf B}|^2/2$ is the total pressure consisting of magnetic and gas pressure; $c_s$ is the isothermal sound speed and $E$ is the total energy density
\begin{equation}
 E = \frac{P}{\gamma -1} + \frac{1}{2} \rho v^2 + \frac{|{\bf B}|^2}{2}
\end{equation} 

We solve the equations in a frame of reference such that the WD is at the origin, and the donor star rotates around the WD on a circular Keplerian orbit. The position and velocity vectors in this frame are centered on WD are ${\bf r} = (R, \phi, z)$ and ${\bf v} = (v_R, v_\phi, v_z)$ respectively. Since we solve the equations in a non-inertial frame, we must include an indirect gravitational force \citep{1987gady.book.....B}. The total gravitational potential in this frame is 
\begin{equation}
\Phi_{tot} = - \frac{G M_1}{|{\bf r}|_{z=0}} - \frac{G M_2}{|{\bf r} - {\bf R_2}|_{z=0}} + G M_2 \frac{ ({\bf R_2} \cdot {\bf r})_{z=0} }{R_2^3}, 
\label{eq:phitot}
\end{equation}
which consists of the potentials of the WD and donor star, and the indirect term due to movement of the origin where $M_1$ and $M_2$ are the masses of the WD and the donor star respectively, and $R_2$ is the position of the donor star. We define the mass ratio of the binary to be $q = M_2 / M_1$. Throughout this work, we adopt $q=0.3$ which is a typical value for dwarf novae \citep{Hellier2001}. Since we are studying unstratified disks as in \citet{2012ApJ...749..189S} and \citet{2001ApJ...554..534H}, the gravitational potential does not include the vertical component of gravity.

%%%%%%%%%%%%%%%%%%%%%%%%
\subsection{Units \& Scaling}
\label{subsec:units-scaling}
To set up initial conditions that are in general agreement with the properties of typical CV systems, we begin by listing the properties of a well-studied CV system: SS Cygni. This system has a 1.2 $\Msol$ WD and a 0.7 $\Msol$ companion star in orbit with a period of 6.6 hr, which implies a binary separation of 2.24 $\Rsol$. The mean interval time between outbursts of SS Cygni is 40 days, with typical time of decay from outbursts being 2.4 days. The mass inflow rate from the companion star is of order $10^{-9} \Msol /$ yr \citep{cannizzo93}. According to thermal limit models \citep[e.g.][]{cannizzo93}, at a radius of $2 \times 10^{10}$ cm from the WD (around the mid-radius area of the disk), the typical surface density of the WD accretion disk is $\approx 200$ g cm$^{-2}$, and the midplane temperature is $\approx 3000$ K which implies the local Mach number of $\mathcal{M} \sim 280$ (where $\mathcal{M}$ is ratio of kinematic velocity and sound speed of the gas). While there is a wide variation of these parameters among CV systems, we take the SS Cygni system as a reference system.

 In this work, the equations are solved in dimensionless form. We define the unit of mass such that $G M_0 = GM_1 + GM_2 =1$, and the unit of length to be the separation of the binary $a_0=1$. Therefore, the orbital frequency of the binary $\Omega_0 = \sqrt{GM_0 / a_0^3} = 1$.  The unit of time is the inverse of the angular frequency of the binary orbit $t_0 = \Omega_0^{-1}=1$ which makes one orbital period of the binary $P/t_0 = 2\pi$. The donor star orbits on a unit circle at frequency $\Omega_0$, so ${\bf R_2} = (cos(t), sin(t), 0)$.
 
If our internal units were scaled to the SS Cygni system, then our unit of length is $a_0 = 2.24 R_\odot$, our unit of mass is $M_0 \sim 2.44 \times 10^{-9} \Msol$ and our unit of time is 1.05 hr which yields a binary orbit of 6.6 hrs. If we express the properties of SS Cygni in our internal units, then the binary stars are separated by $1 R_0$, the mean interval time between outbursts is 914 $t_0$, and outbursts decay on the timescale of 54 $t_0$; the mass inflow rate from the companion star is $4.9 \times 10^{-5} M_0 / t_0$; the reference radius $2 \times 10^{10}$ cm is about 0.13 $R_0$, and at this radius, the surface density of the disk is around 1 $M_0/R_0^2$.

CV disks may be very cold according to both eclipse mapping observations and 1D analytical models. From brightness map reconstruced by eclipse mapping, the surface temperature of CV disks spans from 8000K (outer radius) to 40000K (inner radius) during outbursts, and from 2000K (outer radius) to 6000K (inner radius) during quiescence \citep{1985MNRAS.214..307H, 1986MNRAS.219..629W, 1992A&A...260..213R, 1998MNRAS.298.1079B}. The corresponding Mach numbers are $50 - 200$ during outbursts, and $200 - 600$ during quiescence (note uncertainties exist in evaluation of WD mass and CV mass ratio). It is worth noting that the midplane temperature of CV disk which essentially affect disk structures may be a few decades higher than the surface temperature from eclipse mapping observations. According to 1D thermal limit models \citep[e.g.][]{cannizzo93, Gammie-Menou1998, 2000ARep...44...89V}, SS Cygni has midplane Mach number of $\sim 280$ in quiescence and $\sim 60$ in outburst at a radius of $2 \times 10^{10}$ cm from the WD (around the mid-radius area of the disk). Such extreme Mach numbers are very challenging for numerical simulations. Previous numerical work on CV disks typically have adopted Mach numbers from less than 10 to 20 \citep[e.g.][]{2000MNRAS.316..906M, 2008A&A...487..671K}. In this work, we take an initial sound speed of 0.1 $R_0/t_0$ throughout the disk, corresponding to $\mathcal{M} \sim 60$ at the inner boundary. It is worth noting that although in isothermal simulation the Mach number will stay nearly constant over time, in adiabatic simulations the Mach number becomes smaller as the disk evolves and heats up. We discuss the effect of unphysically low Mach numbers on angular momentum transport by spiral waves  more thoroughly in \S \ref{sec:result-noinflow}.

%%%%%%%%%%%%%%%%%%%%%%%%%
\subsection{Initial and Boundary Conditions}

In cylindrical coordinates, our computational domain spans $(R, \phi, z) \in [0.02, 0.62] \times [0, 2\pi] \times [-2H_0,2H_0]$ for 3D MHD simulations and $(R, \phi) \in [0.02, 0.62] \times [0, 2\pi]$ for 2D hydro simulations where $H_0 = c_s / \Omega=0.02$ is the fiducial thermal scale height. The outer radial boundary at $R=0.62$ is the radius of the first Lagrangian point $L_1$ for a binary with mass ratio $q=0.3$, and the inner radial boundary 0.02 is approximately four times the surface radii of the WD. Our fiducial simulations have  $384 \times 704 \times 32$ cells for 3D MHD models and $384 \times 704$ cells for 2D hydro cases. Parameters of all models in this work are summarized in Table \ref{tab:parameter}.

\begin{deluxetable*}{ccccc}
\tablecolumns{5} 
\tabletypesize{\scriptsize}
\tablewidth{0pc}
\tablecaption{Model Parameters \label{tab:parameter}}
\tablehead{ 
\colhead{Model} & \colhead{Inflow} & \colhead{EOS}  & \colhead{Domain} & \colhead{Resolution}
}
\startdata
MHD & Inflow & adiabatic $\gamma=1.1$ & $[0.02, 0.62] \times [0, 2\pi] \times [-0.04, 0.04]$  &  $384 \times 704 \times 32$ \\
\hline
Hydro & No Inflow & adiabatic $\gamma=1.1$ & $[0.02, 0.62] \times [0, 2\pi]$ &  $384 \times 704$\\
Hydro & No Inflow & adiabatic $\gamma=1.2$ & $[0.02, 0.62] \times [0, 2\pi]$ &  $384 \times 704$\\
Hydro & No Inflow & adiabatic $\gamma=1.3$ & $[0.02, 0.62] \times [0, 2\pi]$ &  $384 \times 704$\\
Hydro & No Inflow & isothermal $c_s=0.1$ & $[0.02, 0.62] \times [0, 2\pi]$ &  $384 \times 704$ \\ 
Hydro & No Inflow & isothermal $c_s=0.3$ & $[0.02, 0.62] \times [0, 2\pi]$ &  $384 \times 704$ \\
\hline
Hydro & Inflow & adiabatic $\gamma=1.1$ & $[0.02, 0.62] \times [0, 2\pi]$ &  $384 \times 704$\\
Hydro & Inflow & adiabatic $\gamma=1.2$ & $[0.02, 0.62] \times [0, 2\pi]$ &  $384 \times 704$\\
Hydro & Inflow & adiabatic $\gamma=1.3$ & $[0.02, 0.62] \times [0, 2\pi]$ &  $384 \times 704$\\
Hydro & Inflow & isothermal $c_s=0.1$ & $[0.02, 0.62] \times [0, 2\pi]$ &  $384 \times 704$\\
Hydro & Inflow & adiabatic $\gamma=1.1$ & $[0.02, 0.62] \times [0, 2\pi]$ &  $192 \times 352$\\
Hydro & Inflow & adiabatic $\gamma=1.1$ & $[0.02, 0.62] \times [0, 2\pi]$ &  $768 \times 1408$
\enddata
\end{deluxetable*}

The grid spacing is increased logarithmically in the radial direction ($\Delta R \propto R$) and is uniform in the azimuthal and vertical directions. Spiral arms excited in binary systems are nearly self-similar \citep{1987A&A...184..173S}, so that the radial spacing between shocks becomes smaller at small radii. By using logarithmic grid spacing in radius, we can resolve the spiral shocks equally well at any radii. Moreover, with a logarithmic grid, we are able to ensure $\Delta R /R = \Delta \phi = const$, i.e. the grid cells are square in physical size at all radii which reduces numerical diffusion due to highly distorted grid cells.

In the 3D MHD simulation, we initially put a small low-density disk  with $\Sigma = 0.01$ in the radial range of $R \in [0.02, 0.1]$ and set $\Sigma=10^{-5}$ in all the other regions ($R \in [0.1, 0.62]$) to represent vacuum. The initial velocity is Keplerian where $v_R=v_z=0$, $v_{\phi} = \sqrt{GM_1/R^3}$. The initial pressure is $P=10^{-4}$ throughout the disk. We then inject gas stream from the L1 point, letting the disk form self-consistently. We use adiabatic EOS with the specific heat ratio $\gamma=1.1$. 

In the 2D hydrodynamical simulations, we use two sets of initial conditions, with one set simulating a CV disk without an accretion stream (no-inflow case), and the other set simulating a CV disk with an accretion stream from the L1 point (with-inflow case). For the no-inflow case, we start with uniform gas surface density $\Sigma_0=1$ throughout the disk to $R=0.62$. The initial gas pressure is $P=0.01$ for adiabatic runs and sound speed is $c_s = 0.1$ for isothermal runs. For the with-inflow case, we initially set floor density in the whole computational domain $\Sigma_0 = 10^{-5}$ to represent vacuum. The initial gas pressure is $P=10^{-4}$ for adiabatic runs and sound speed $c_s = 0.1$ for isothermal runs. The initial velocity is Keplerian, with $v_R=0$, $v_{\phi} = \sqrt{GM_1/R^3}$. 

We use periodic boundary conditions in the azimuthal and vertical directions. The boundary conditions in the radial direction depend on whether or not there is gas inflow at the location. At the L1 region which spans $\phi \in [-0.1, 0.1]$ at outer boundary and consists of 10 azimuthal cells with the fiducial resolution, the gas surface density is $\Sigma_0=1$, radial velocity is $v_R=-0.01$, and azimuthal velocity is set to be $v_\phi=R\Omega_0$ to make the L1 region corotate with the donor star. The gas pressure is $P=0.01$ for adiabatic runs and sound speed $c_s = 0.1$ for isothermal runs. All the variables are constant over time at the L1 ghost region. At all other regions, we use "free outflow, no inflow" boundary condition, which copies variables from the last active cells to the ghost cells except radial velocities $v_R$ and azimuthal velocities $v_\phi$. Azimuthal velocities in ghost zones are set to their local Keplerian values. For radial velocities, if $v_R$ of the last active cell is outflowing (i.e. $v_R>0$ at the outer boundary, or $v_R<0$ at the inner boundary), we copy its value to the ghost cells; otherwise, we set $v_R=0$ at the ghost cells to avoid unwanted mass inflow.  

For our MHD runs, the boundary conditions for the magnetic field are also treated separately in the L1 region and other regions. At non-L1 regions, we simply copy the magnetic field components of the last active cells to ghost cells. At the L1 region, we provide seed magnetic field flowing in with the gas. Magnetic field loops are set by a magnetic vector potential 
\begin{equation}
{\bf A}({\bf R}) = \sqrt{\frac{2P_{gas}}{\beta}} \frac{2R_{loop}}{\pi} \cos(\frac{\pi}{2} \frac{|{\bf R}-{\bf R}_{ref}(t)|}{R_{loop}}) {\bf \hat{z}},
\end{equation}
where $\beta$ is magnetic pressure to gas pressure ratio, ${\bf R}_{ref}(t) = (R=R_0 + v_Rt, \phi=\Omega_0 t, z)$ is a reference point that has the same velocity with the gas inflow at L1 point. The size of magnetic loops is $R_{loop} \sim 0.06$. The magnetic field components can then be calculated by 
\begin{equation}
{\bf B} = \nabla \times {\bf A},
\end{equation}
which forms field loops in the disk plane advecting with the background inflow at L1 region. In this work, we adopt $\beta = 100$. One problem we ran into with this boundary condition of field loop injection is that the seed magnetic filed leaks out of the gas stream since the L1 region rotates with the companion star and the ghost cells representing L1 region keeps changing. We deal with this problem by doing the MHD models in the frame that corotates with the companion star in which the L1 region stays static. Coriolis force and centrifugal force in the rotating frame is added accordingly to the momentum and energy conservation equations (Eq. \ref{eq:momentumcons} and \ref{eq:energycons}).

%We assume the outer boundary is crossing a static clockwise magnetic field loop in the $R\phi$ plane centered at $[R, \phi] = [R_{max}, 0]$. Considering the L1 region is symmetric about $\phi=0$ by design, this field loop at L1 ghost region satisfies the solenoidal constraint. The three components of magnetic field are given by
%\begin{subequations}
%\begin{align}
%B_R &= -\sqrt{\frac{2P_{gas}}{\beta}} \cos(\frac{\pi}{2} \frac{\phi}{0.1}) \text{sgn}(\phi), \\
%B_\phi &= B_z = 0.
%\end{align}
%\end{subequations}

%Therefore, the mass transfer rate through L1 region is $|\Sigma_0 v_R R \delta \phi| =1.24 \times 10^{-3}$ in code units. With the unit scaling we discussed in \S \ref{equation-to-be-solved}, this value yields a mass transfer rate of $~1.e-8 M\odot/yr$. [Discuss how this inflow rate is compared with $10^{-9} M_\odot$/yr.]

%%%%%%
\subsection{Angular Momentum Budget}
\label{sec:ang-mom-budget}

It is critical to understand the angular momentum budget of infalling gas in our simulations. Thus, we use terms from the angular momentum conservation equation as diagnostics of the flow. For inviscid gas, the angular momentum conservation equation in cylindrical coordinates is
\begin{eqnarray}
\label{eq:angmom}
 \partial_t <\rho R v_\phi&>&= -\frac{1}{R} \partial_R (R^2<\rho v_R v_\phi> \nonumber \\
&-& R^2 <B_R B_\phi> ) + <\bf{R} \times \bf{F}_{ext}>,
\end{eqnarray}
where $\bf{F}_{ext} = -\nabla \Phi_{tot}$ is the external gravitational force. The notation $<X>$ represents the vertical and azimuthal integration of variable $X$ within a ring of radial width $\Delta R$ at radius $R$: 
\begin{eqnarray}
<X> &=& \int_{z_{min}}^{z_{max}} \int_{0}^{2\pi} X d\phi dz \Delta R \nonumber \\
&=& \sum_k \sum_j X_{i,j,k}  \Delta \phi_j \Delta z_k \Delta R_i.
\end{eqnarray}

Once the disk reaches steady state, it is almost Keplerian throughout the simulation time, so the evolution can be more clearly seen in the perturbed angular momentum $\rho R \delta v_\phi = \rho R (v_\phi-v_K)$. Substituting $v_\phi = v_K + \delta v_\phi$ and multiplying $R$ on both sides, equation \ref{eq:angmom} becomes
\begin{eqnarray}
&& \partial_t <\rho> (R v_K) R +  \partial_t <\rho R \delta v_\phi>R \nonumber \\
&=& - \partial_R <R\rho v_R> (R v_K)  - <R\rho v_R> \partial_R (R v_K) \nonumber \\
&& - \partial_R (R^2<\rho v_R \delta v_\phi> ) +  \partial_R (R^2<B_R B_\phi> ) \nonumber \\
&&+  <\mathbf{R} \times \mathbf{F_{ext}}>R .
\end{eqnarray}
Note that the first terms on the left and right side cancel due to mass conservation equation (Eq.\ref{eq:massconservation}). The conservation equation for perturbed angular momentum becomes
\begin{eqnarray}
\label{eq:angmom}
 &&\partial_t <\rho R \delta v_\phi> R \nonumber \\
 &=& - <R\rho v_R> \partial_R (R v_K) - \partial_R (R^2<\rho v_R \delta v_\phi> ) \nonumber \\
&&  + \partial_R (R^2<B_R B_\phi> ) + <\mathbf{R} \times \mathbf{F}_{ext}>R .
\end{eqnarray}

The left hand side is the time derivative of perturbed angular momentum volume-integrated within the ring, which is nearly zero if the disk reaches quasi-steady state. We define
\begin{equation}
AM_t (R) = \partial_t <\rho R \delta v_\phi> R.
\end{equation}
On the right hand side of Eq.\ref{eq:angmom}, the first term is related with the total mass accretion rate passing through this ring, thus we define
\begin{equation}
AM_{\dot{M}} (R) = \dot{M} \partial_R (R v_K)  \Delta R,
\end{equation}
where $\dot{M} = -<R\rho v_R>/ \Delta R = -2\pi \overline{R\Sigma v_R}$. The second term is radial advection of the angular momentum flux volume-integrated within the ring, thus
\begin{equation}
AM_{FH} (R) = - \partial_R (R^2 F_H),
\end{equation}
where $T_{H} = <\rho v_R \delta v_\phi>$ is the Reynolds stress. The third term is the radial gradient of Maxwell stress volume-integrated within the ring,
\begin{equation}
AM_{TM} (R) = -\partial_R (R^2 T_M),
\end{equation}
where $T_M = - B_R B_\phi$ is the Maxwell stress. Finally the last term is the torque exerted by the donor star volume-integrated within the ring
\begin{equation}
T(R) = <\mathbf{R} \times \mathbf{F}_{ext}>R.
\end{equation}
For linear wave propagation, the angular momentum flux advection term $AM_{FH}$ should exactly cancel external torques $T$ and lead to zero mass accretion rate. In this scenario, the torques are all used to restore the spiral structures, i.e. all the angular momentum added via the torques is immediately carried away by advective angular momentum fluxes. However, if the waves steepen into shocks, part of the angular momentum is lost by shock dissipation, and net mass accretion occurs. In terms of angular momentum conservation, the shock dissipation can be quantified by the difference between the external torques and the angular momentum flux advection, which is the sum of the last two terms in Eq.\ref{eq:angmom}. It is clear from the equation that when the disk reaches steady state, dissipation equates the mass accretion term, 
\begin{equation}
\label{eq:dissipation}
- AM_{\dot{M}} (R) \approx AM_{FH} (R) + AM_{TM} + T(R)
\end{equation}
which indicates dissipation is the source of accretion in the case of inviscid hydrodynamics.

%%%%%%%%%%%%%%%%%%%%%%%%%%%%%%%%%%%%%%%%%%%%%%%%%%%%%%%%%%%%%%%%
%%%%%%%%%%%%%%%%%%% Results: MHD %%%%%%%%%%%%%%%%%%%%%%%%%%%%%%%
%%%%%%%%%%%%%%%%%%%%%%%%%%%%%%%%%%%%%%%%%%%%%%%%%%%%%%%%%%%%%%%%

\section{The MHD Model}
\label{sec:result-mhd}

We begin by describing the evolution of our fiducial MHD simulation. Parameters of this model are summarized in Table \ref{tab:parameter}. Initially gas containing toroidal magnetic field loops from the companion star flows at constant rate through the L1 point into the Roche lobe of WD which is nearly vacuum initially. Since the gas has non-zero angular momentum relative to the WD, the gas stream follows an elliptical trajectory. After the trajectory closes the ellipse and crosses the initial inflow stream, the gas shocks, and the gas stream is circularized at a radius of $R \sim 0.19$, and a disk begins to form. Due to the nonaxisymmetric potential of the binary stars, two spiral arms are excited in the disk. The toroidal field loops in the gas stream are stretched in the azimuthal direction and MRI turbulence quickly develops. 

%\begin{table}[htbp]
%   \centering
%   \caption{Model Parameters} % requires the topcapt package
%     \begin{tabular}{@{} ccccc @{}} % Column formatting, @{} suppresses leading/trailing space
%      \toprule
%      Model & Inflow & EOS & Domain & Resolution \\
%      & & & $R \times \phi (\times z)$ & $R \times \phi (\times z)$\\
%      \midrule
%      MHD & Inflow & adiabatic $\gamma=1.1$ & $[0.02, 0.62] \times [0, 2\pi] \times [-0.04, 0.04]$ &  $384 \times 704 \times 32$ \\
%      Hydro & No Inflow & adiabatic $\gamma=1.1$ & $[0.02, 0.62] \times [0, 2\pi]$ &  $384 \times 704$\\
%      Hydro & No Inflow & adiabatic $\gamma=1.2$ & $[0.02, 0.62] \times [0, 2\pi]$ &  $384 \times 704$\\
%      Hydro & No Inflow & adiabatic $\gamma=1.3$ & $[0.02, 0.62] \times [0, 2\pi]$ &  $384 \times 704$\\
%      Hydro & No Inflow & isothermal $c_s=0.1$ & $[0.02, 0.62] \times [0, 2\pi]$ &  $384 \times 704$\\
%      Hydro & Inflow & adiabatic $\gamma=1.1$ & $[0.02, 0.62] \times [0, 2\pi]$ &  $384 \times 704$\\
%      Hydro & Inflow & adiabatic $\gamma=1.2$ & $[0.02, 0.62] \times [0, 2\pi]$ &  $384 \times 704$\\
%      Hydro & Inflow & adiabatic $\gamma=1.3$ & $[0.02, 0.62] \times [0, 2\pi]$ &  $384 \times 704$\\
%      Hydro & Inflow & isothermal $c_s=0.1$ & $[0.02, 0.62] \times [0, 2\pi]$ &  $384 \times 704$\\
%      \bottomrule
%   \end{tabular}
%%   \caption{Values of thermal scale height at various radius assuming isothermal EOS with $c_s=0.1$.}
%   \label{tab:parameter}
%\end{table}

\begin{figure}
\centering
\includegraphics[width=0.49\textwidth]{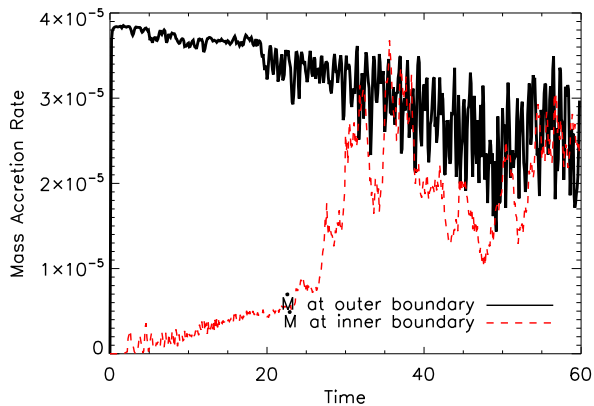}
\caption{Evolution of mass accretion rates at inner (red dashed line) and outer (black solid) boundaries for the MHD model. During $0<t<30$, the accretion rate at the inner boundary increases with time as spiral shocks and MRI develop. Meanwhile, net mass inflow at the outer boundary decreases because gas outflow increases as the disk gets hotter while the mass injection rate at the L1 point stays constant. After time $30$, the accretion rates at inner and outer boundaries match and the disk reaches steady state.}
\label{fig:mhd-history-mdot}
\end{figure}

Angular momentum transport is driven by both the spiral shocks (See \S \ref{sec:result-noinflow} for more detailed discussion) as well as Maxwell stress associated with MRI turbulence, and leads to mass accretion. In Fig. \ref{fig:mhd-history-mdot} we plot the time evolution of azimuthally integrated mass accretion rate at the inner boundary (red line) and the outer boundary (black line). At the inner boundary, the mass accretion rate increases rapidly at early times. This is due to two factors. On the one hand, the mass supply rate at the outer boundary is larger than the mass accretion rate at the inner boundary during $0<t<30$. Thus, mass accumulates in the disk and the spiral shocks become stronger which causes faster mass accretion. On the other hand, magnetic field builds up in the disk and MRI turbulence becomes more vigorous which also adds to the mass accretion rate. At the outer boundary, the mass accretion rate is a combination of gas injection at the L1 region and gas outflow at all other regions. Although the gas injection rate stays roughly constant, the outflow rate grows as the disk gets hotter and gas is expelled due to outward transport of angular momentum. The increase in disk temperature is due to our use of adiabatic EOS with specific heat ratio $\gamma=1.1$. The disk is heated up by the dissipation of both spiral shocks and MRI turbulence. At $t \sim 30$, the mass accretion rate at inner and outer boundaries match and the disk reaches quasi-steady state.

\begin{figure}
\centering
\includegraphics[width=0.45\textwidth]{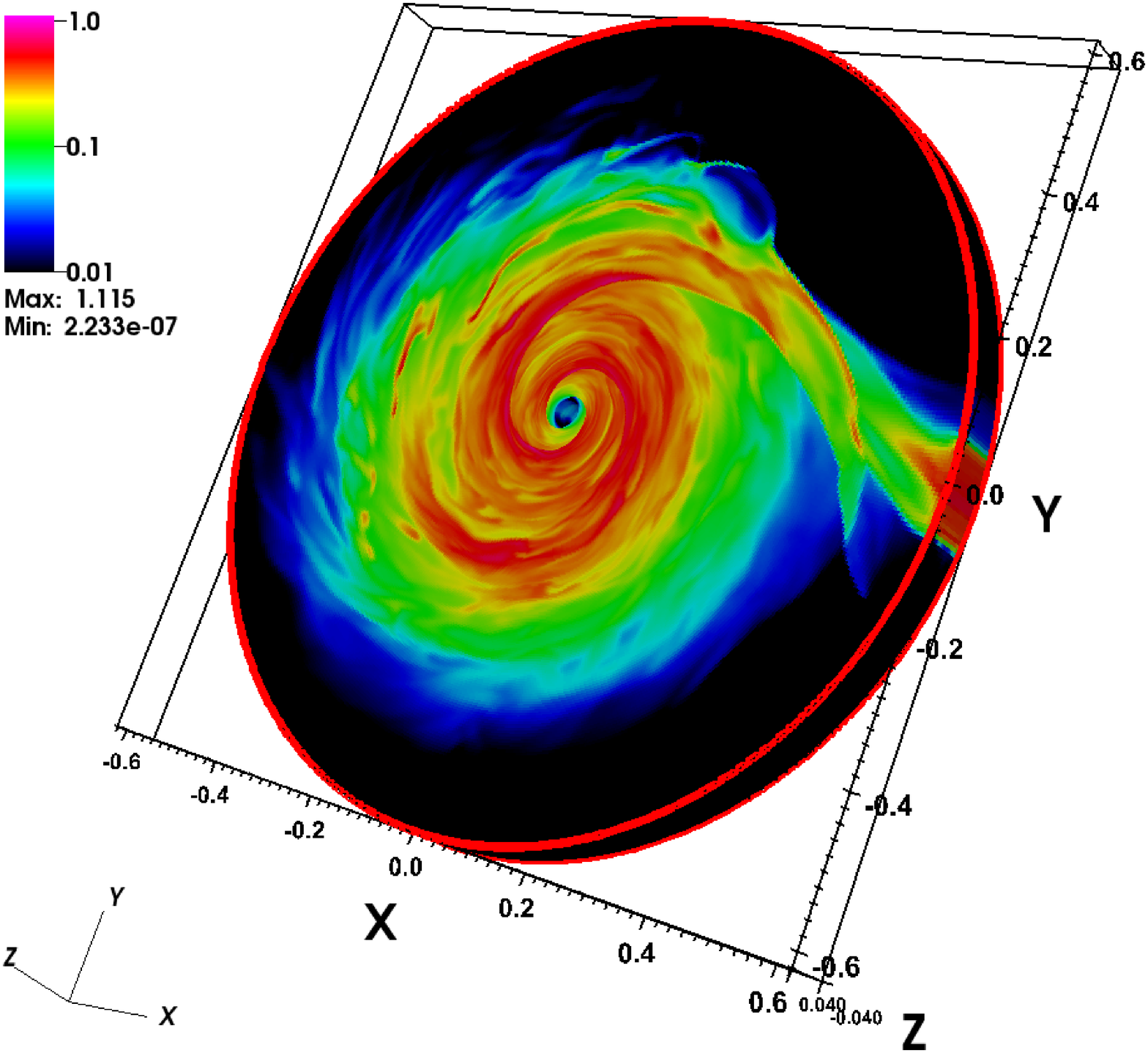}
\includegraphics[width=0.45\textwidth]{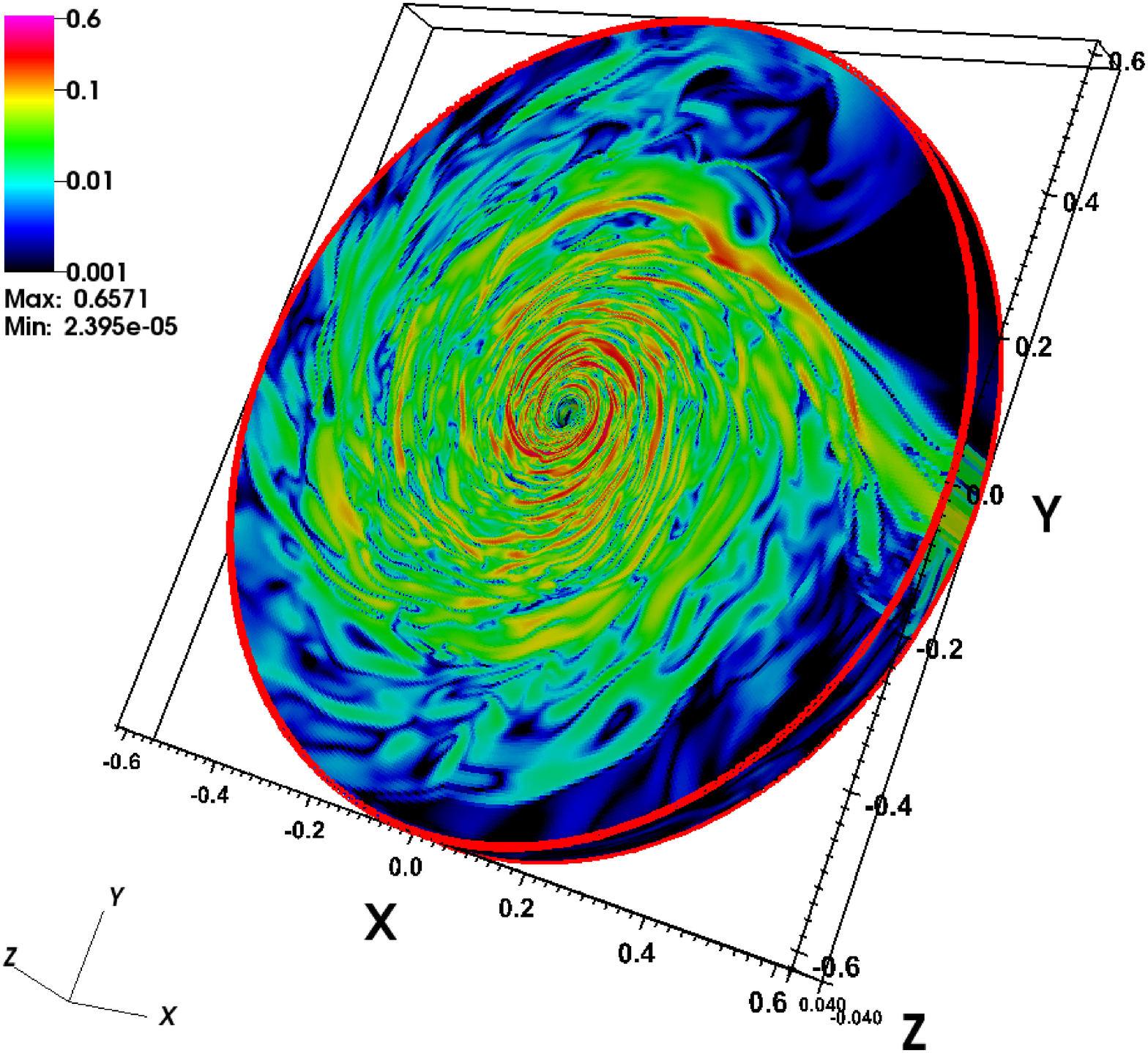}
\caption{Snapshots of gas density (upper panel) and strength of magnetic field (lower panel) in the MHD model at $t=43$ ($\sim 7$ binary orbits). The two-armed spiral pattern is evident with saturated MRI turbulence present in the disk.}
\label{fig:mhd-snapshot}
\end{figure}

In Fig. \ref{fig:mhd-snapshot} we show snapshots of the density (upper panel) and strength of the magnetic field ($|{\bf B}|$, lower panel) in the steady-state disk at $t=43$. The two-armed spiral pattern is clearly evident even with the existence of saturated MRI turbulence. The magnetic $\beta$ averaged around this time period is in the range of [300 - 1000] with variations along radius.
%The white lines on the plots are the Roche lobe surface of the WD.

\begin{figure}
\centering
\includegraphics[width=0.45\textwidth]{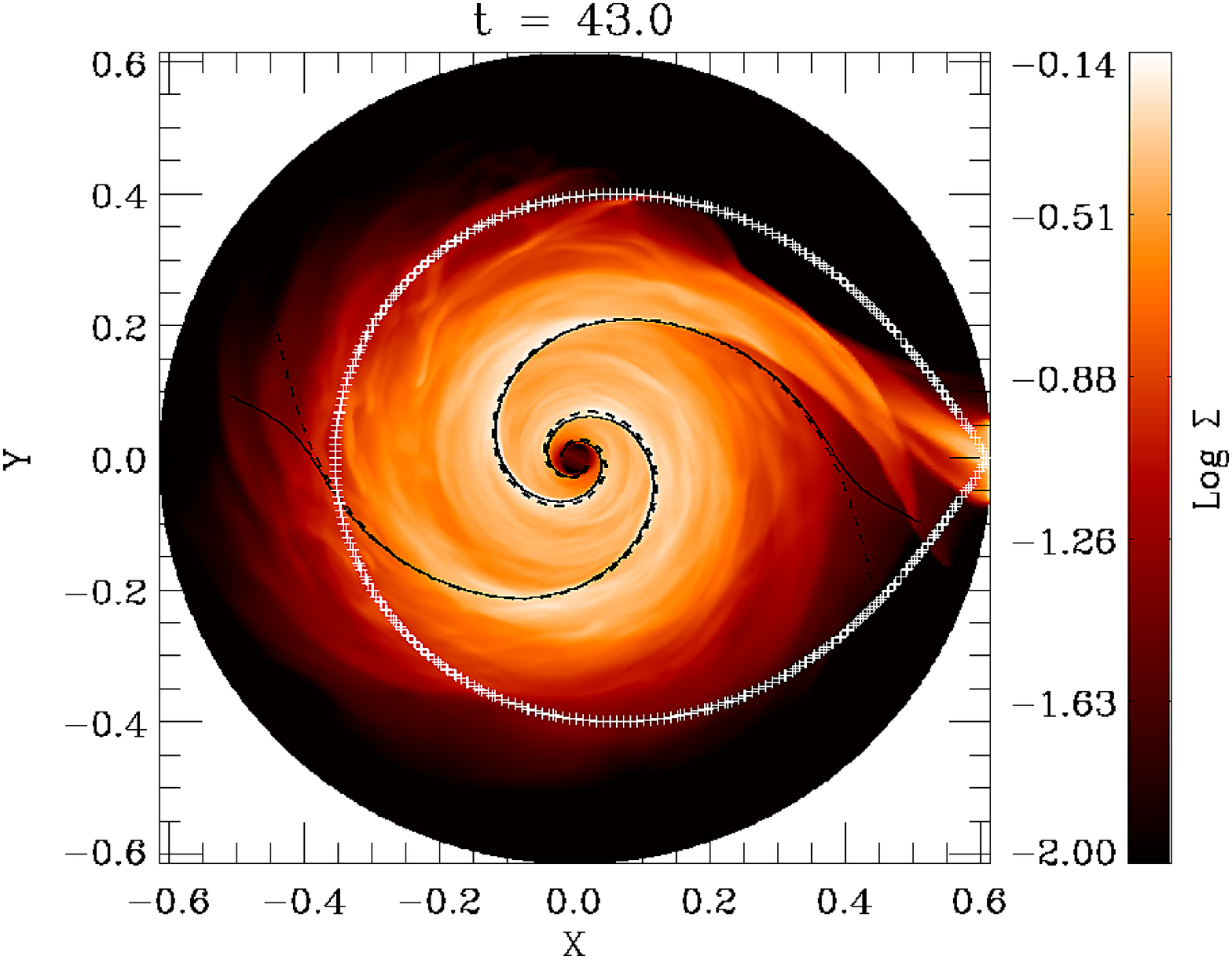}
\includegraphics[width=0.45\textwidth]{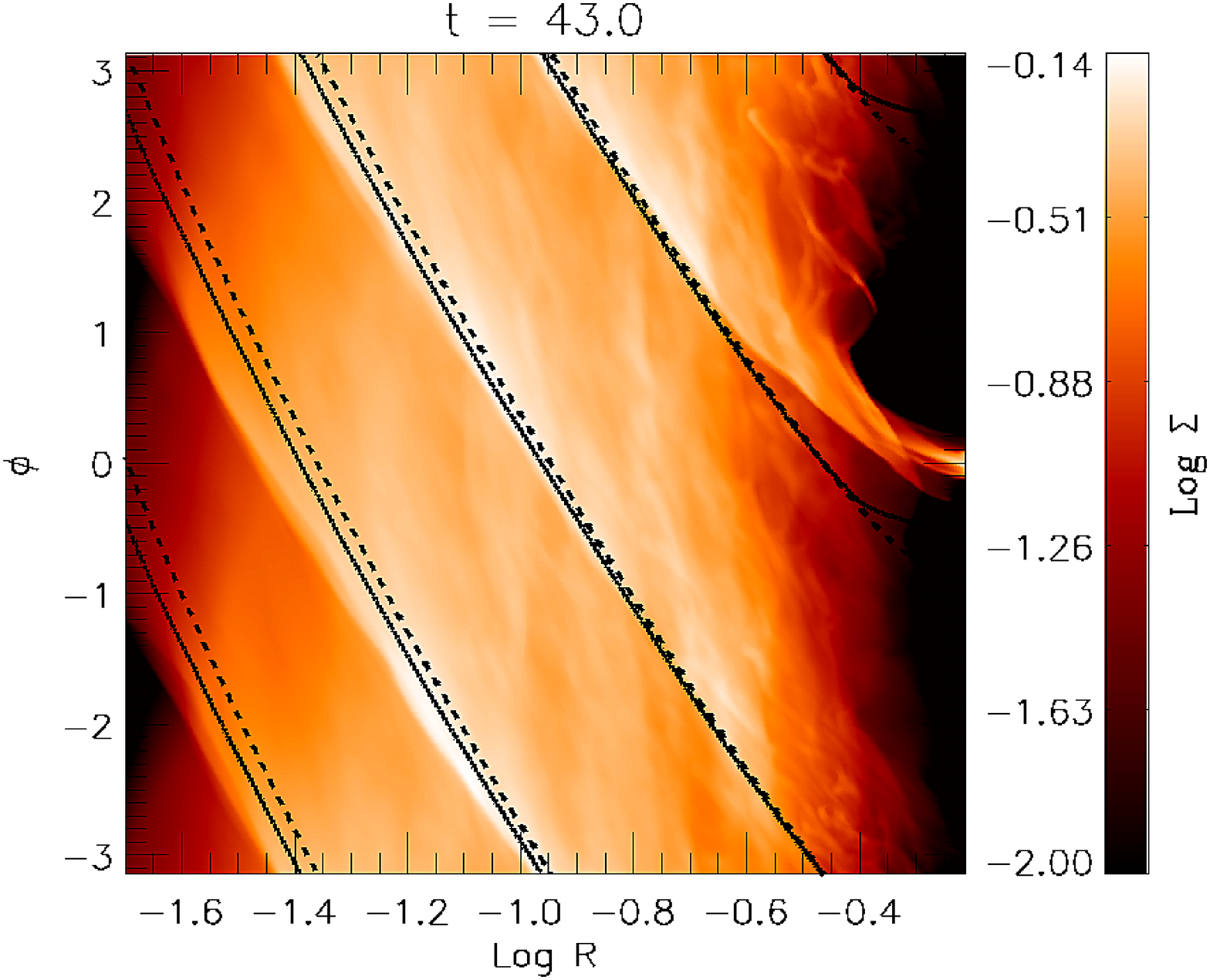}
\caption{Surface density of the MHD model in polar coordinates (upper panel) and in $\log R - \phi$ coordinates (lower panel). The black lines show a fit to the spiral structures using linear dispersion relation with the fast magnetosonic speed $\sqrt{c_s^2+v_A^2}$ (solid line) and sound speed $c_s$ (dashed line). The former gives a better fit.}
\label{fig:mhd-dispersion}
\end{figure}

As we will discuss in more detail in \S \ref{sec:wavepropagation}, the patterns of spiral arms are due to resonant wave propagation excited by the tail of $m=2$ Lindblad resonance and can be understood with linear wave theory. The major difference of this MHD model from the previous hydrodynamical models \citep{1994MNRAS.268...13S, 2000NewA....5...53B, 2000MNRAS.316..906M, 2008A&A...487..671K} and those we present in \S \ref{sec:wavepropagation} is that density waves propagate as fast magnetosonic waves instead of sound waves. In Fig. \ref{fig:mhd-dispersion} we show the surface density of the disk in polar coordinates (upper panel) and in $\log R - \phi$ coordinates (lower panel) repectively. The black lines show a fit to the spiral arm pitch angles using the linear dispersion relation with both the fast magnetosonic speed $\sqrt{c_s^2 + v_A^2}$ (solid line) and sound speed $c_s$(dashed line). Although the Alfven speed is small compared to the sound speed in this model ( $v_A/c_s \approx 0.05 - 0.15$), it makes a clear difference; the dispersion relation for fast magnetosonic waves fits the spiral arms better. This is consistent with work by \citet{2013ApJ...768..143Z} discussing spiral shocks in protoplanetary disks.

To compare the importance of angular momentum transport due to spiral arms and MRI, we compare the time averaged Reynolds stress $T_R$ and Maxwell stress $T_M$ scaled by gas pressure:
% equation of stresses
\begin{equation}
\label{eq:alphar}
\alpha_R  (R)= \frac{2}{3} \frac{<\rho \delta v_R \delta v_\phi>_{\phi, z, t}}{<P>},
\end{equation}
where $ \delta v_R(R, \phi, z, t) = v_R(R, \phi, z, t) - <v_R>_{\phi, z} (R, t)$, and
\begin{equation}
\label{eq:alpham}
\alpha_M (R) = \frac{2}{3}\frac{<- B_R B_\phi>_{\phi, z, t}}{<P>}.
\end{equation}
The constant factor $2/3$ is to be consistent with the definition of $\alpha$ in standard $\alpha$ disk model \citep{1973A&A....24..337S}. 

\begin{figure}
\centering
\includegraphics[width=0.49\textwidth]{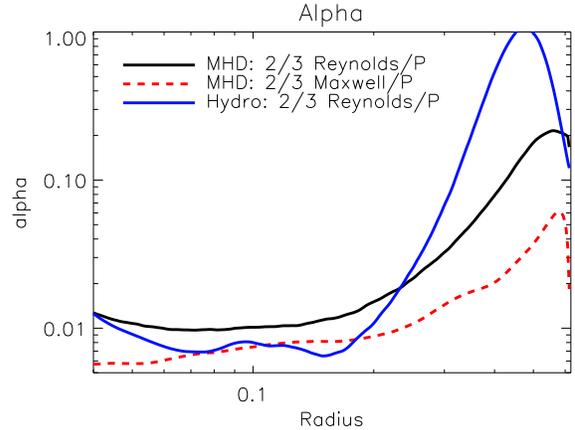}
\caption{Time-averaged (over time 35 to 60) radial profiles of effective $\alpha_R$ (solid black line, Eq.\ref{eq:alphar}) and $\alpha_M$ (dashed red line, Eq.\ref{eq:alpham})of the MHD model, and $\alpha_R$ (solid blue line) of an equivalent 2D hydro model (run"Hydro-inflow-adiabatic $\gamma=1.1$" in Table \ref{tab:parameter}). $\alpha_R$ dominates over $\alpha_M$ in the MHD model, which indicates spiral waves play a major role in angular momentum transport. $\alpha_R$ in the MHD model is larger than that in the equivalent hydro model within the region of $R<0.25$ where the majority of the disk resides, which implies that spiral shocks may be enhanced by MRI.}
\label{fig:mhd-stress}
\end{figure}

Fig. \ref{fig:mhd-stress} shows $\alpha_R$ and $\alpha_M$ averaged over time 35 to 60 of the MHD model (black and red lines), as well as $\alpha_R$ from an equivalent 2D hydro model (run ``Hydro-inflow-adiabatic $\gamma=1.1$" in Table \ref{tab:parameter}) with the same configuration (blue line). In the MHD model, $\alpha_R$ is larger than $\alpha_M$, implying that spiral shocks may be the dominant source of angular momentum transport. This result is in stark contrast to previous local shearing-box MHD simulations \citep{1995ApJ...440..742H} or global MHD simulations \citep{2001ApJ...554..534H, 2002ApJ...573..749B, 2012ApJ...749..189S} of disks under axisymmetric gravitational potentials, where the Maxwell stress is usually several times larger than the Reynolds stress. The major reason of this difference is that, in CV disks, the spiral shocks excited by the non-axisymmetric potential induce more efficient redistribution and outward transport of angular momentum than that induced by pure hydrodynamical turbulence in axisymmetric disks. The Reynolds stress here not only includes angular momentum transport by local turbulence, but more importantly includes transport by spiral shock dissipation.

Note that $\alpha_R$ in the MHD model is larger than that in the hydro model at $R<0.25$ where the majority of the disk resides. This suggests that somehow the MRI enhances the strength of spiral shocks. A possible reason is that the MRI induces viscous spreading of the disk. The increased size of the disk induces much larger excitation of spiral density waves due to overlap with resonances (see \S \ref{sec:waveexication}). Even a subtle increase of the disk size can greatly enhance the strength of spiral shocks in the disk.

An important caveat of this result is that, in this particular MHD model with adiabatic EOS, the disk temperature is quite high, giving an average Mach number in quasi-steady state of around 5-10. This is consistent with Mach numbers of previous hydrodynamical simulations of CV disks \citep[e.g.][]{2000MNRAS.316..906M, 2000NewA....5...53B, 2008A&A...487..671K}. However, in a realistic CV disk at outburst, the Mach number should be around 60 according the thermal limit cycle models \citep{cannizzo93, Gammie-Menou1998}. As we will discuss in \S \ref{sec:wavepropagation}, as the disk temperature decreases, the pitch angle of spiral arms becomes smaller (spirals become tighter) and the strength of spiral shocks decreases. Therefore, results from this particular model should be taken with caution. In a model with more realistic temperature, the spiral shocks may play a less important role.

Nevertheless, it is clear that in order to understand the relative importance of spiral shocks and MRI in angular momentum transport in CV disks, and their interaction during outbursts, first we need to understand thoroughly the excitation, propagation and dissipation of spiral shocks. In the following sections, we focus on 2D hydro models to study the formation of spiral shocks and its mechanism of angular momentum transport in such disks.

%%%%%%%%%%%%%%%%%%%%%%%%%%%%%%%%%%%%%%%%%%%%%%%%%%%%%%%%%%%%%%%%%
%%%%%%%%%%%%%%%% Results Hydro %%%%%%%%%%%%%%%%%%%%%%%%%%%%%%%
%%%%%%%%%%%%%%%%%%%%%%%%%%%%%%%%%%%%%%%%%%%%%%%%%%%%%%%%%%%%%%%%%
\section{The Hydrodynamics of CV Disks}
\label{sec:result-noinflow}

%% Our hydro models
In this section, we use 2D hydrodynamical simulations of CV disks to investigate the excitation, propagation and non-linear dissipation of spiral density waves. We use two sets of hydrodynamical models. In the first set of models, we start with uniform gas density filling the whole computational domain, and do not include an accretion stream from the L1 point (the ``no-inflow model). The purpose of excluding the accretion stream in this model is to separate the spiral structures excited by the non-axisymmetric potential from those associated with the impacts of the accretion flow. In the second set of models, we start with near vacuum represented by a floor density, and inject gas at the L1 point (the ``with-inflow model"). The parameters of all models are listed in Table \ref{tab:parameter}. In both models, the spiral arms are quickly excited within one orbit of the binary, and reach quasi-steady state at around 7 orbits. There is no strict steady states in most of our models, because the no-inflow models have net accretion but no mass supply, while the with-inflow models have much slower mass accretion rates than the mass inflow rates except the $\gamma=1.3$ case where the accretion and inflow rates match and a steady state is reached. 

\begin{figure*}
\centering
\includegraphics[width=0.49\textwidth]{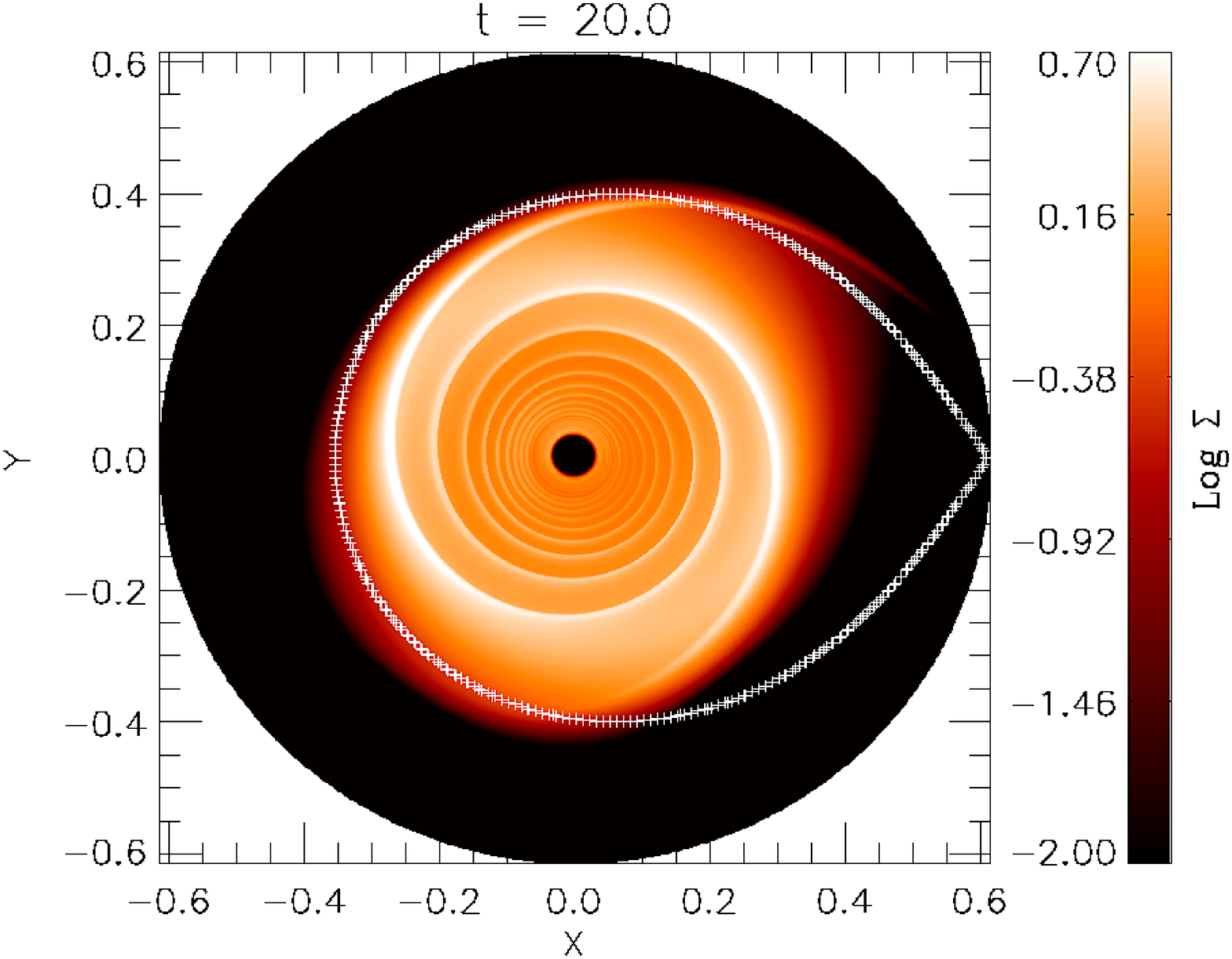}
\includegraphics[width=0.49\textwidth]{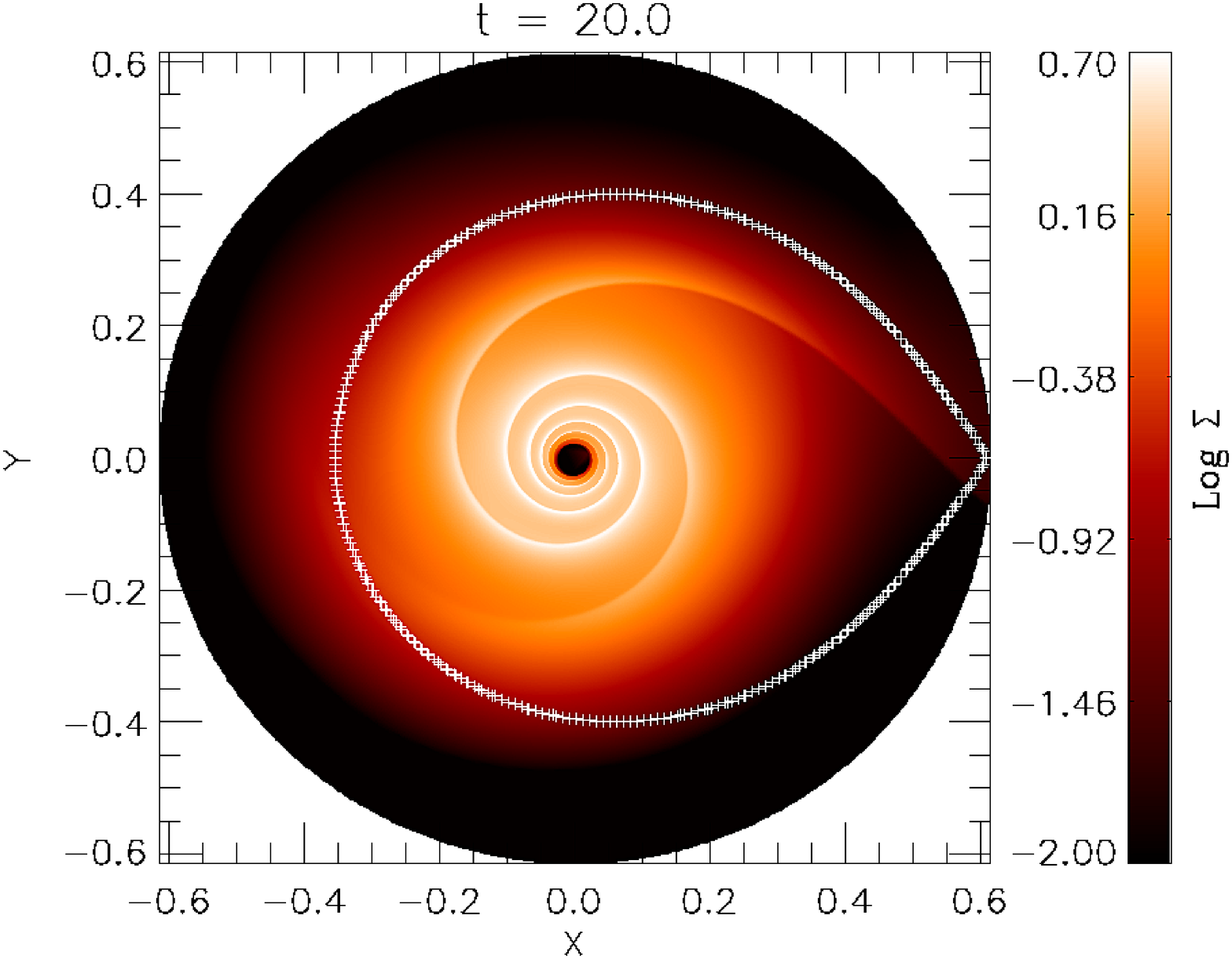}
\includegraphics[width=0.49\textwidth]{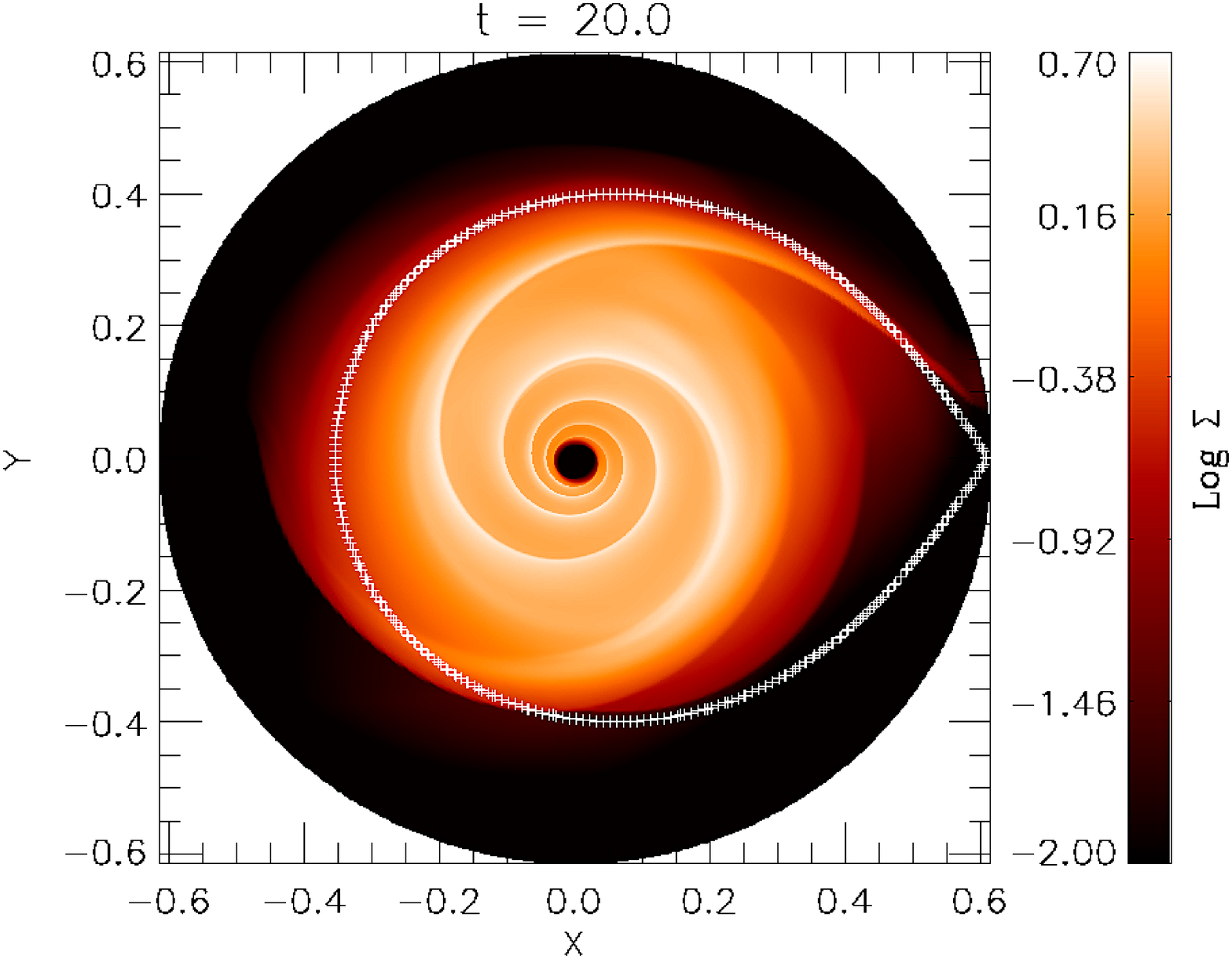}
\includegraphics[width=0.49\textwidth]{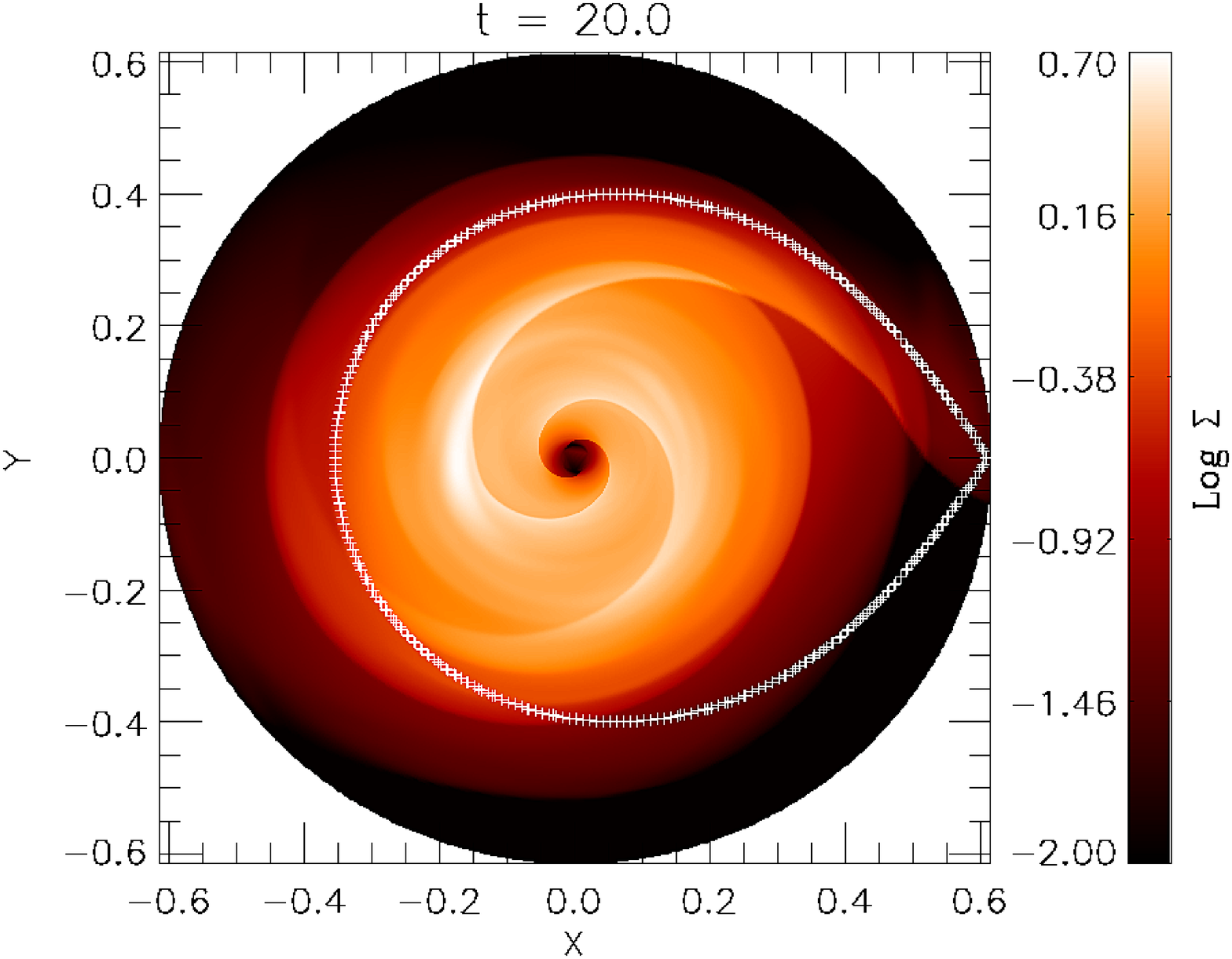}
\includegraphics[width=0.49\textwidth]{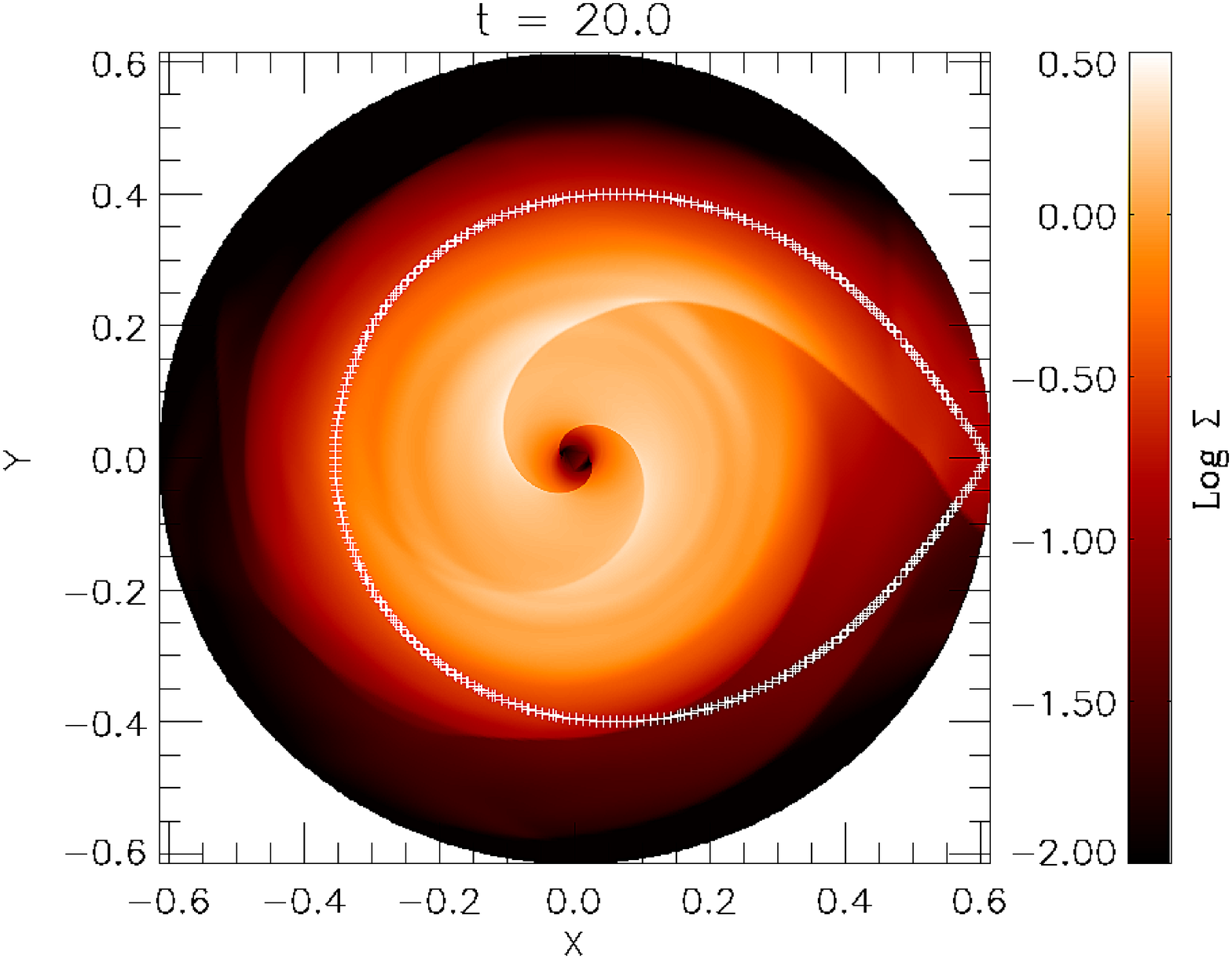}
\caption{Snapshots of surface density of the no-inflow hydro models at $t =20$($\sim 3$ binary orbits). The top panels show isothermal models with sound speed $c_s=0.1$ and $c_s=0.3$ respectively. The bottom three panels show adiabatic models with specific heat index $\gamma = 1.1, 1.2, 1.3$. White lines show the Roche lobe surface of the WD. The spiral patterns are wound more openly as $c_s$ or $\gamma$ increases, which can be explained by linear dispersion relation (see \S \ref{sec:wavepropagation}).}
\label{fig:noinflow-den2d}
\end{figure*}

\begin{figure*}
\centering
\includegraphics[width=0.49\textwidth]{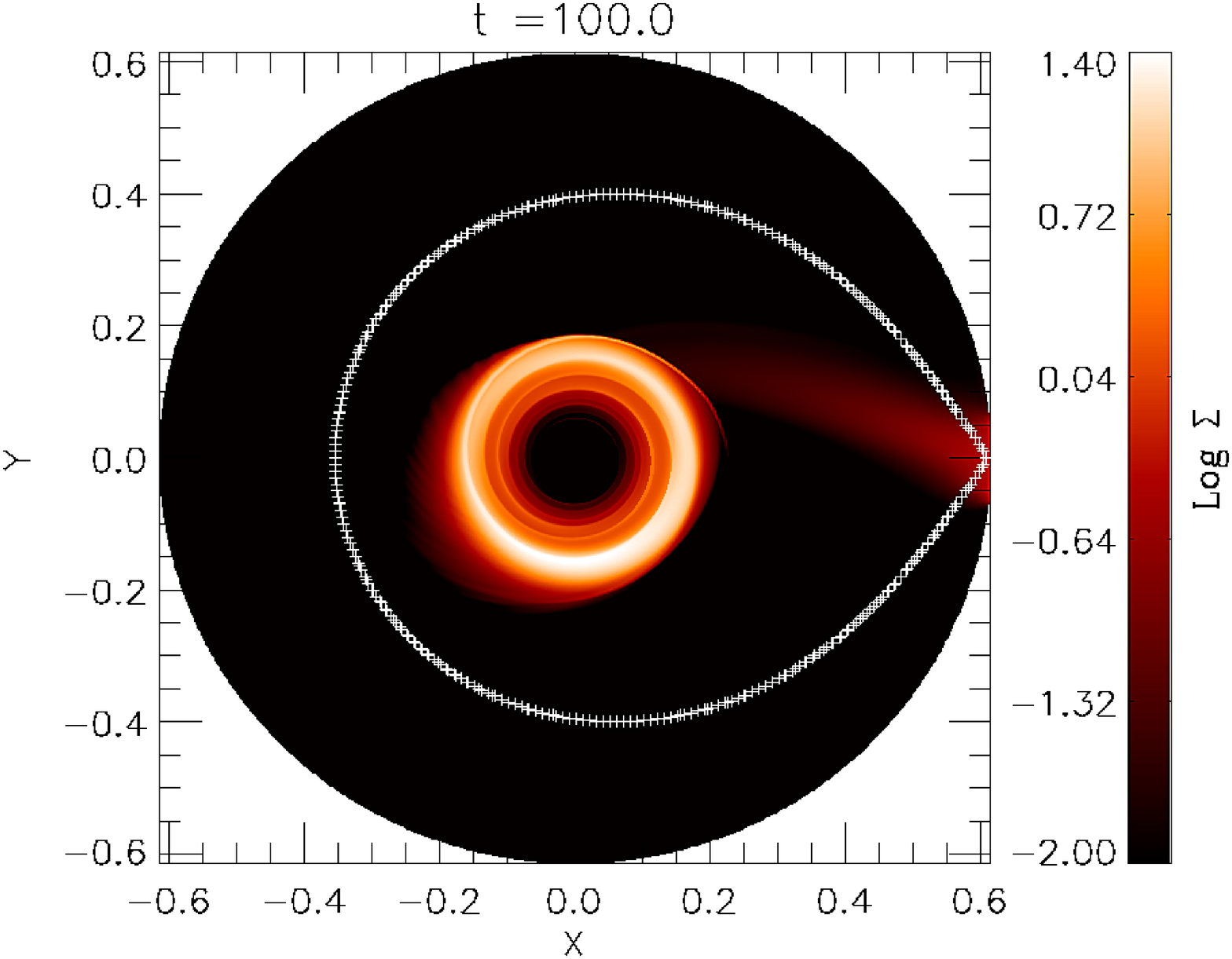}
\includegraphics[width=0.49\textwidth]{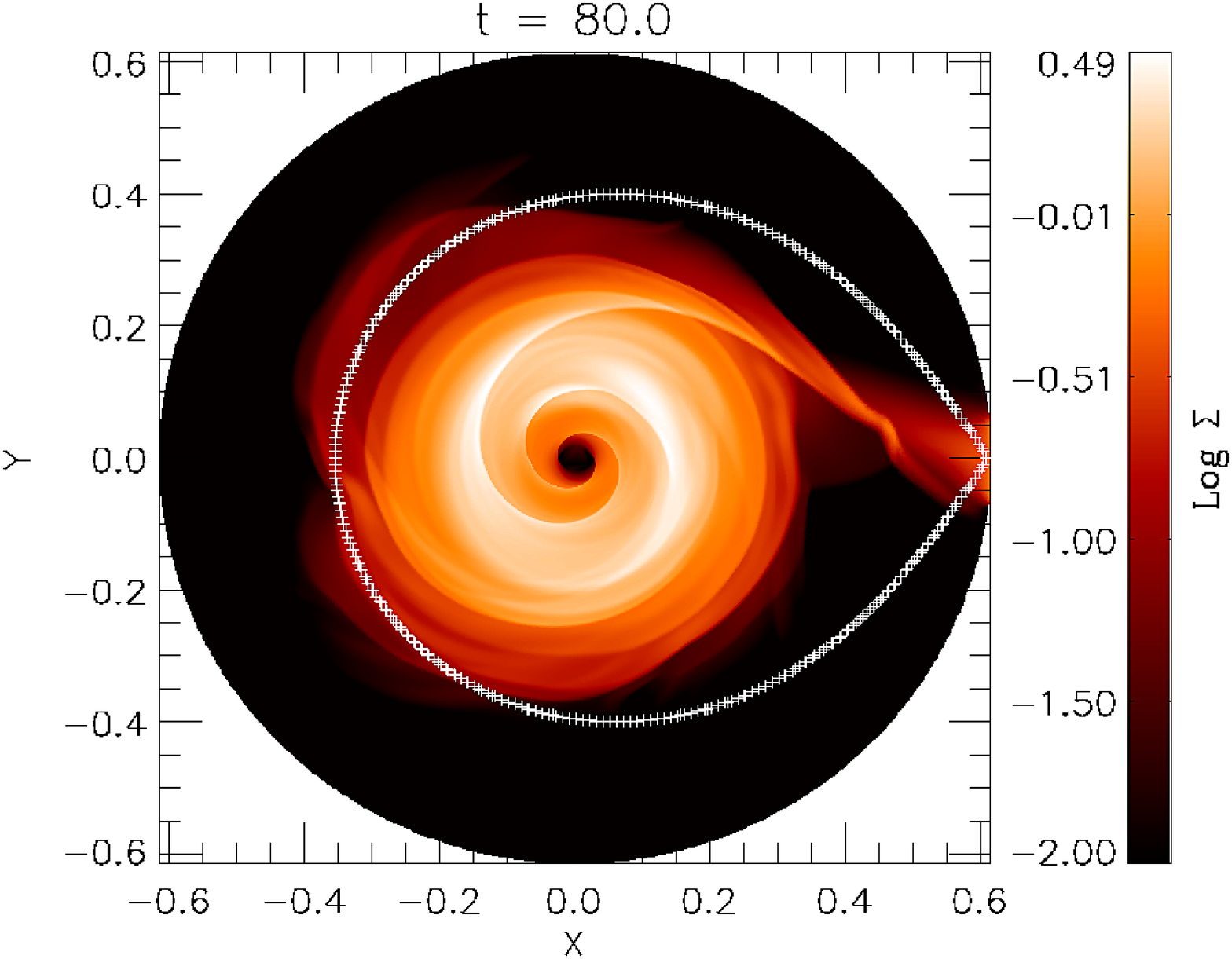}
\includegraphics[width=0.49\textwidth]{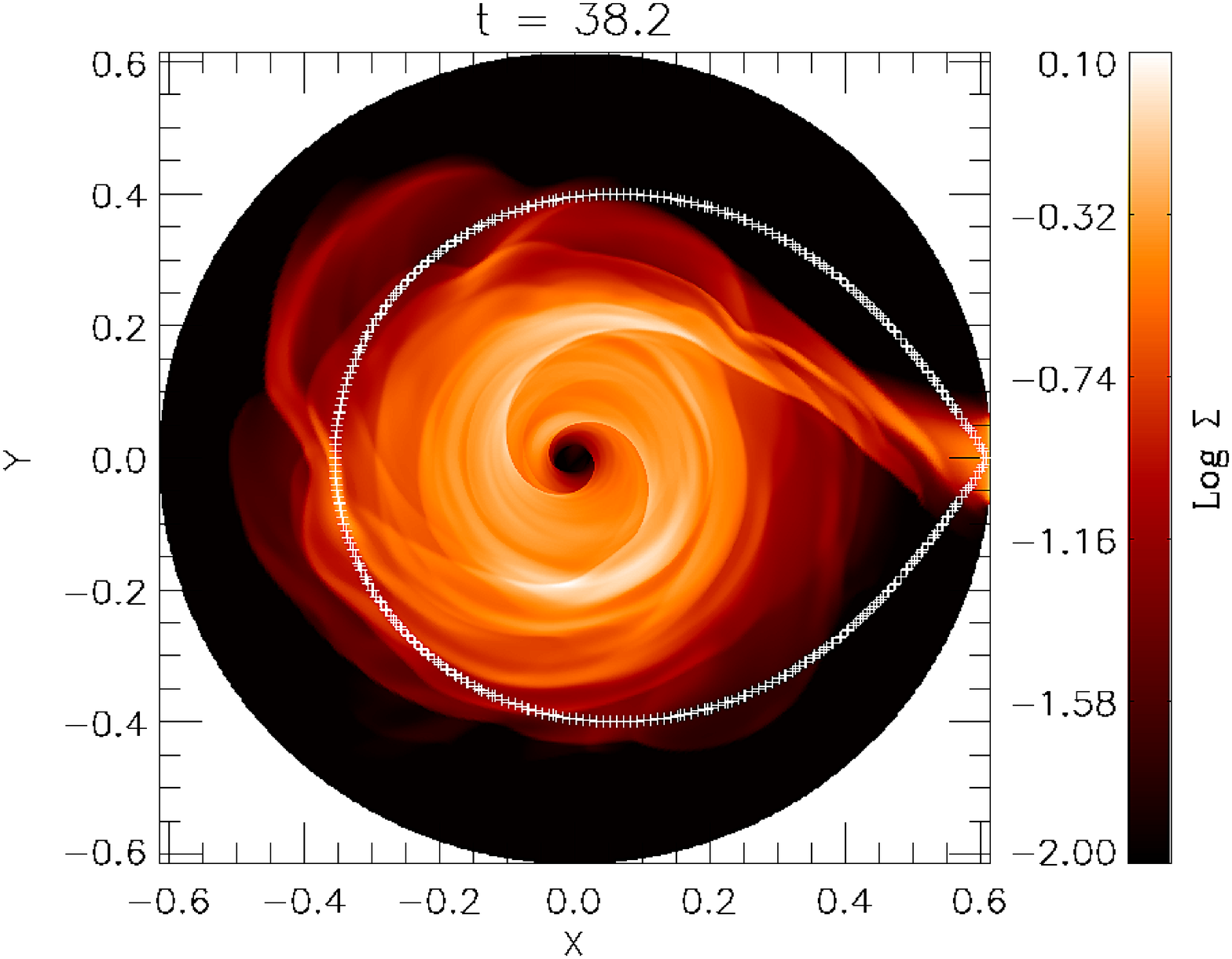}
\includegraphics[width=0.49\textwidth]{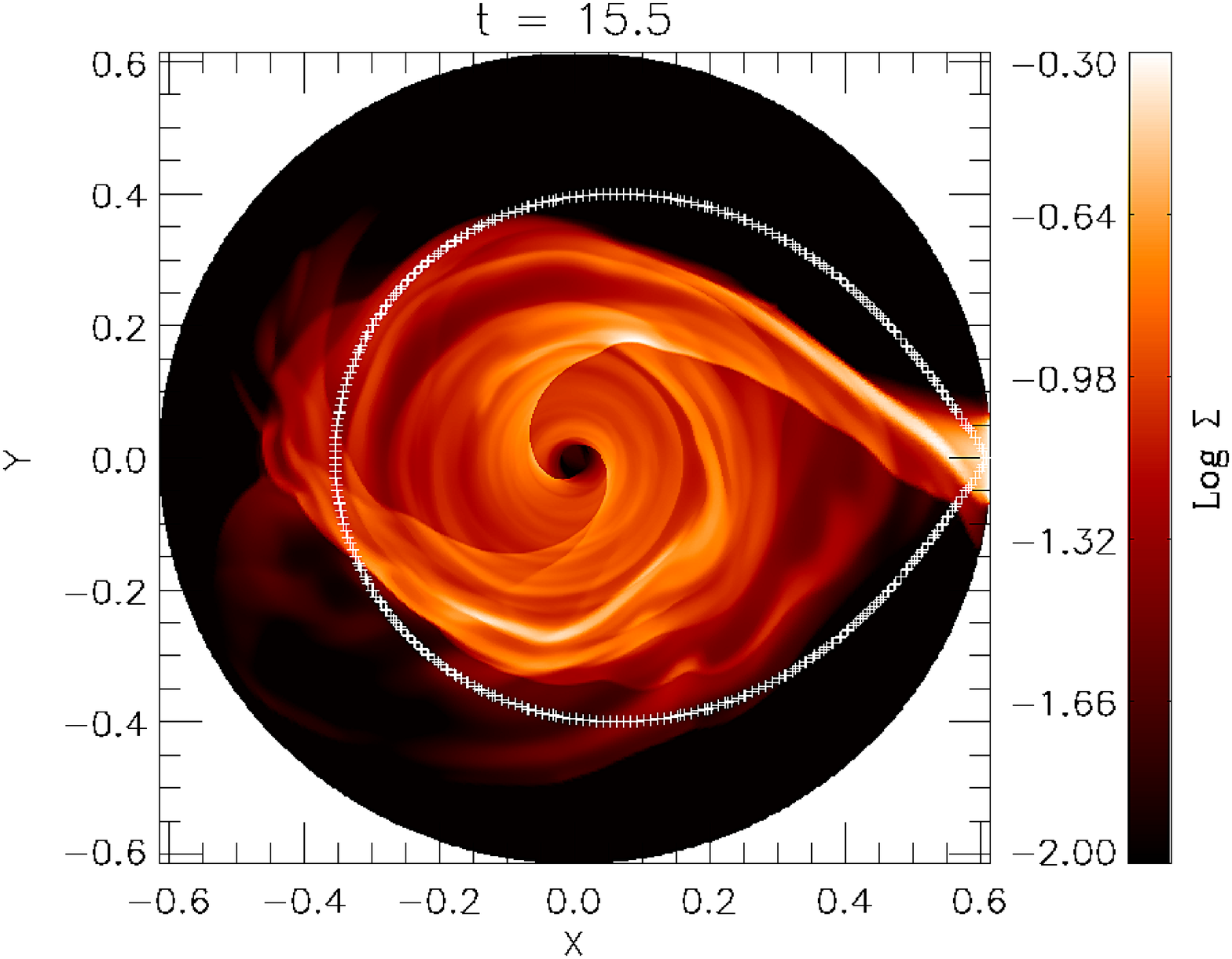}
\caption{Snapshots of surface density of the with-inflow hydro models. The top left panel shows an isothermal model with sound speed $c_s=0.1$. The rest three  panels show adiabatic models with specific heat index $\gamma = 1.1, 1.2, 1.3$ respectively. White lines show the Roche lobe surface of the WD. Similar to the no-inflow models, the spiral patterns are wound more openly as $\gamma$ increases, which can be explained by linear dispersion relation (see \S \ref{sec:wavepropagation}). Compared with no-inflow models, the with-inflow models have disks  that are more turbulent and smaller in size.}
\label{fig:inflow-den2d}
\end{figure*}

In Fig. \ref{fig:noinflow-den2d} and Fig. \ref{fig:inflow-den2d} we show the surface density profiles of quasi-steady states of the no-inflow and with-inflow models respectively. The panels in each figure use a different EOS: isothermal EOS with sound speed $c_s=0.1$ (and $c_s=0.3$ in no-inflow model) and adiabatic EOS with specific heat ratio $\gamma=1.1, 1.2, 1.3$. A clear trend is that spiral patterns are wound more openly as the disk temperature or $\gamma$ is increased. Moreover, comparing the two models shows that the major effects of the accretion stream are: 1) the with-inflow disks are more turbulent; 2) the with-inflow scenarios have smaller disk sizes. The former is likely driven by hydrodynamical instabilities associated with thin shock waves \citep[e.g.][]{1983ApJ...274..152V}. The latter is due to different initial conditions in the two scenarios: in the no-inflow model we start with uniform Keplerian flow that includes gas with a larger initial specific angular momentum; whereas in the with-inflow model, the disk forms self-consistently from vacuum with the size of the disk depending on processes that transport angular momentum outward, e.g. spiral shock dissipation. The with-inflow model with isothermal EOS has the smallest disk because it has the weakest spiral shock dissipation. In the following sections, we perform detailed investigations of the spiral patterns from the perspectives of wave excitation (\S \ref{sec:waveexication}), propagation (\S \ref{sec:wavepropagation}), shock dissipation (\S \ref{sec:shockdissipation}) and energy budget (\S \ref{sec:energybudget}) respectively, and also present a convergence study in \S \ref{sec:convergence}. 

%Spiral arms have been observed in CV disks (e.g. in IP Peg, Harlaftis et al. 1999) using the techniques of eclipse mapping and Doppler tomography. 

% snapshots of no-inflow models

% snapshots of with-inflow models

%%%%%%%%%%
\subsection{Wave Excitation}
\label{sec:waveexication}

%% Difference between CV disk and protoplanetary disk
Linear theory predicts that density waves are excited at the Lindblad resonances in the context of protoplanetary disks, and the satellite exerts a torque on the disk only in the immediate vicinity of its Lindblad resonances \citep{1980ApJ...241..425G}. However, in a CV disk, since the mass ratio of the binary stars is of order unity, strong tidal effects truncate the disk far from the Lindblad resonances. For example, in our simulations where the mass ratio is $q=0.3$, the position of the inner 2:1 resonance of the companion star is at $R=0.57 a_0$ ($a_0$ being the binary separation) while higher order resonances are further out, whereas the disk is usually truncated at around $R \sim 0.35 a_0$. 

%% Resonances have finite widths
In this case, could density waves still be excited by Lindblad resonances? \citet{2012ApJ...747...24R} found the torque density exerted by the satellite is not localized in a narrow region around each Lindblad resonance, but rather spreads out over a finite width. The spatial distribution of the perturbations in a homogeneous disk can be approximated with Airy functions, where the perturbation wavelength is related to the local scale height, and amplitude drops exponentially with distance from the resonance. 

%% Savonije, Papaloizou \& Lin (1994) conducted linear response calculation of wave excitation in binary systems. 
In a realistic CV disk, the density profile is not homogeneous due to tidal truncation. The total torque exerted on the disk should then be the convolution of the perturbation for a homogeneous disk, the realistic surface density profile, as well as the gravitational force profile which can change the perturbation amplitude on a length scale much larger than the wave perturbation. \citet[SPL94 hereafter]{1994MNRAS.268...13S} presented linear response calculations of tidal interaction in binary systems, and found the total torque exerted on the disk (i.e the angular momentum transferred away from the disk) in a WKB approximation is (see their Equation 13):
\begin{equation}
T_{tot} \propto \left | \int_{R_{min}}^{R_{max}} \Sigma_0 r S X dr \right |^2
\end{equation}
where $\Sigma_0$ is the unperturbed surface density profile, X is the wave perturbation in the disk without forcing by the companion star, and S is related to the $m=2$ component of the gravitational force from the companion star.

%% Prediction of linear theory on how torques scale with Mach number and disk size.
The total torque exerted on the disk strongly depends on two factors: Mach number $\mathcal{M}$ and size of the disk. If the disk has a large Mach number, the wave perturbations have shorter wavelengths. Considering the large length scale over which the binary gravitational force varies, cancellation effects tend to lead to a smaller residual total torque than that exerted on a disk with smaller Mach numbers. On the other hand, if the disk is smaller in size, then most of the gas can only overlap with the tail of the wave perturbations with smallest amplitudes far away from the Lindblad resonance, and a much smaller total torque is exerted as a result.

%% Numerical test and results 
We have performed idealized numerical tests similar to SPL94 and confirmed the trends. Without adding an accretion stream from the L1 point, we start from a truncated Keplerian disk that initially is in radial hydrostatic equilibrium, and allow it to evolve self-consistently with isothermal equation of state and a binary mass ratio of $q=0.3$. In Table \ref{table:torques} we show the total torques measured from quasi-steady states of our simulations with various sound speeds and disk sizes. The values are time-averaged and normalized by disk mass. With the disk sound speed fixed, the torque drops quickly when the disk size decreases from 0.3 to 0.2. With the initial disk size fixed instead, the torques drops when the disk sound speed decreases (i.e. Mach number increases). While these trends are qualitatively consistent with linear response theory mentioned above, there are complications with such comparisons. For example, we do not find a linear scaling between the total torque and $1/\mathcal{M}$ as SPL94 found. In our simulations when $\mathcal{M}$ decreases, the disk is more radially extended due to the higher gas temperature. The extension makes the disk outer edge closer to the Lindblad resonance, and so leads to a larger torque, which is an effect independent of that from decrease of $\mathcal{M}$. SPL94 found the linear scaling instead because they cut their computational domain according to the size of disk they set initially, while we always cover the whole domain within the radius of L1 point.

\begin{deluxetable}{ccc}
\tablecolumns{3} 
\tabletypesize{\small}
\tablewidth{20pc}
\tablecaption{Total torques\label{table:torques}}
\tablehead{ 
\colhead{Sound Speed $c_s$} & \colhead{Initial Disk Size} & \colhead{Total Torque} \\ 
\colhead{(1)} & \colhead{(2)} & \colhead{(3)}
}
\startdata
0.1 & $\sim 0.3$   & $7.2\times 10^{-5}$ \\
0.1 & $\sim 0.25$ & $2.5\times 10^{-5}$ \\
0.1 & $\sim 0.2$   & $3.8\times 10^{-6}$ \\
\hline
0.03 & $\sim 0.25$   & $\sim 1\times 10^{-7}$ \\
0.1 & $\sim 0.25$   & $2.5\times 10^{-5}$\\
0.3 & $\sim 0.25$   & $2.3\times 10^{-4}$ 
\enddata
\end{deluxetable}

%%%%%%%%%%
\subsection{Wave Propagation and Spiral Structures}
\label{sec:wavepropagation}

%%%
In Fig. \ref{fig:noinflow-den2d} we show snapshots of the CV disk at quasi-steady states at $t=20$ with various equation of states (EOS): isothermal EOS with $c_s =0.1$ and $c_s=0.3$, and adiabatic EOS with specific heat ratio $\gamma=1.1, 1,2, 1.3$. We observe the trend that pitch angles of spiral arms increase, i.e. the spirals are more openly wound, as the gas specific heat ratio $\gamma$ increases. This is consistent with previous numerical work such as SPL94 and \citet{2000MNRAS.316..906M}. \citet{1987A&A...184..173S} analytically solved equations of motion of a thin disk with a disturbance at outer edge and with self-similar shocks, and obtained a unique relation between the opening angles of the shock and the specific heat ratio $\gamma$ (see his Eq. 61):
\begin{equation}
\tan^2 \tilde{\theta} = \frac{2}{\gamma -1} - 2 \frac{n_r +1}{\gamma} -1
\end{equation} 
where $\tilde{\theta}$ is the angle between the shock and radial direction, which is complementary to the opening angle of the shocks, and $n_r$ is the number of shocks assumed. This relation shows the shocks are more openly wound as $\gamma$ increases, which is consistent with our results. However, this relation also indicates $\tilde{\theta}=90^{\circ}$ (i.e. zero opening angle) when $\gamma =1$ which is equivalent to an isothermal EOS. Instead, we find the spirals have non-zero opening angles in our isothermal simulations with $c_s =0.1$ and $c_s=0.3$.

The spiral patterns can be understood from the perspective of wave propagation \citep{2002MNRAS.330..950O}. The spiral over-dense regions in the disk are wave fronts excited at Lindblad resonances. Therefore, if the non-linear shock dissipation is weak, the spiral patterns in our simulations should follow the linear dispersion relation for tightly wound hydrodynamic waves in a two-dimensional gas disk:
\begin{equation}
\label{eq:dispersionrelation}
[m(\Omega - \Omega_p)]^2 = \kappa^2 + c_s^2 k^2,
\end{equation}
where $m$ is the azimuthal wavenumber, $\Omega(r)$ is the local angular frequency of gas, $\Omega_p$ is the pattern angular frequency, $\kappa(r)$ is the epicyclic frequency, and $k(r)$ is the radial wavenumber. In our case, the disk is nearly Keplerian, so that $\kappa \approx \Omega$; The pattern angular frequency is determined by the rotation of the companion star, so that $\Omega_p = \Omega_0 = 1$.

If we assume any quantity X of the linear waves follows the form of 
\begin{equation}
X(r, \phi, t) = \Re [C(r) \exp[i \Phi(r, \phi, t)]],
\end{equation}
where $(r,\phi,t)$ is the position and time of interest, $C$ is the amplitude of the wave and $\Phi$ is the phase of the wave 
\begin{equation}
\Phi (r,\phi,t) = \int k(r) dr + m(\phi - \Omega_p t).
\end{equation}
Then one spiral arm at a specific $t$, which consists of all the positions with the same phase $\Phi$, should satisfy the following relation
\begin{equation}
\frac{d \phi}{d r} = - \frac{k}{m} = - \frac{1}{c_s} \sqrt{(\Omega - \Omega_p)^2 - \kappa^2 / m^2}
\end{equation}
Therefore, the pitch angles of a trailing spiral arm satisfy
\begin{eqnarray}
\label{eq:angle}
tan (\theta) &=& -\frac{dr}{r d\phi}=\frac{c_s}{R \sqrt{(\Omega - \Omega_p)^2 - \kappa^2 / m^2}} \nonumber \\
&=& \frac{1}{\mathcal{M}} \frac{1}{\sqrt{(1-\Omega_p/\Omega)^2 - 1/m^2}}.
\end{eqnarray}
This relation indicates that the local pitch angles should only depend on the local gas temperature and azimuthal velocity.

\begin{figure*}
\centering
\includegraphics[width=0.33\textwidth]{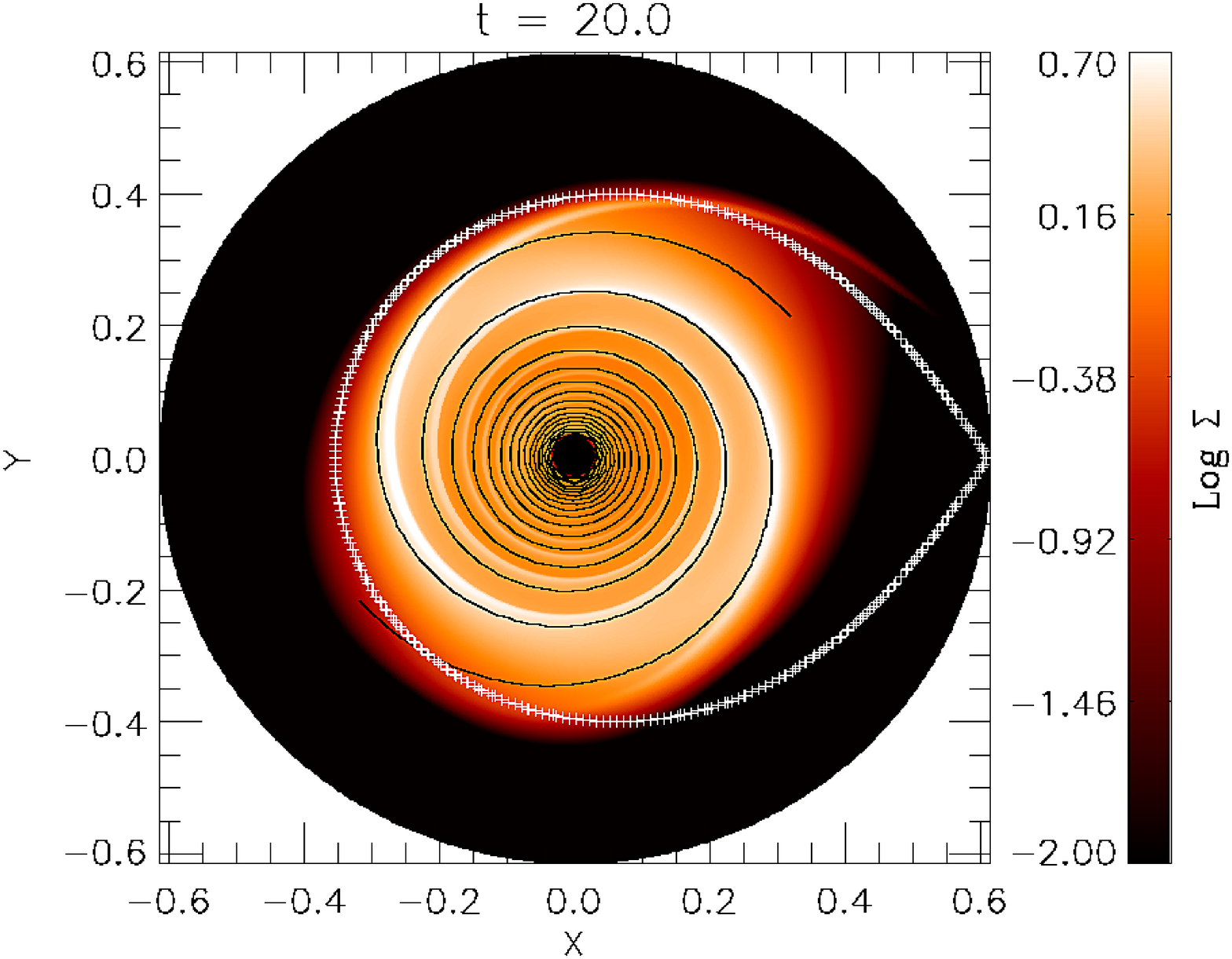}
\includegraphics[width=0.33\textwidth]{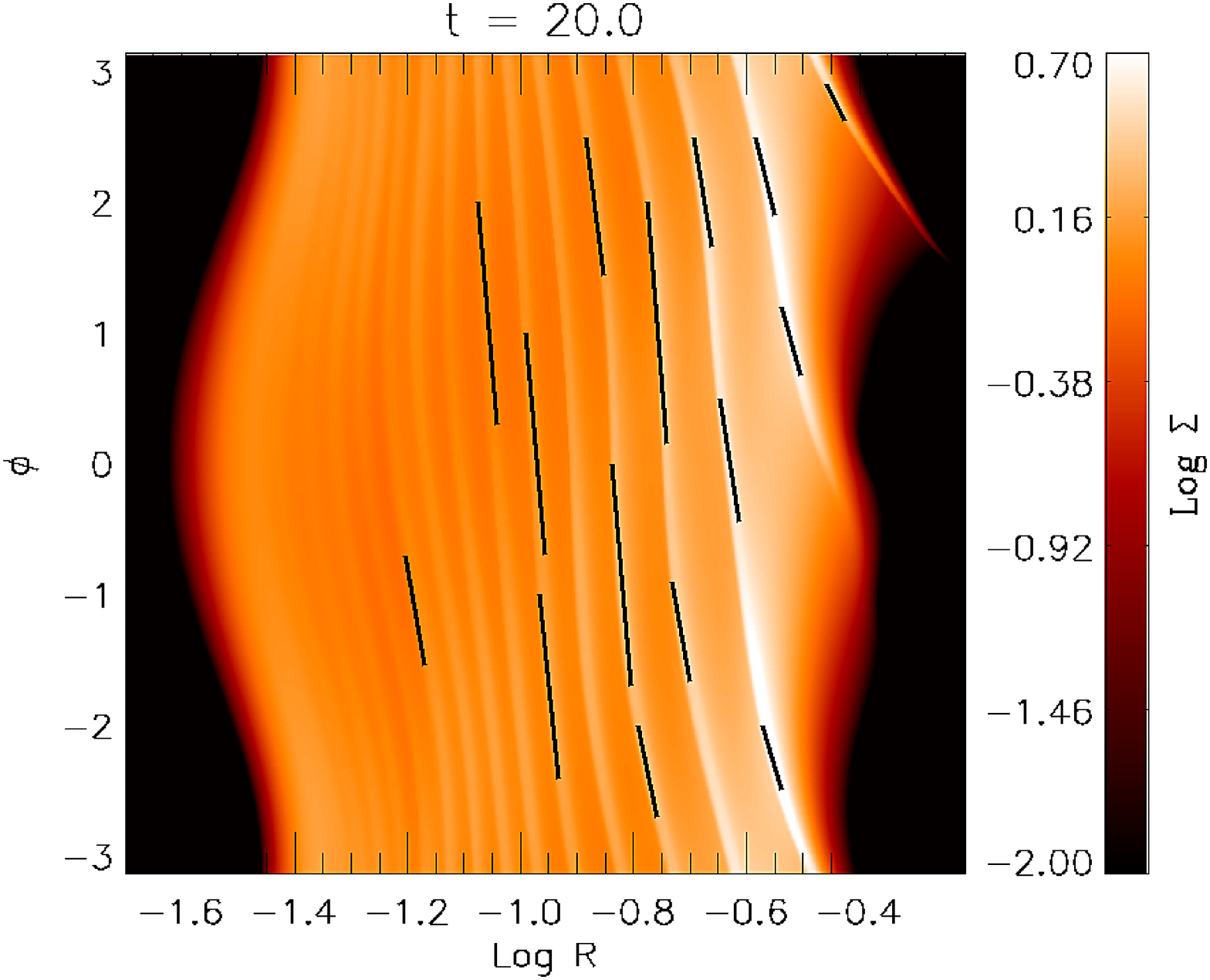}
\includegraphics[width=0.33\textwidth]{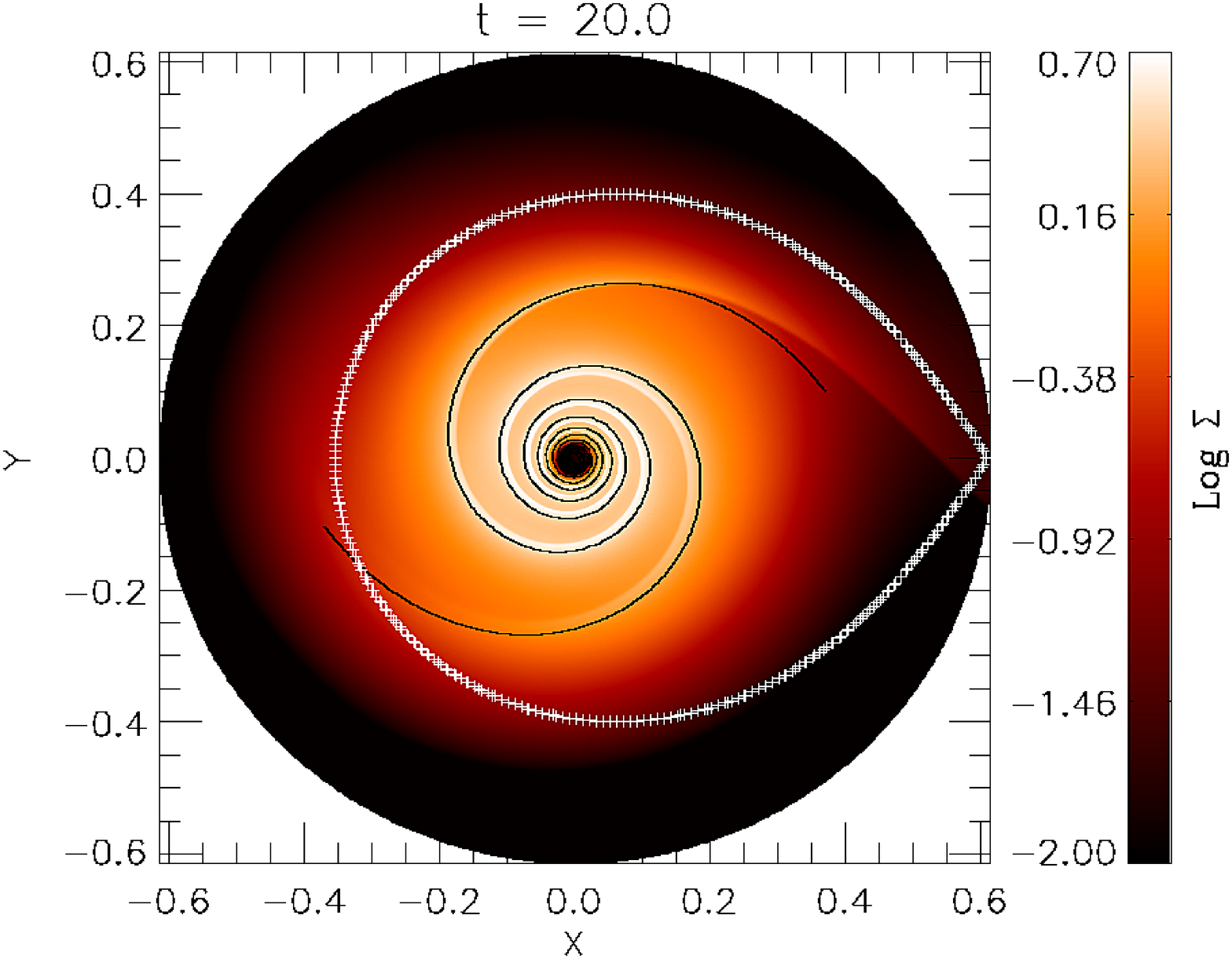}
\includegraphics[width=0.33\textwidth]{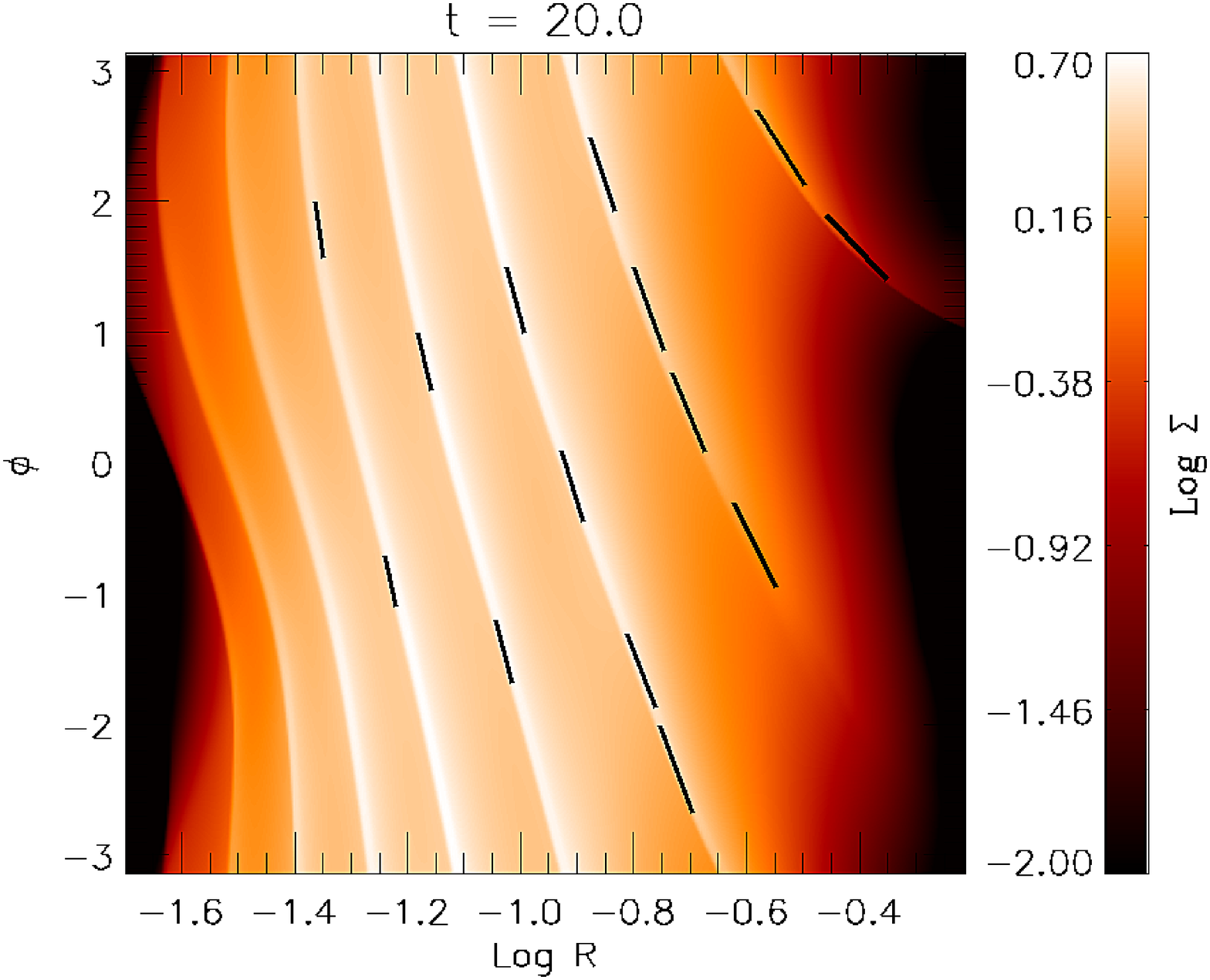}
\includegraphics[width=0.33\textwidth]{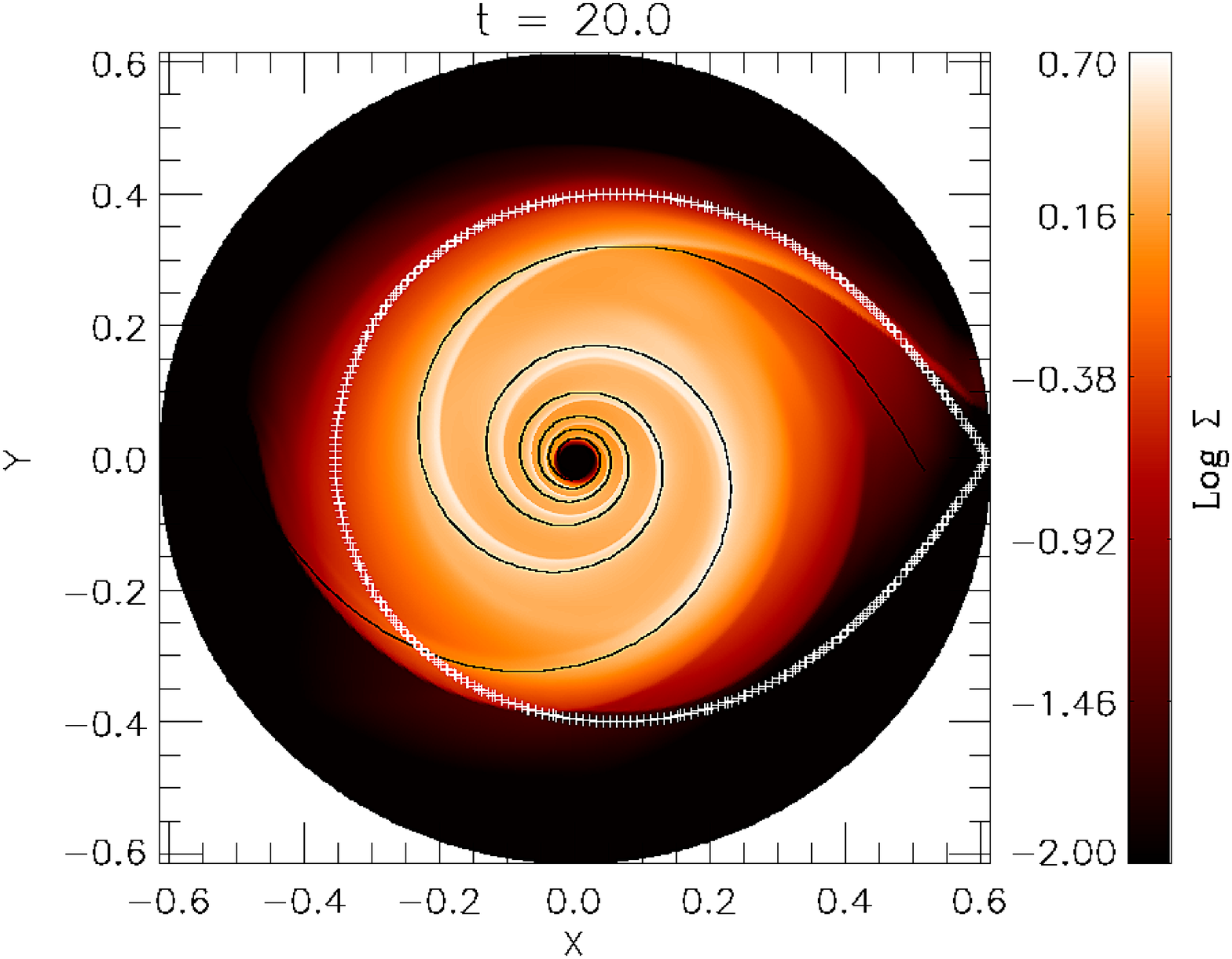}
\includegraphics[width=0.33\textwidth]{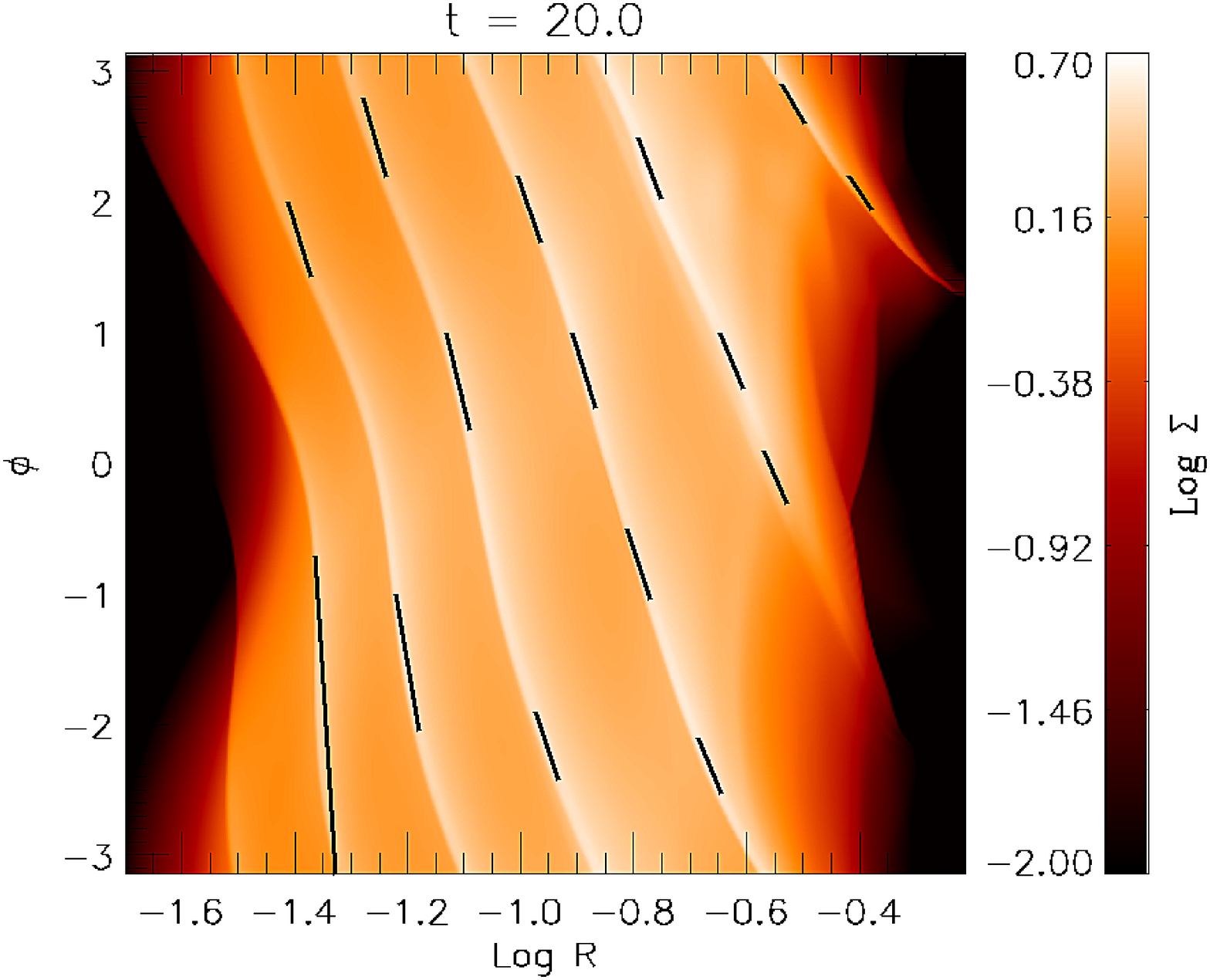}
\includegraphics[width=0.33\textwidth]{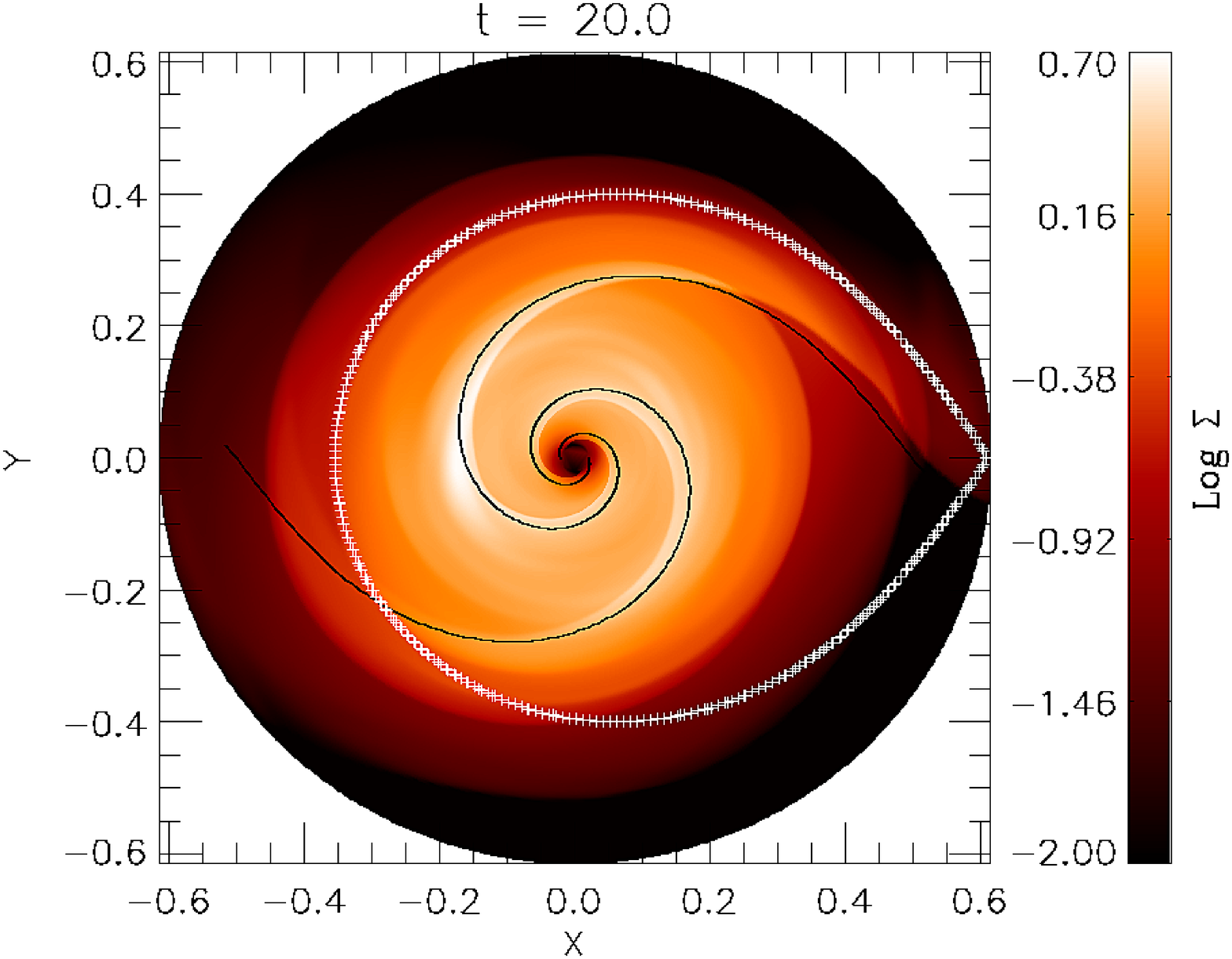}
\includegraphics[width=0.33\textwidth]{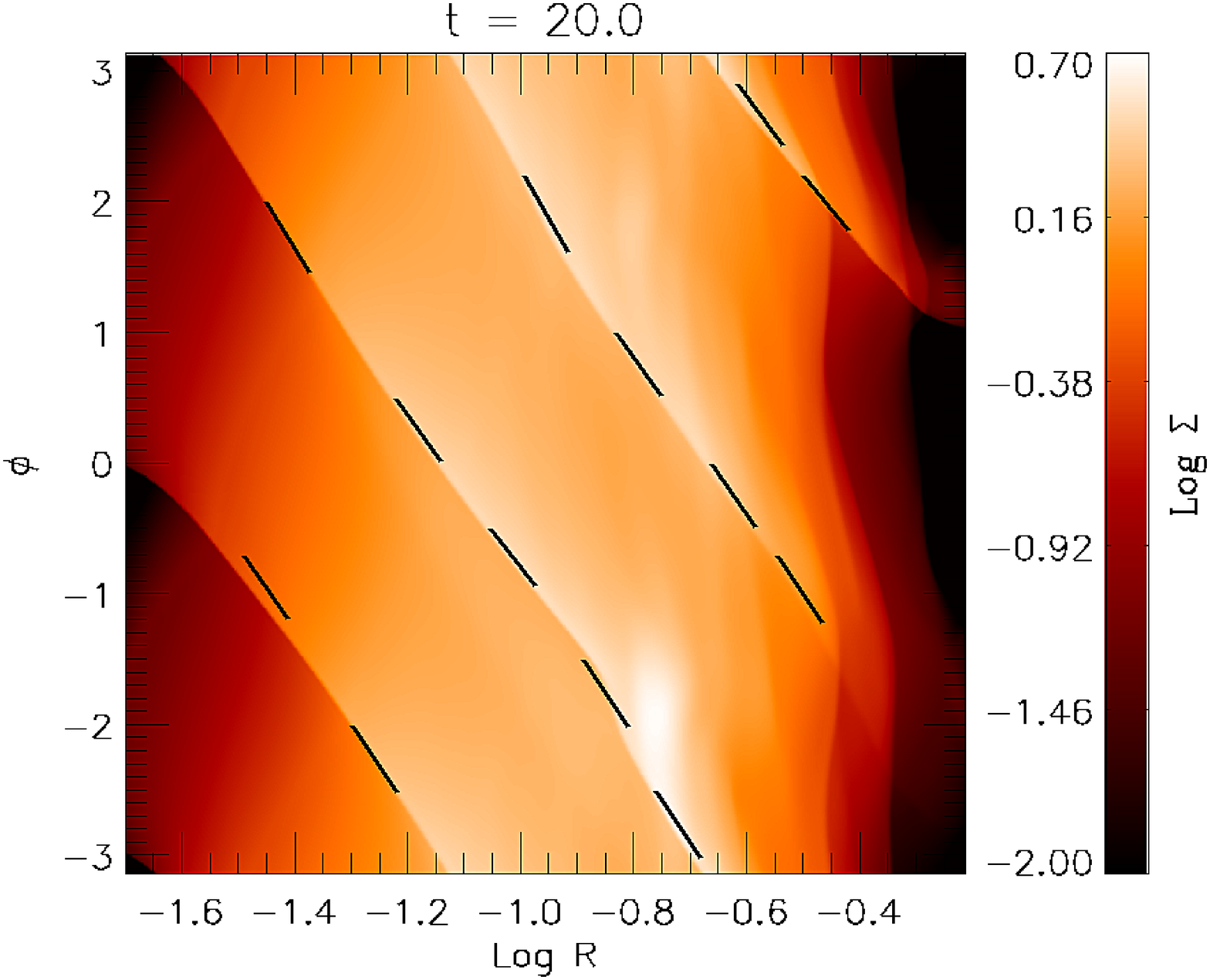}
\includegraphics[width=0.33\textwidth]{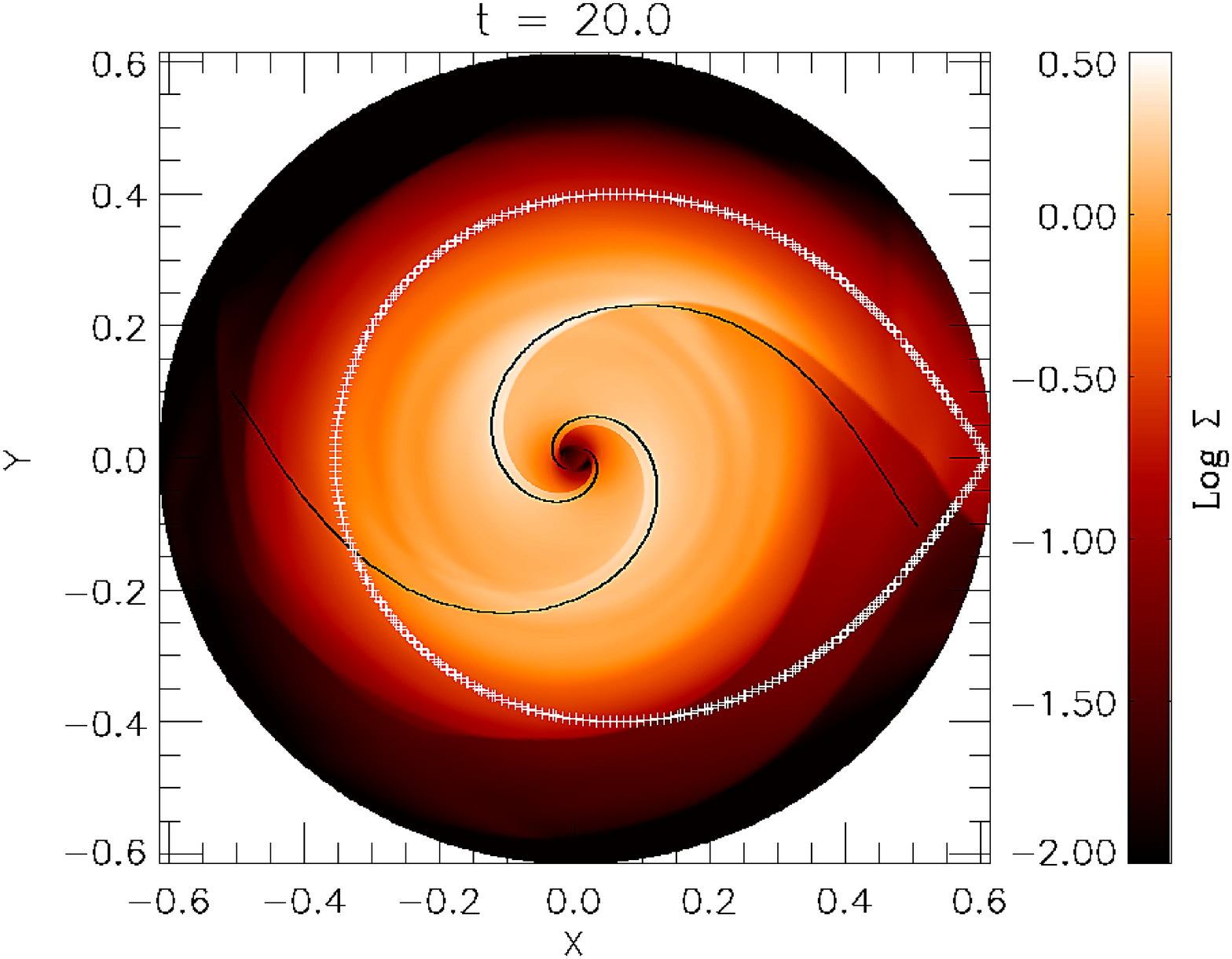}
\includegraphics[width=0.33\textwidth]{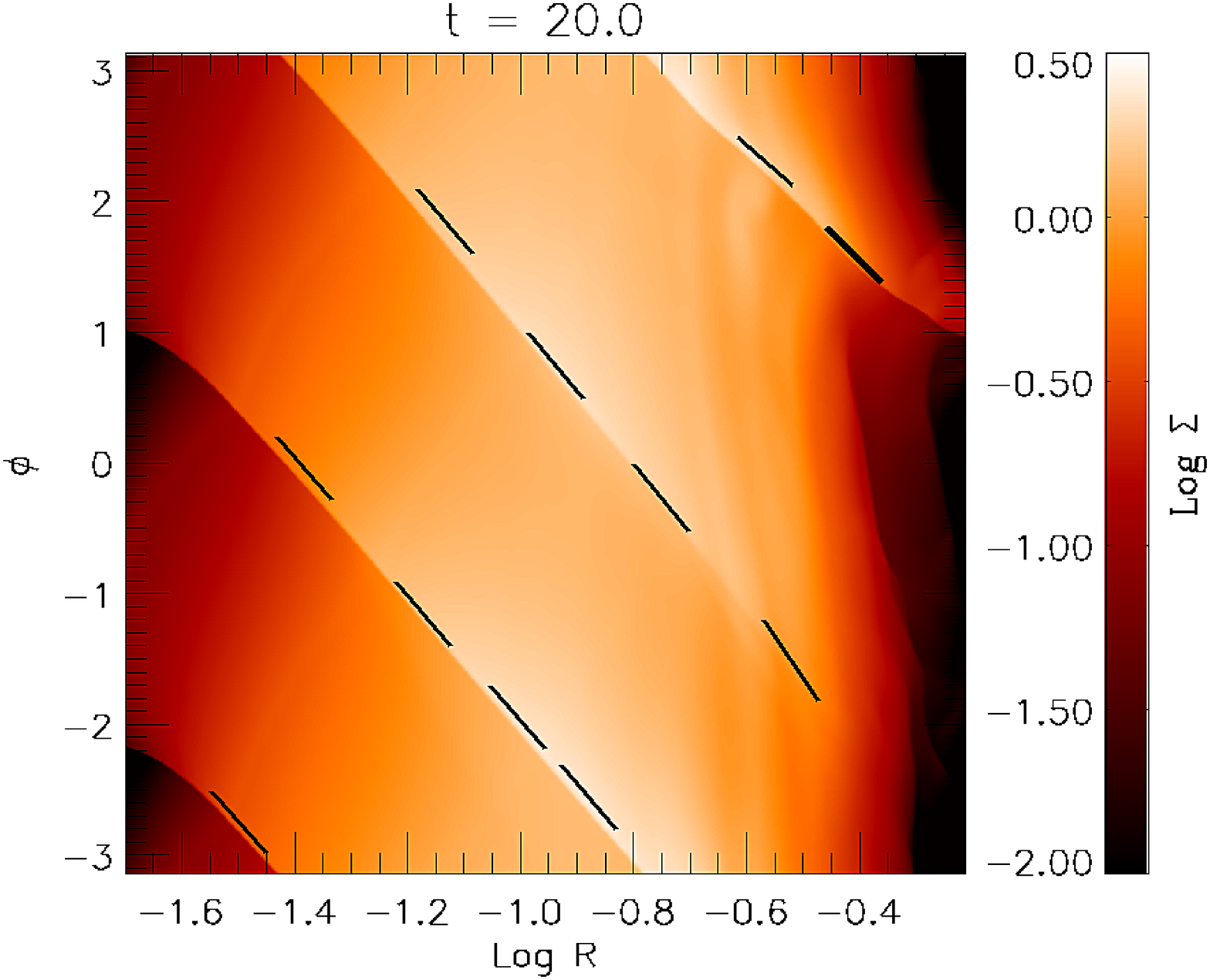}
\caption{{\it Left}: Surface density of no-inflow hydro models in polar coordinates. Black lines show the fitting of spiral arm structures with linear dispersion relation. {\it Right}: Surface density of no-inflow hydro modes in $\log R - \phi$ coordinates. Short black dashes show the sample locations for measuring local pitch angles of spiral arms. The top two rows show isothermal models with sound speed $c_s=0.1$ and $c_s=0.3$ respectively. The bottom 3 panels show adiabatic models with specific heat index $\gamma = 1.1, 1.2, 1.3$.}
\label{fig:noinflow-dispersion}
\end{figure*}

\begin{figure*}
\centering
\includegraphics[width=0.33\textwidth]{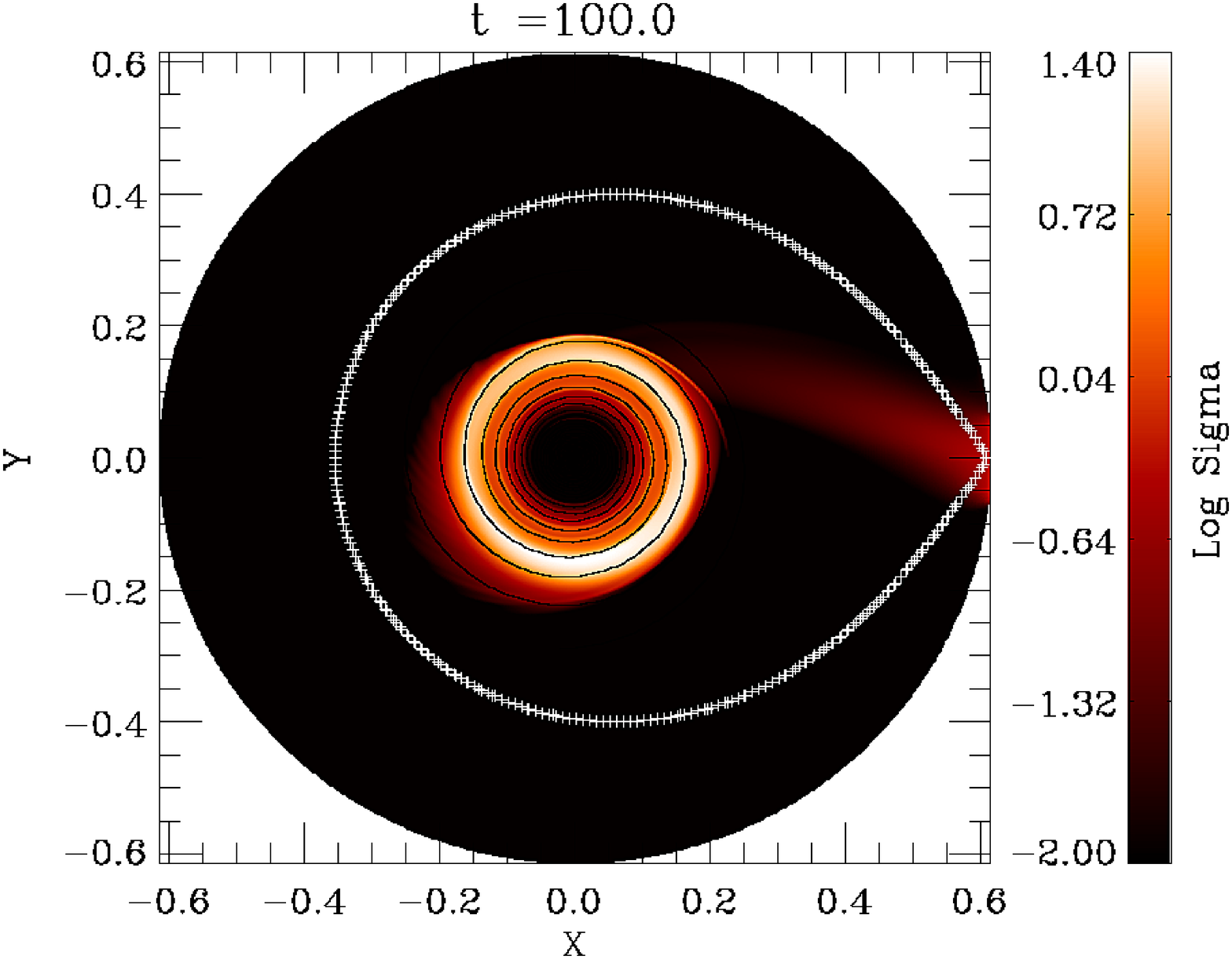}
\includegraphics[width=0.33\textwidth]{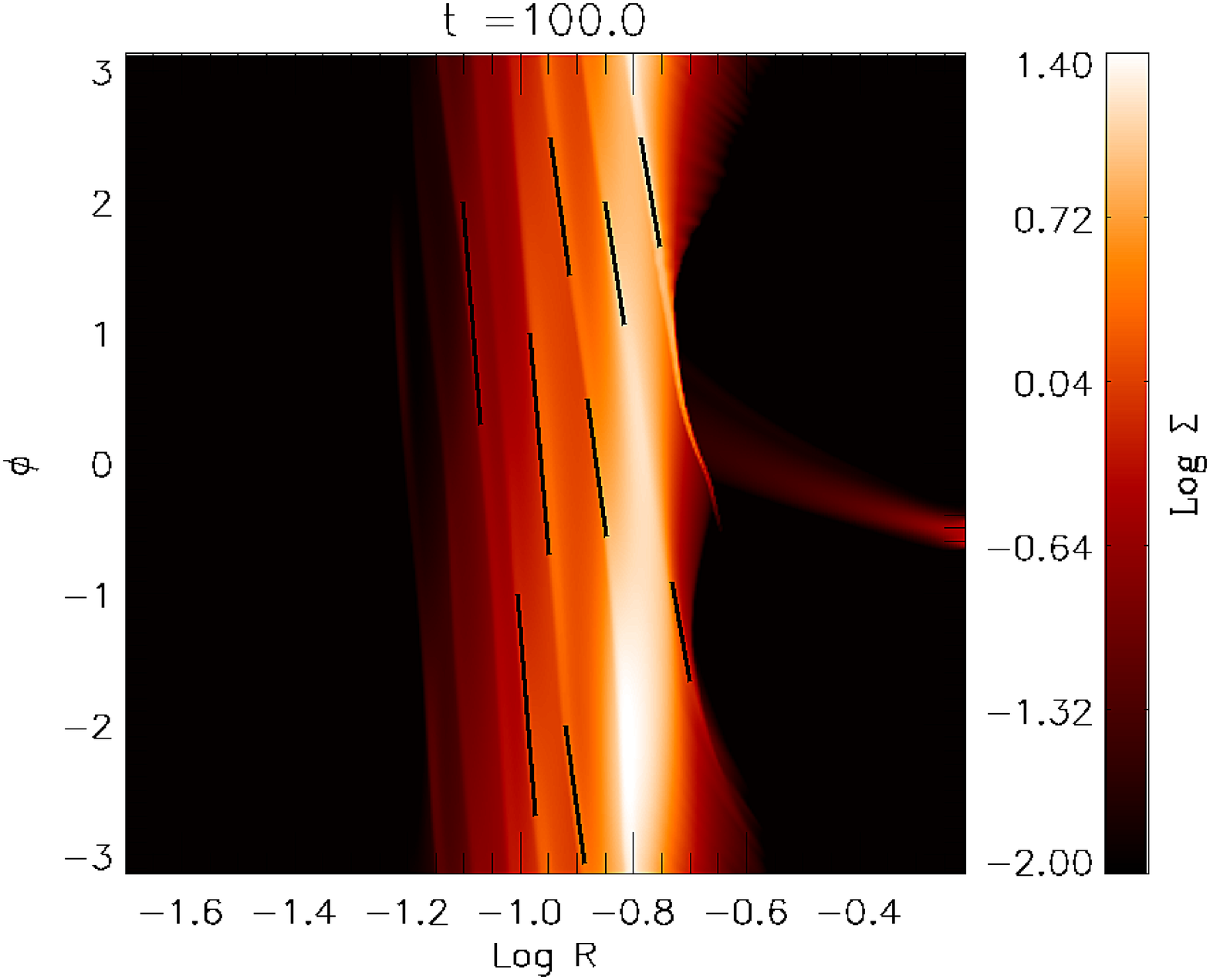}
\includegraphics[width=0.33\textwidth]{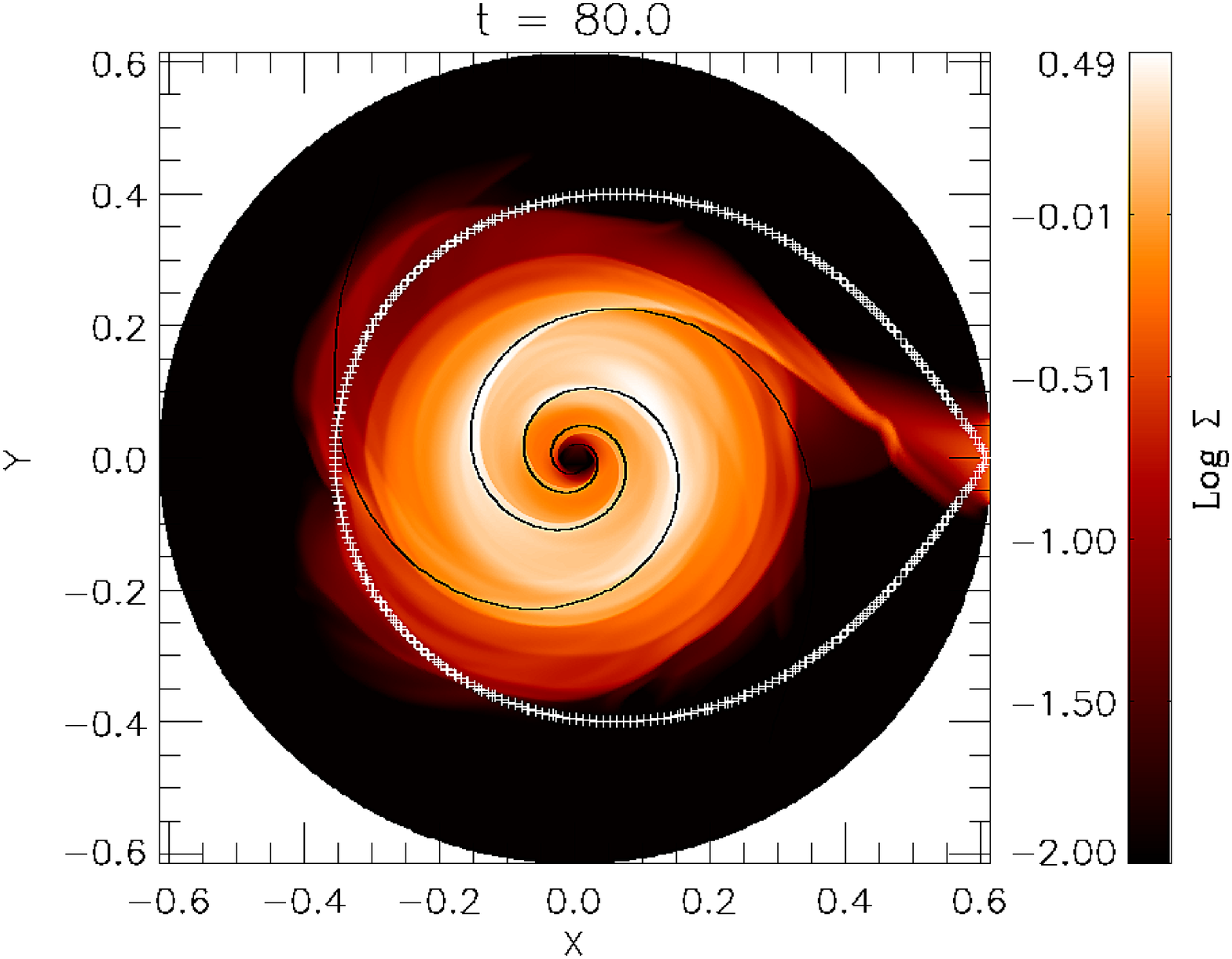}
\includegraphics[width=0.33\textwidth]{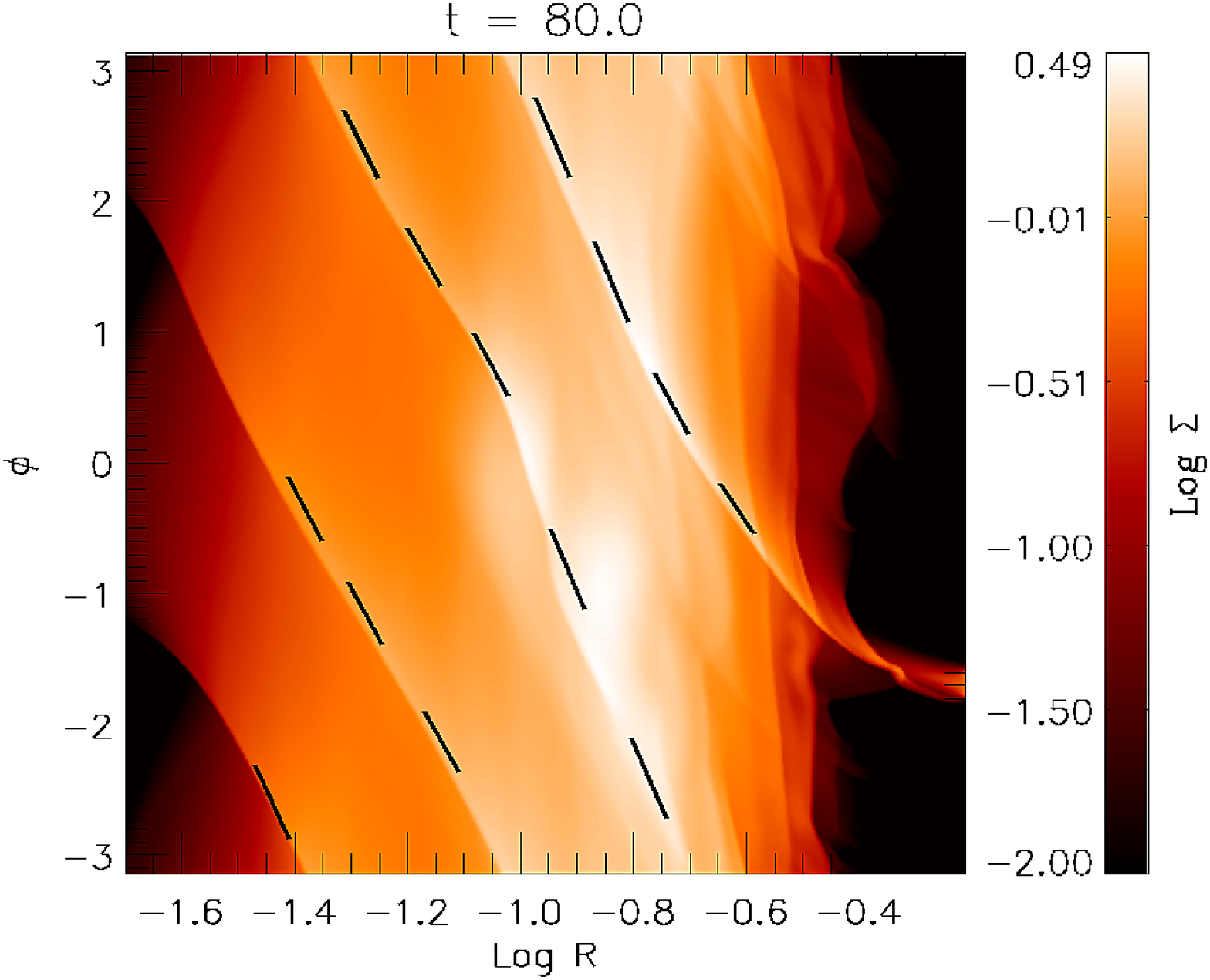}
\includegraphics[width=0.33\textwidth]{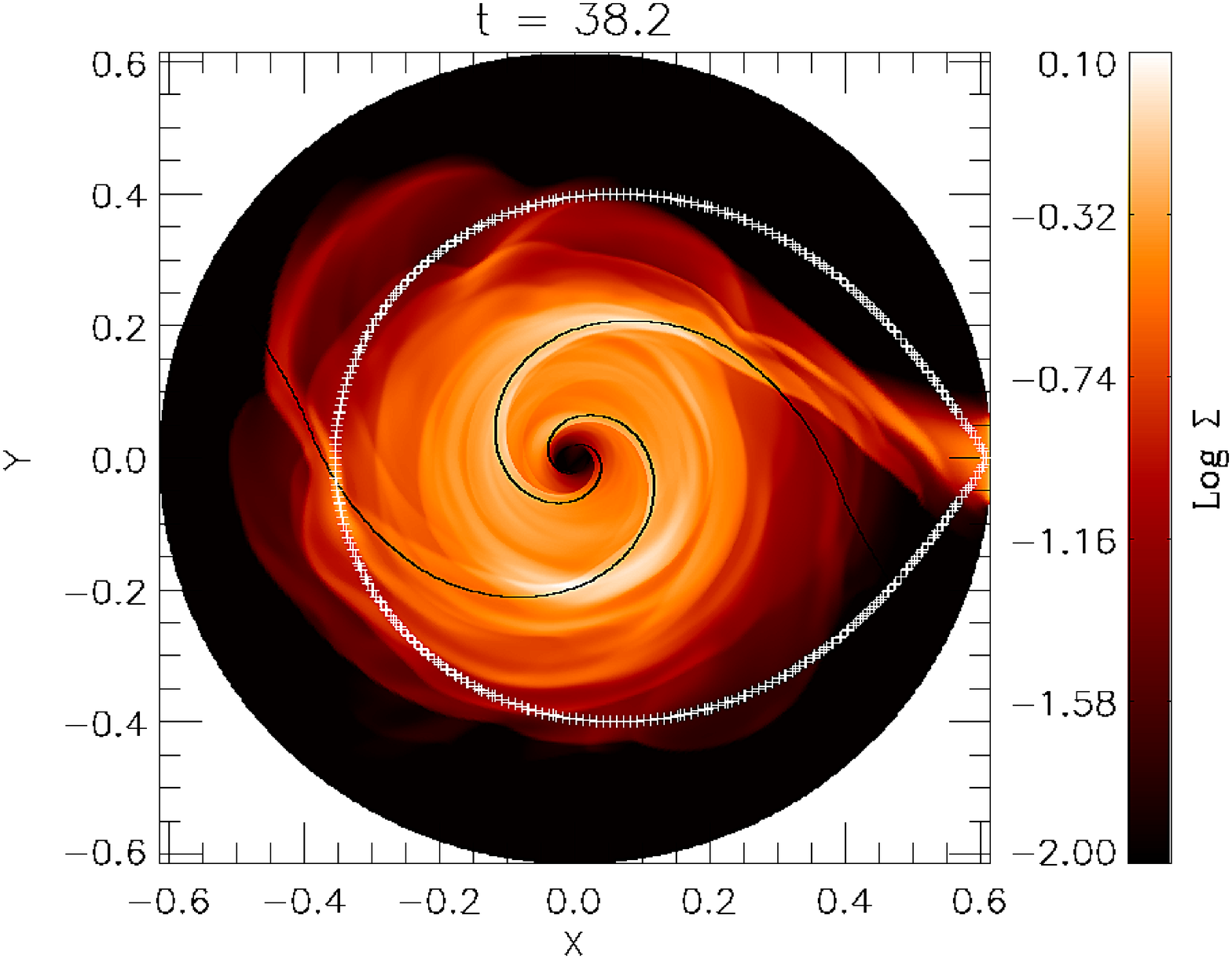}
\includegraphics[width=0.33\textwidth]{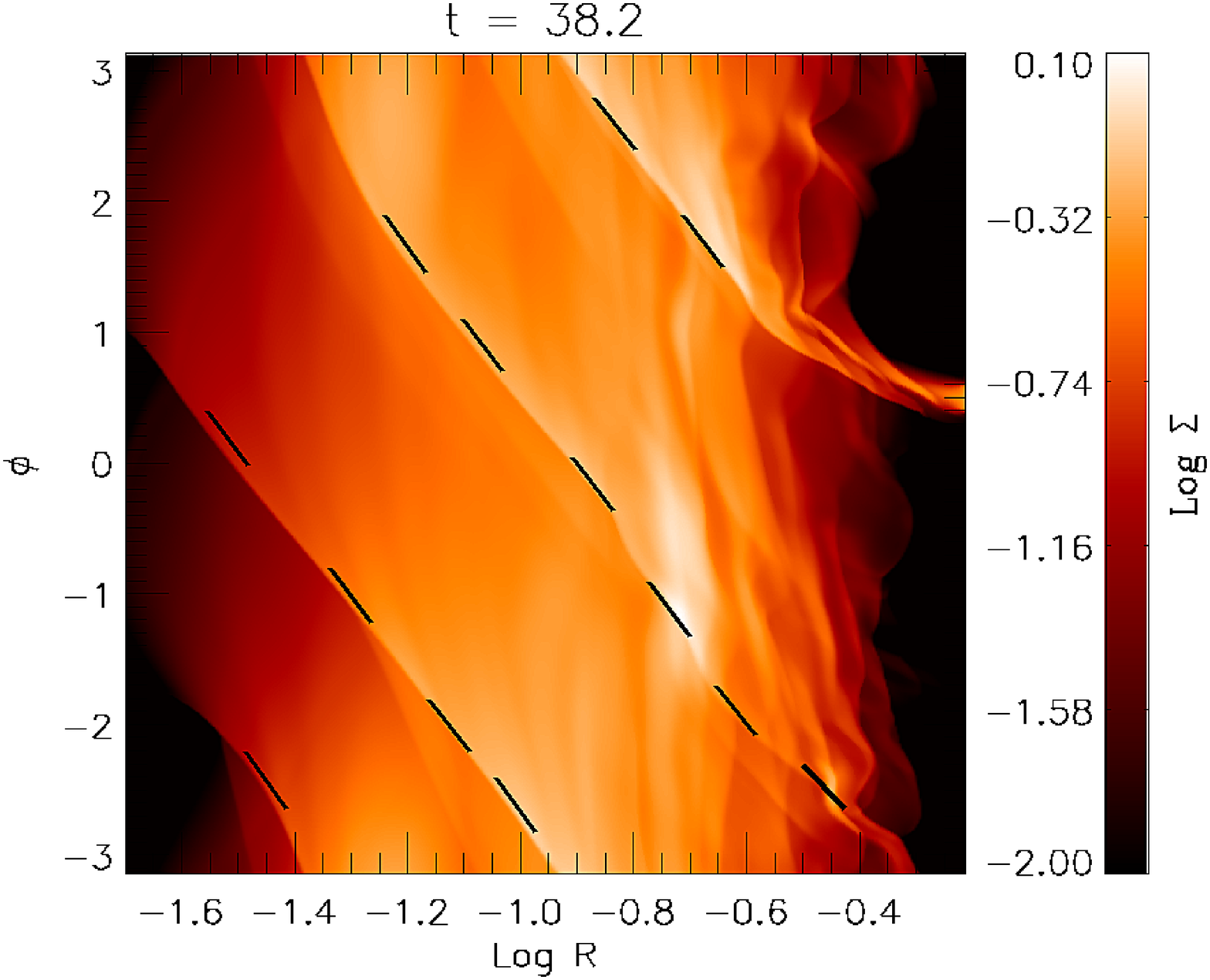}
\includegraphics[width=0.33\textwidth]{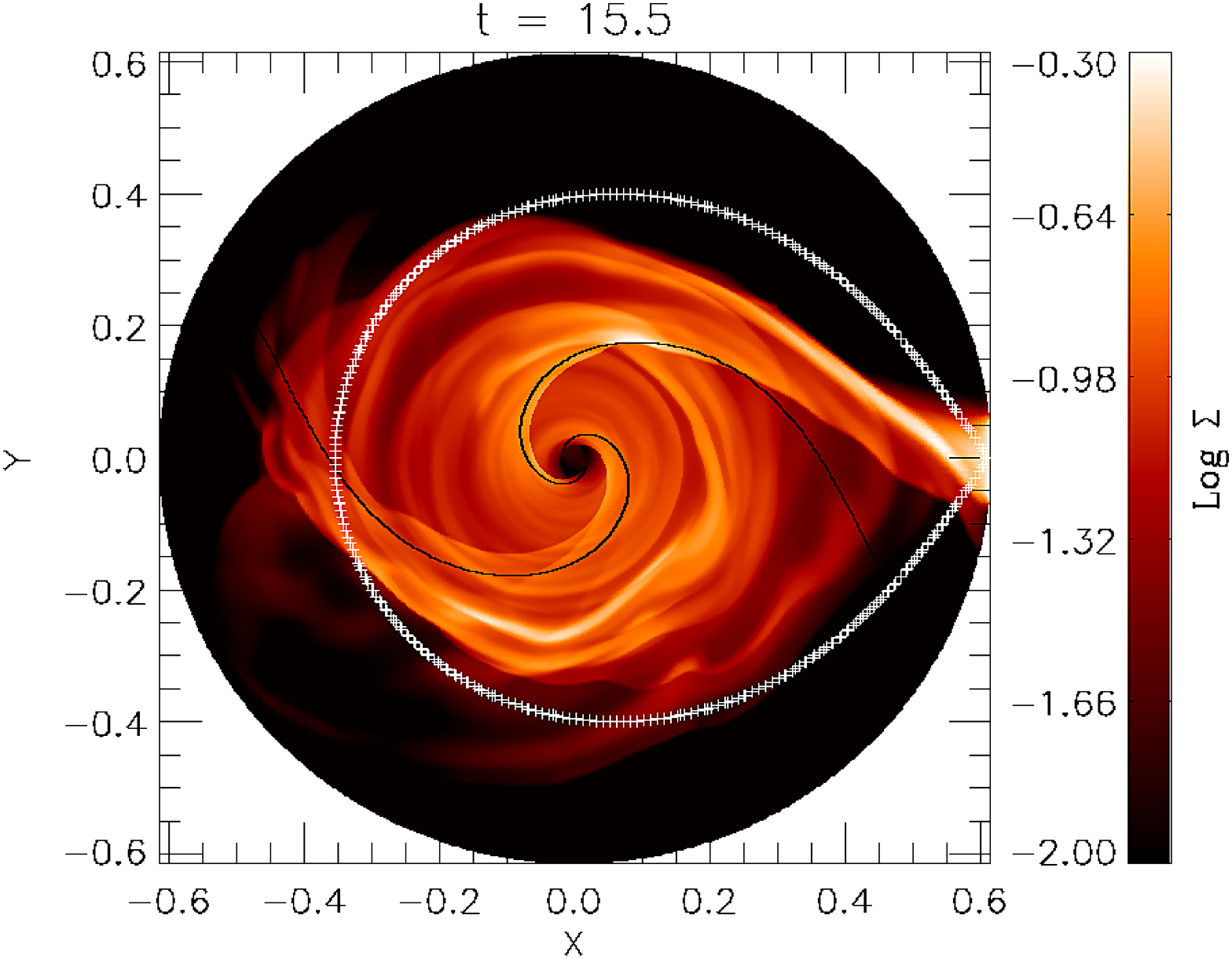}
\includegraphics[width=0.33\textwidth]{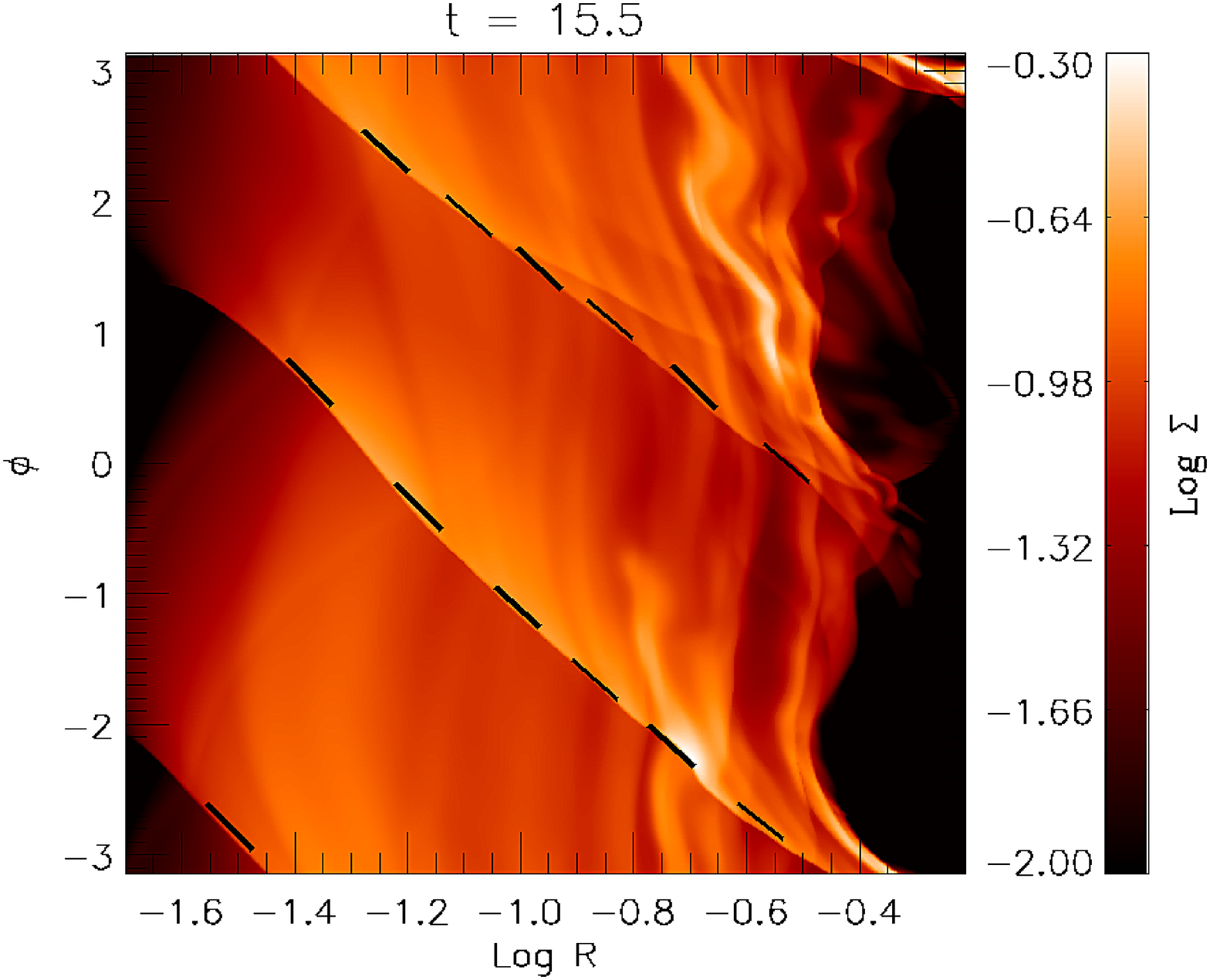}
\caption{The same with Fig. \ref{fig:noinflow-dispersion} for with-inflow hydro models. The top row shows isothermal model with sound speed $c_s=0.1$. The bottom three rows show adiabatic models with specific heat index $\gamma = 1.1, 1.2, 1.3$.}
\label{fig:inflow-dispersion}
\end{figure*}

We show the spiral arm patterns fitted with Eq. \ref{eq:angle} and $m=2$ in Fig. \ref{fig:noinflow-dispersion} (no-inflow models) and Fig. \ref{fig:inflow-dispersion} (with-inflow models). On the left panels of each figure, the black lines, which are the positions of wave fronts calculated from Eq. \ref{eq:angle}, overlap with the spiral patterns from our simulations quite well in most of the disk area. However, deviations are observed at the outer edge of the disk which are due to the non-linear effects in the disk: 1) the pressure gradient at the outer disk edge is significant which we do not include in the linear dispersion relation (Eq. \ref{eq:dispersionrelation}); 2) the impact between the L1 accretion stream and background flow excites a bow shock (the ``hot spot" in observations) which does not follow the linear wave dispersion relations.

\begin{figure*}
\centering
\includegraphics[width=0.49\textwidth]{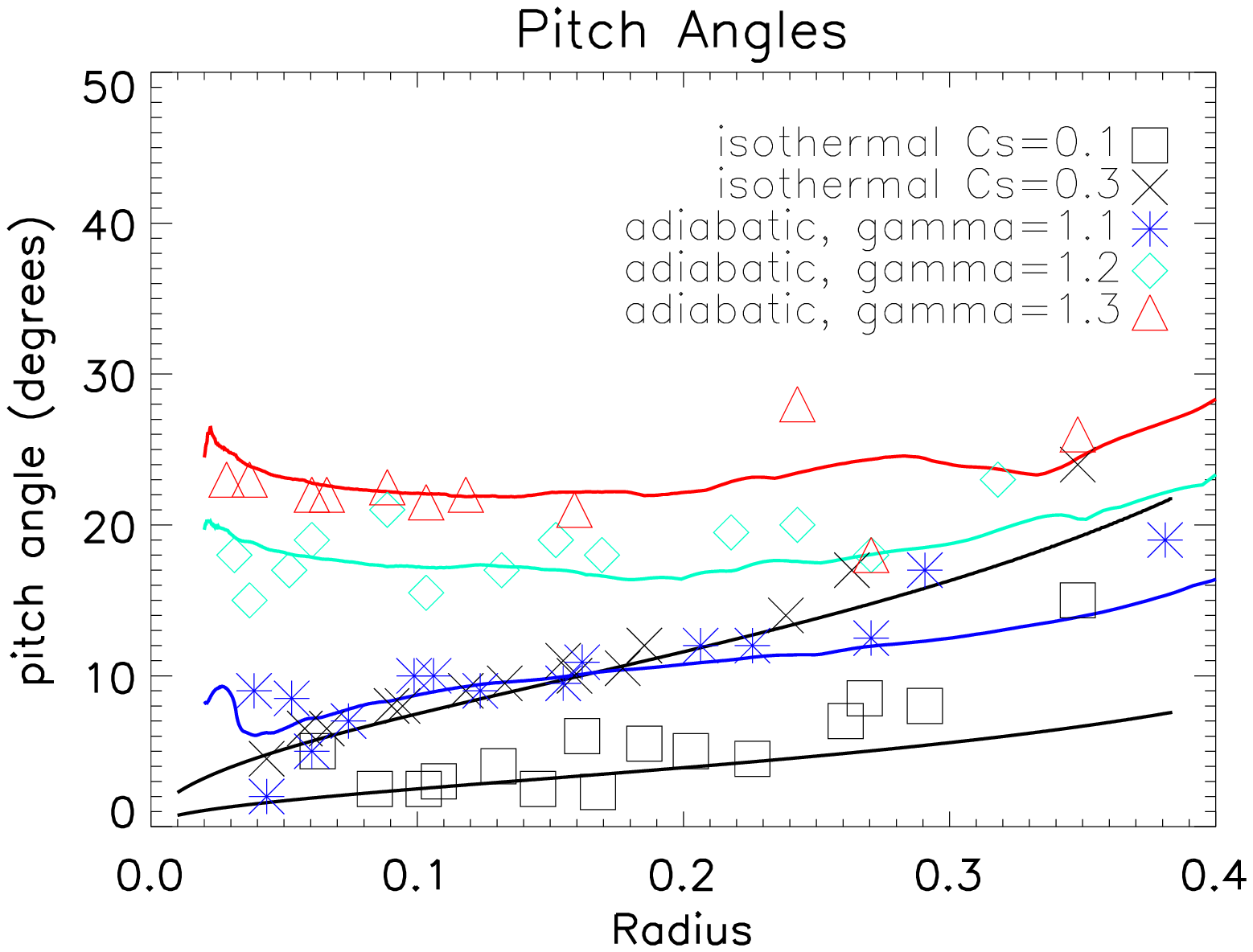}
\includegraphics[width=0.49\textwidth]{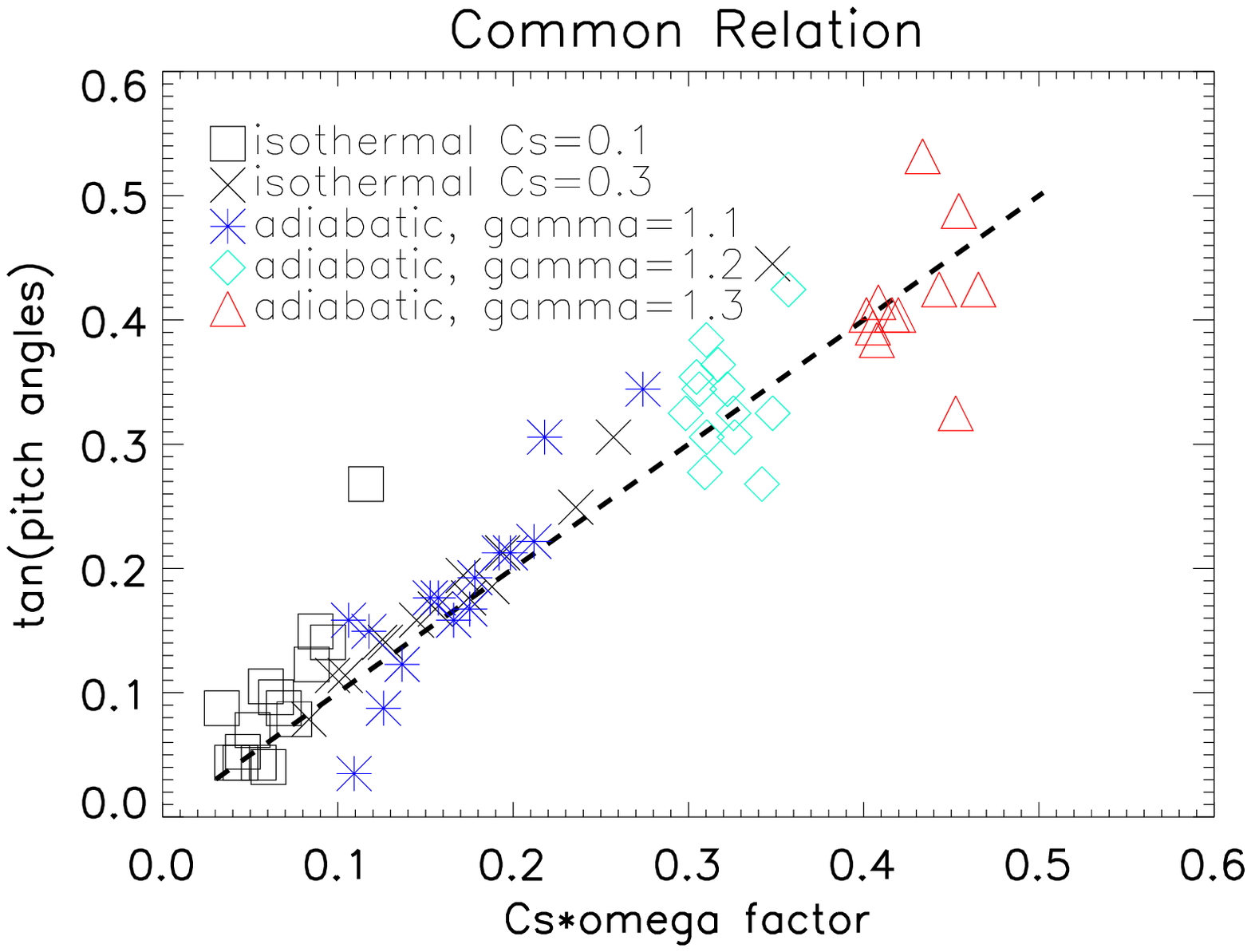}
\includegraphics[width=0.49\textwidth]{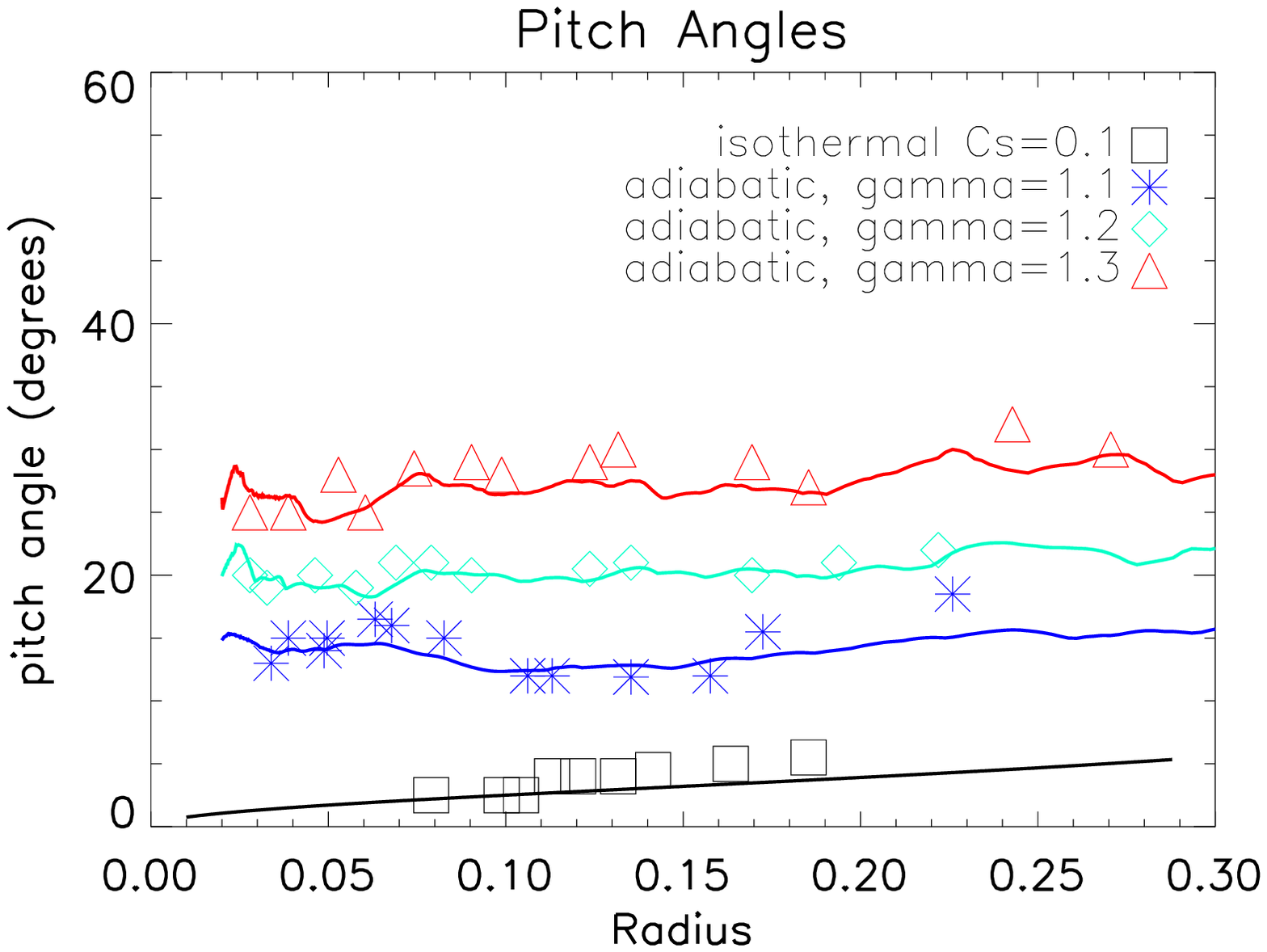}
\includegraphics[width=0.49\textwidth]{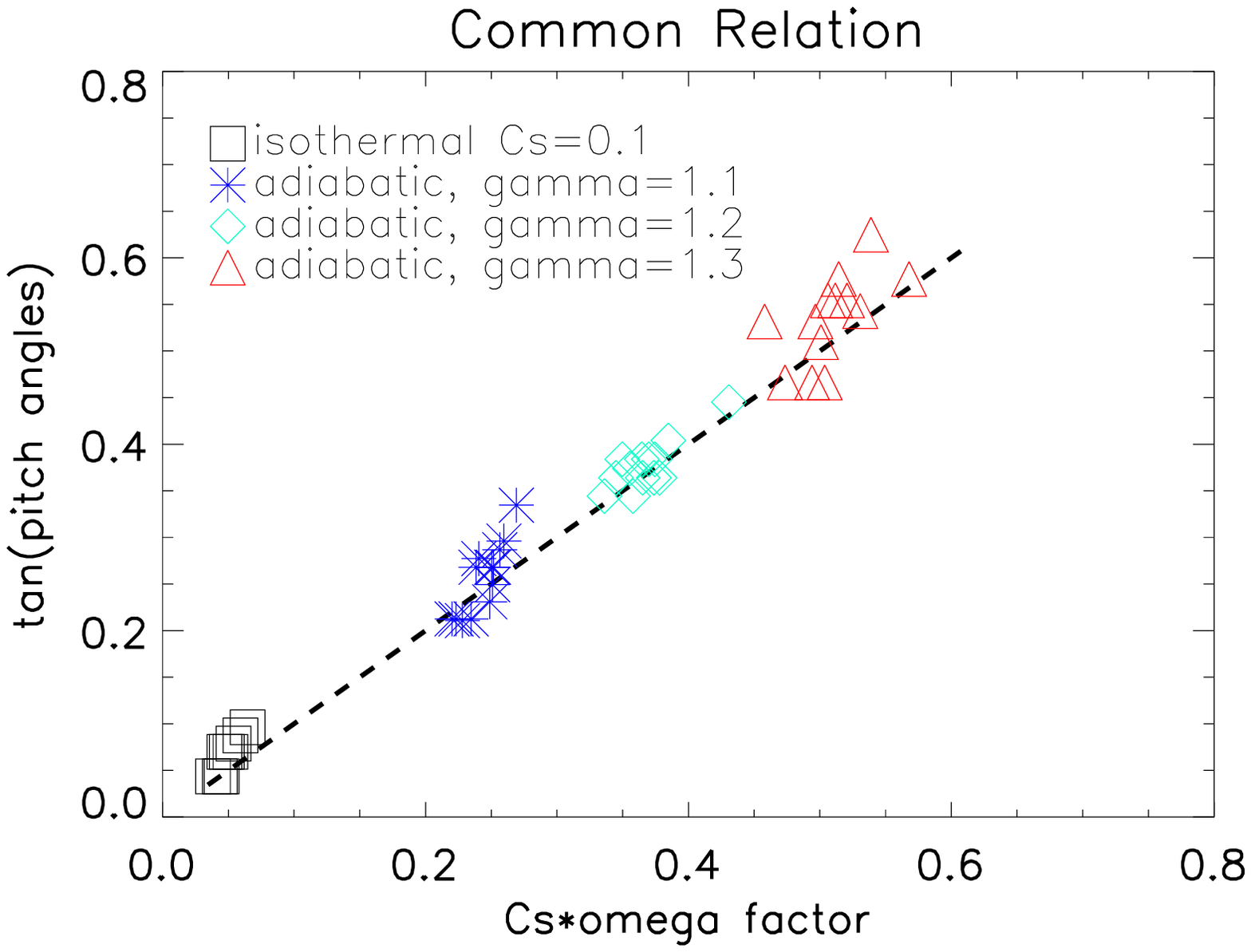}
\caption{Linear wave dispersion relation fitting pitch angles of the spiral patterns for no-inflow models (upper panels) and with-inflow models (lower panels). Left panels show radial profiles of pitch angles, and right panels show the relation between tangent of pitch angles and the right hand side term in Eq. \ref{eq:angle} related with local Mach numbers. Individual points represent the pitch angles measured at the location of short dashes in Fig. \ref{fig:noinflow-dispersion} and \ref{fig:inflow-dispersion}. Solid lines show fitting using linear dispersion relation (Eq. \ref{eq:angle}) with $m=2$.}
\label{fig:pitch-angle}
\end{figure*}

To better understand how well the pitch angles satisfy the dispersion relation, we locally measure the angles at the sampling positions marked as short black dashes on the right panel of Fig. \ref{fig:noinflow-dispersion} and \ref{fig:inflow-dispersion}. We plot these measured angles versus radius on the left panels of Fig. \ref{fig:pitch-angle}, and the relation in Eq. \ref{eq:angle} on the right panels of Fig. \ref{fig:pitch-angle}. The pitch angles follow the dispersion relation very well. Most of the outlying points are due to non-linear effects at outer edge of the disk as mentioned above. 

Therefore, instead of a unique relation between the opening angles of the shock $\theta$ and the specific heat ratio $\gamma$, we find the more underlying relation is between $\theta$ and the local gas temperature or $\mathcal{M}$: $\theta$ increases as $\mathcal{M}$ decreases. The trend of increasing $\theta$ with increasing $\gamma$ observed in our simulations as well as in previous numerical work  \citep[e.g.][]{2000MNRAS.316..906M} is because a larger $\gamma$ usually results in a disk that heats up faster. 

%%%%%%%%%%%%%%%%%%%%%%%%%%%%%%%%%%%%%%%%%%%%%%%%%%%%%%%%%

%%%%%%%%%%%%%%%%%%%%%
\subsection{Angular Momentum Transport by Spiral Shocks}
\label{sec:shockdissipation}

\begin{figure}
\centering
\includegraphics[width=0.49\textwidth]{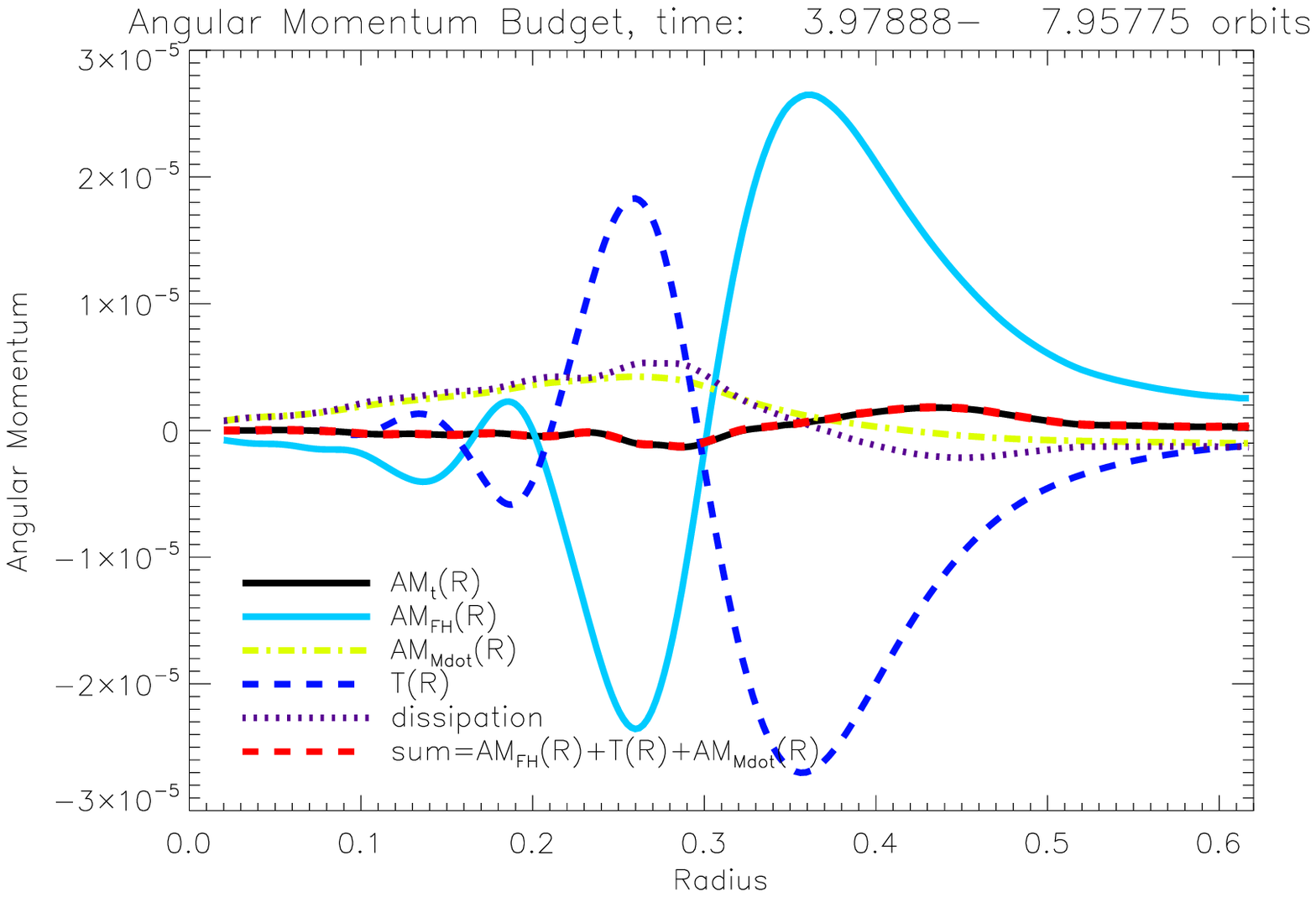}
\includegraphics[width=0.49\textwidth]{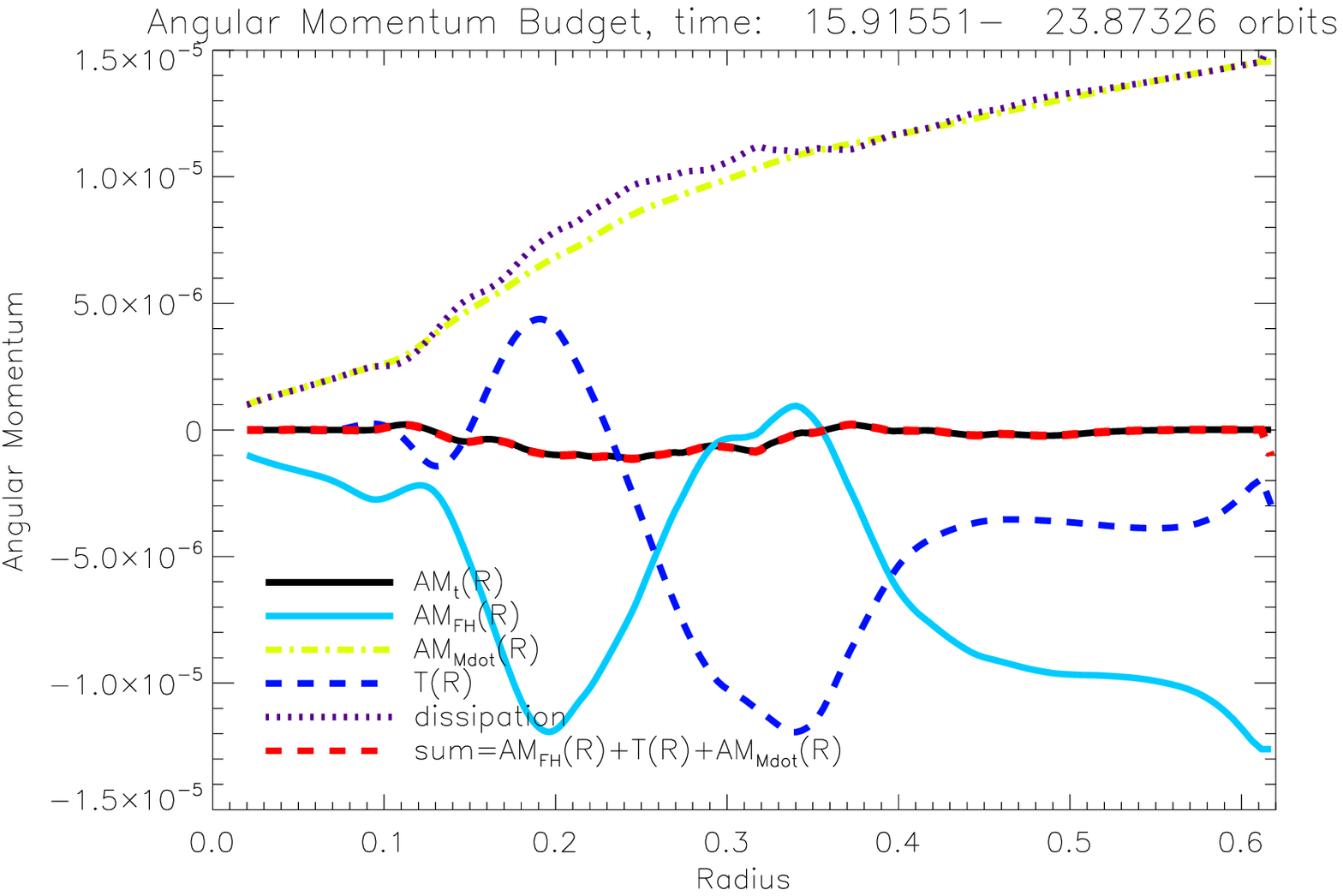}
\caption{Angular momentum budget terms (see \S \ref{sec:ang-mom-budget} for definition) of an adiabatic disk with $\gamma=1.1$ for no-inflow (top) and with-inflow(bottom) hydro models. All quantities are averaged over every timestep during $25<t<50$ for no-inflow models and $100<t<150$ for with-inflow models. Solid black line shows the time derivative term $AM_t (R)$. Solid light blue line shows the angular momentum flux term $AM_{FH} (R)$. Dashed dark blue line shows the torque term $T(R)$. Dash-dotted yellow line shows the dissipation term $AM_{FH} (R) + T(R)$. Dotted purple line shows negative of the mass accretion rate term $-AM_{\dot{M}} (R)$. Dashed red line shows the sum of $AM_{FH} (R) + T(R) + AM_{\dot{M}} (R)$. The good match between the dissipation term (dash-dotted yellow line) and the mass accretion term (dotted purple line) indicates that dissipation indeed drives mass accretion. }
\label{fig:noinflow-tab}
\end{figure}

In \S \ref{sec:ang-mom-budget} we analytically studied the angular momentum budget of the disk. In Fig.\ref{fig:noinflow-tab} we show the time-averaged values of the angular momentum terms of no-inflow and with-inflow hydro models using adiabatic EOS with $\gamma=1.1$: The time derivative term $AM_t (R)$ is shown as a solid black line, the angular momentum flux term $AM_{FH} (R)$ as a solid light blue line, the torque term $T(R)$ as a dashed dark blue line, the dissipation term $AM_{FH} (R) + T(R)$ as a dash-dotted yellow line, negative of the mass accretion rate term $-AM_{\dot{M}} (R)$ as a dotted purple line, and lastly the sum of $AM_{FH} (R) + T(R) + AM_{\dot{M}} (R)$ as a red dashed line. 

The perfect match between the black and red lines shows the numerical conservation of angular momentum. The time derivative term $AM_t (R)$ (solid black line) has much smaller values than the other terms, meaning the disk reaches quasi-steady state. The good match between the mass accretion rate term $-AM_{\dot{M}} (R)$ (dotted purple line) and the dissipation term (dash-dotted yellow line) shows proof of the relation in Eq. \ref{eq:dissipation}: local shock dissipation indeed drives mass accretion. The excellent agreement term-by-term between the simulation and simple theory gives us confidence that numerical diffusion of angular momentum is very small, and our numerical methods are accurate. We stress numerical effects further in \S \ref{sec:convergence}. 

\subsubsection{Effective Alpha}
\label{sec:effectivealpha}
In addition to mass accretion rate, another parameter of broad interest is the effective angular momentum transport efficiency, $\alpha_{eff}$. Most previous work on hydrodynamical simulations of accretion disks calculate $\alpha_{eff}$ using ratio of averaged Reynolds stress $\Sigma \delta v_R \delta v_\phi$ to gas pressure. However, this is only suitable in a disk with axisymmetric gravitational potential where no net torque is exerted on the disk, and thus the Reynolds stress acts like a viscous tensor. Instead, in a disk with binary potential like a CV disk, the angular momentum budget in Fig. \ref{fig:noinflow-tab} shows that the Reynolds stress gradient $AM_{FH}(R)$ is mostly canceled by the torques $T(R)$ exerted by the companion star, and only the residual, which is equivalently angular momentum loss due to shock dissipation, is directly related to mass accretion rate. Therefore, it is not appropriate to calculate $\alpha_{eff}$ from scaled Reynolds stress in a CV disk. 

Another point that is worth noticing is that angular momentum transport by spiral shocks is fundamentally different from that by local viscosity. In the former case, transport occurs at the location of spiral shocks where Reynolds stress is generated due to shocks, whereas in the latter case, transport occurs everywhere in the disk due to local turbulence. However, the definition of $\alpha$ in the standard thin disk theory as well as the values of $\alpha$ quoted in the disk instability models (DIM) interpreting CV observations are based on a kinetic viscosity \citep{1973A&A....24..337S}. Then how do we evaluate $\alpha_{eff}$ properly for comparison with observations? We believe the mass accretion rate provides the answer. If we define $\alpha_{eff}$ such that the corresponding kinetic viscosity under the standard $\alpha$ theory induces the same mass accretion rate observed in our simulations, then $\alpha_{eff} = \frac{\dot{M}}{3\pi \Sigma c_s H}$.  This suggests caution is required when comparing values for $\alpha$ measured from observations with models that assume a kinetic viscosity.

\begin{figure}
\centering
\includegraphics[width=0.49\textwidth]{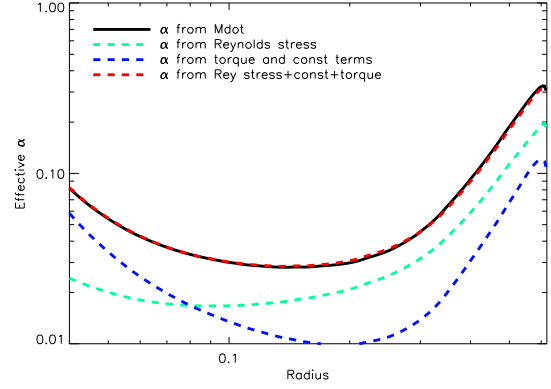}
\caption{$\alpha_{eff}$ calculated in several ways for the adiabatic with-inflow hydro model with $\gamma=1.3$: solid black line from $\dot{M}$, dashed cyan line from Reynolds stress ($2T_R / 3P$), dashed blue line from the torque and constant $C$ terms, dashed red line showing sum of $\alpha_{eff}$ from Reynolds stress, torque and $C$ terms (see Eq. \ref{eq:effalpha} for definition). The quantities are azimuthally- and time-averaged during the period of $t=100 - 150$. Scaled Reynolds stress underestimates the total $\alpha_{eff}$ by a factor of $1.5-4$.}
\label{fig:alphadecompose}
\end{figure}

We can quantitatively prove this statement by re-writing Eq.\ref{eq:angmom} for a 2D steady-state hydrodynamical disk:
\begin{eqnarray}
 &&\frac{\dot{M}}{R} \partial_R (R v_K) \nonumber \\
 &=& \frac{1}{R} \partial_R (R^2<\Sigma v_R \delta v_\phi>_\phi ) - <\bf{R} \times \bf{F}_{ext}>_\phi,
\end{eqnarray}
where $<X> = \int_{0}^{2\pi} X d\phi$, $\dot{M}=- <R\Sigma v_R>_\phi$. Comparing with Eq.\ref{eq:angmom}, here we neglected the time derivative term by assuming the disk reaches steady state, and the Maxwell term in the case of hydrodynamics. Integrating this equation in the radial direction gives
\begin{equation}
\dot{M}  = \frac{ R^2 <\Sigma v_R \delta v_\phi>_\phi - \int R <\mathbf{R} \times \mathbf{F}_{ext}>_\phi dR + C} {R v_K}
\end{equation}
where we assume $\dot{M}$ is a constant over radius in steady state. Under the assumption of a geometrically thin disk, we can adopt the standard $\alpha$ disk theory \citep{1973A&A....24..337S} and write
\begin{eqnarray}
\label{eq:effalpha}
\alpha_{eff} &=& \frac{\dot{M}}{3\pi \Sigma c_s H} \nonumber \\
&=& \frac{2}{3} \frac{T_R}{P} + \frac{ -\int R <\mathbf{R} \times \mathbf{F}_{ext}>_\phi dR + C}{3\pi \Sigma c_s H R v_K}
\end{eqnarray}
where $T_R = <\Sigma v_R \delta v_\phi>_\phi/2\pi$ is the azimuthally averaged Reynolds stress, and $H=c_s/\Omega$ is the local thermal scale height. The constant $C$ is set by the boundary condition. From this equation, we can clearly conclude that in a disk with non-axisymmetric gravitational potential the mass accretion rate cannot be accounted for with scaled Reynolds stress only, but also includes effects from the external torque and boundary condition. In Fig.\ref{fig:alphadecompose} we show the effective alpha calculated from $\dot{M}$ (solid black line), from Reynolds stress ($2T_R / 3P$, dashed cyan line), from the torque and integration constant terms (dashed blue line), and lastly the sum of the latter two $\alpha$ (dashed red line) which perfectly overlap with the black line. The scaled Reynolds stress underestimates the effective $\alpha$ by a factor of $1.5 - 4$. At the outer radius, the external torque exerted by the companion star plays an important role, while at the inner radius, the constant $C$ set by the boundary cannot be neglected. 

\begin{figure*}
\centering
\includegraphics[width=0.49\textwidth]{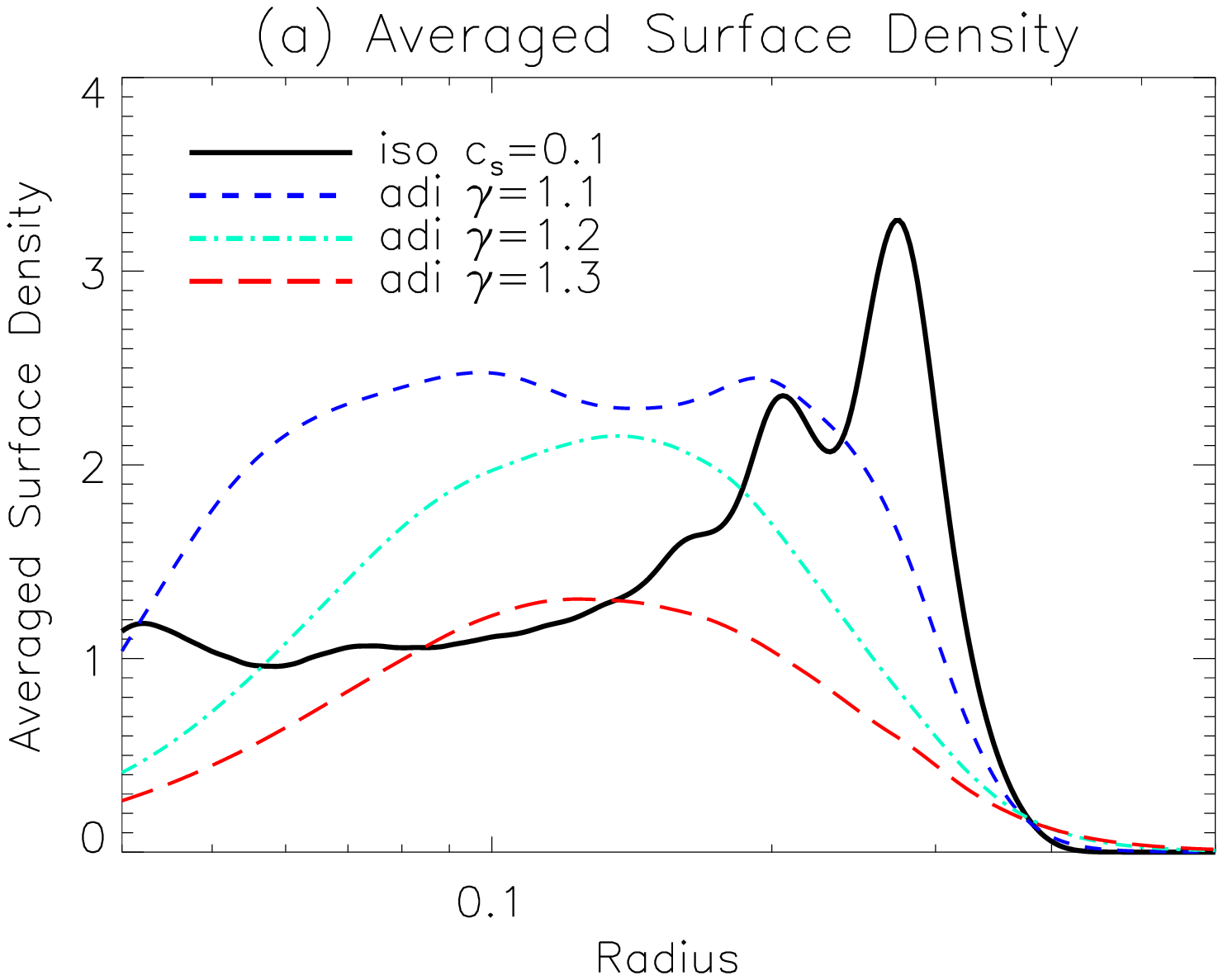}
\includegraphics[width=0.49\textwidth]{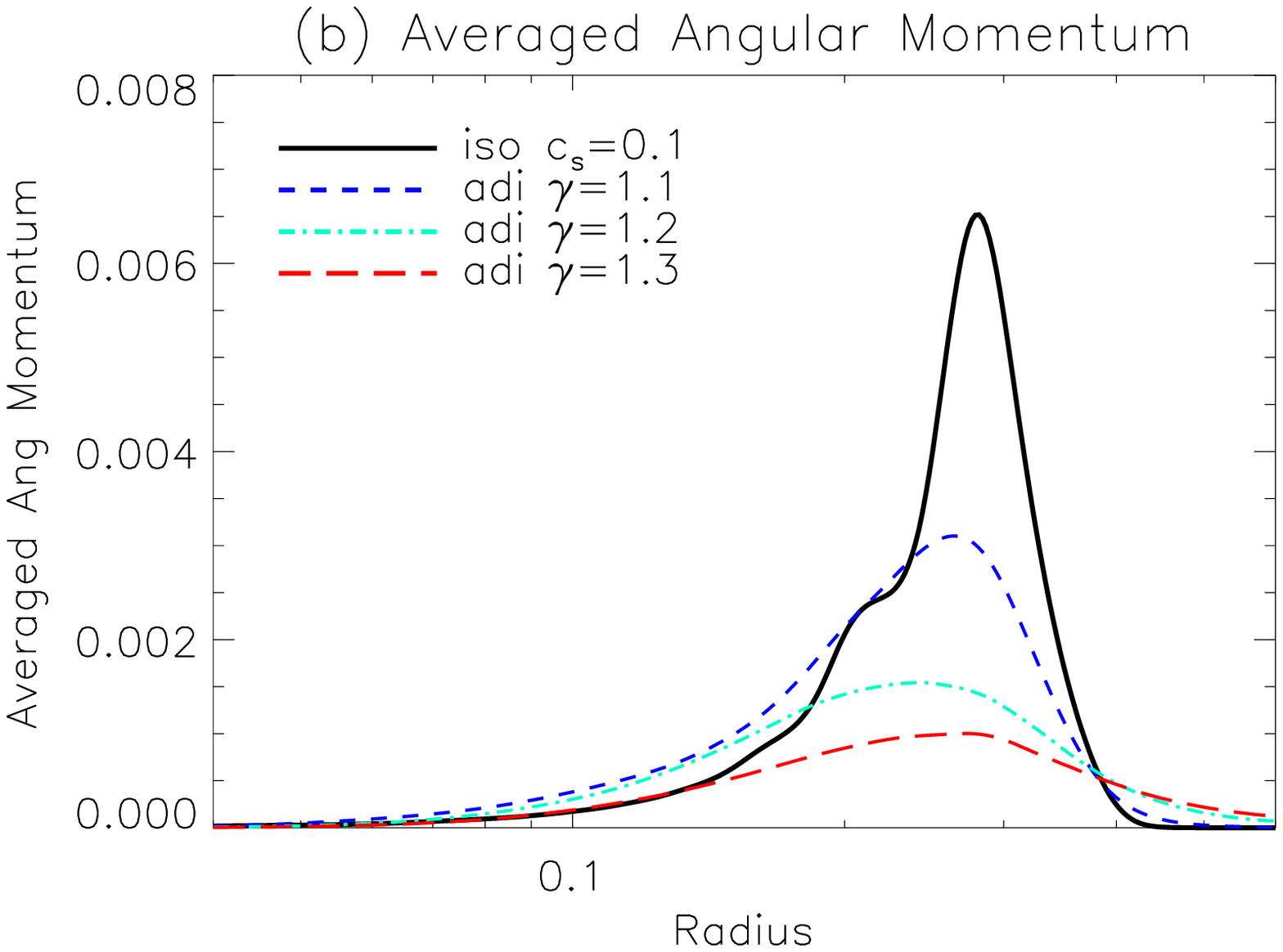}
\includegraphics[width=0.49\textwidth]{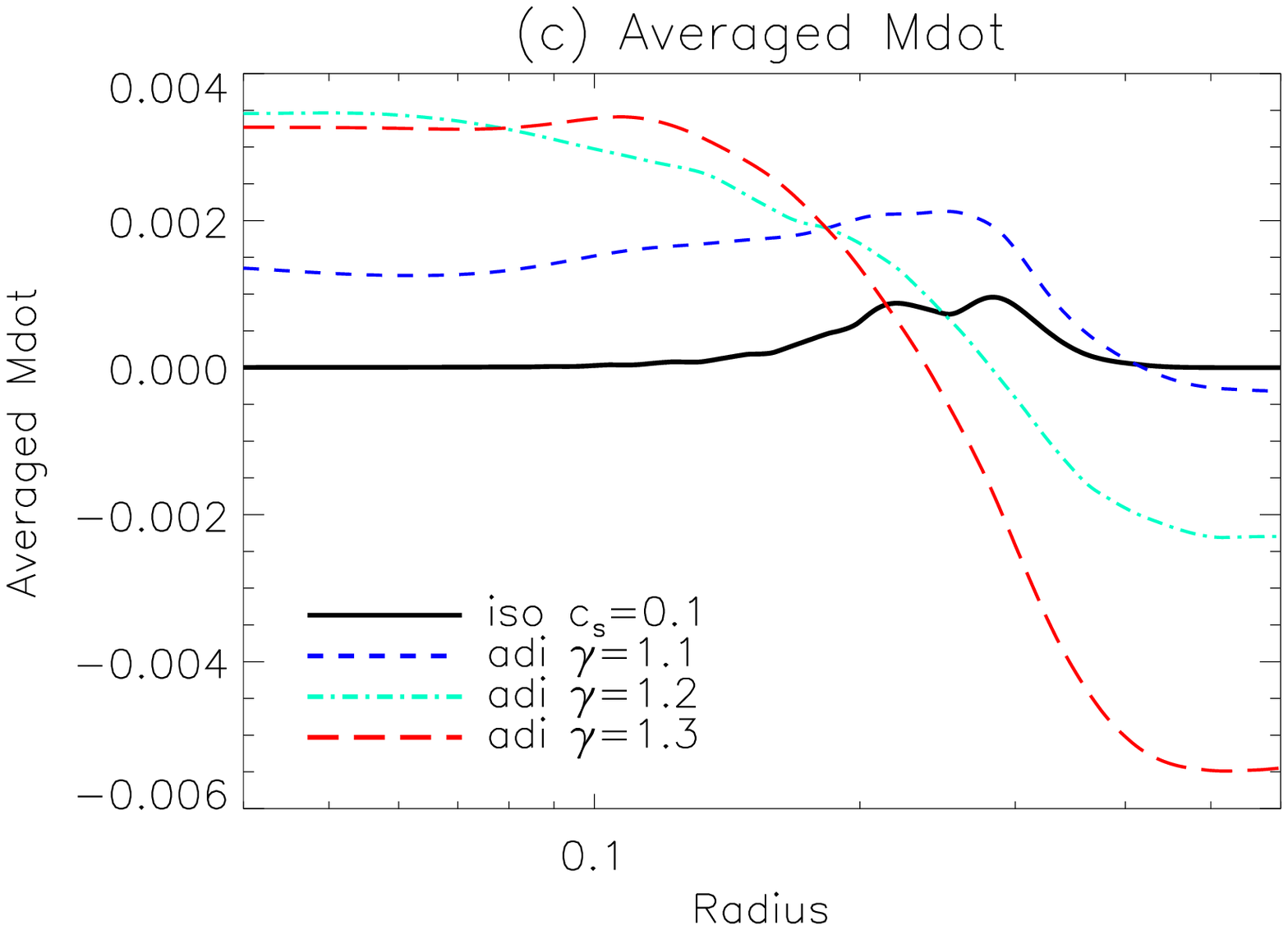}
\includegraphics[width=0.49\textwidth]{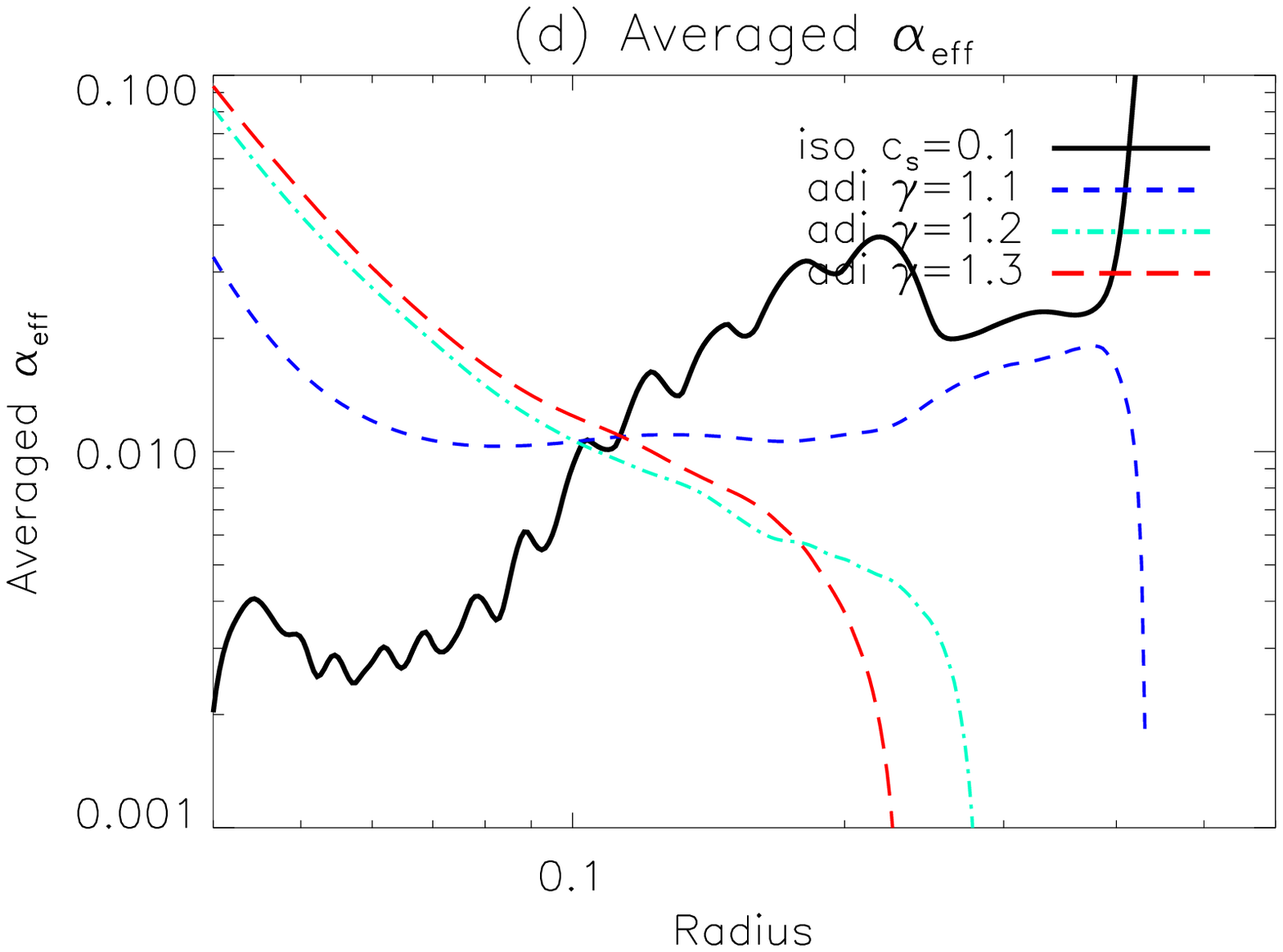}
\includegraphics[width=0.49\textwidth]{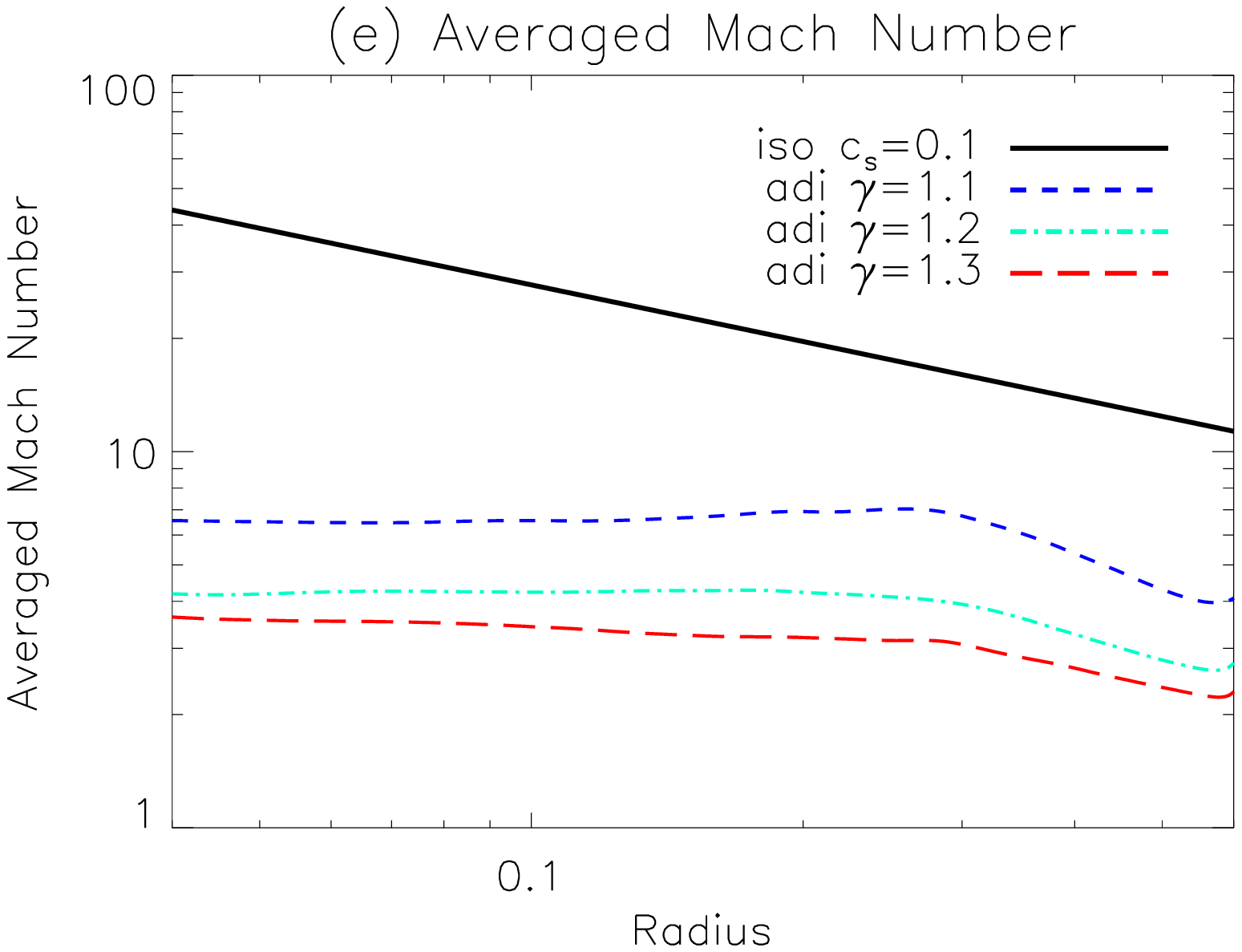}
\caption{Time- and azimuthal- averaged profiles over 25 time units ($\sim$ 4 binary orbits) of surface density(panel a), angular momentum(panel b), mass accretion rate($\dot{M}$, panel c), effective $\alpha$(panel d) for the no-inflow hydro models. The isothermal case has very low accretion rate with $\alpha_{eff} \sim 0.003$ at the inner part of the disk ($R<0.1$), and thus mass pile up beyond radius 0.1. The adiabatic cases have larger accretion rates down to the inner edge of the disk with $\alpha_{eff}$ increasing as radius decreases.}
\label{fig:noinflow-average}
\end{figure*}

\begin{figure*}
\centering
\includegraphics[width=0.49\textwidth]{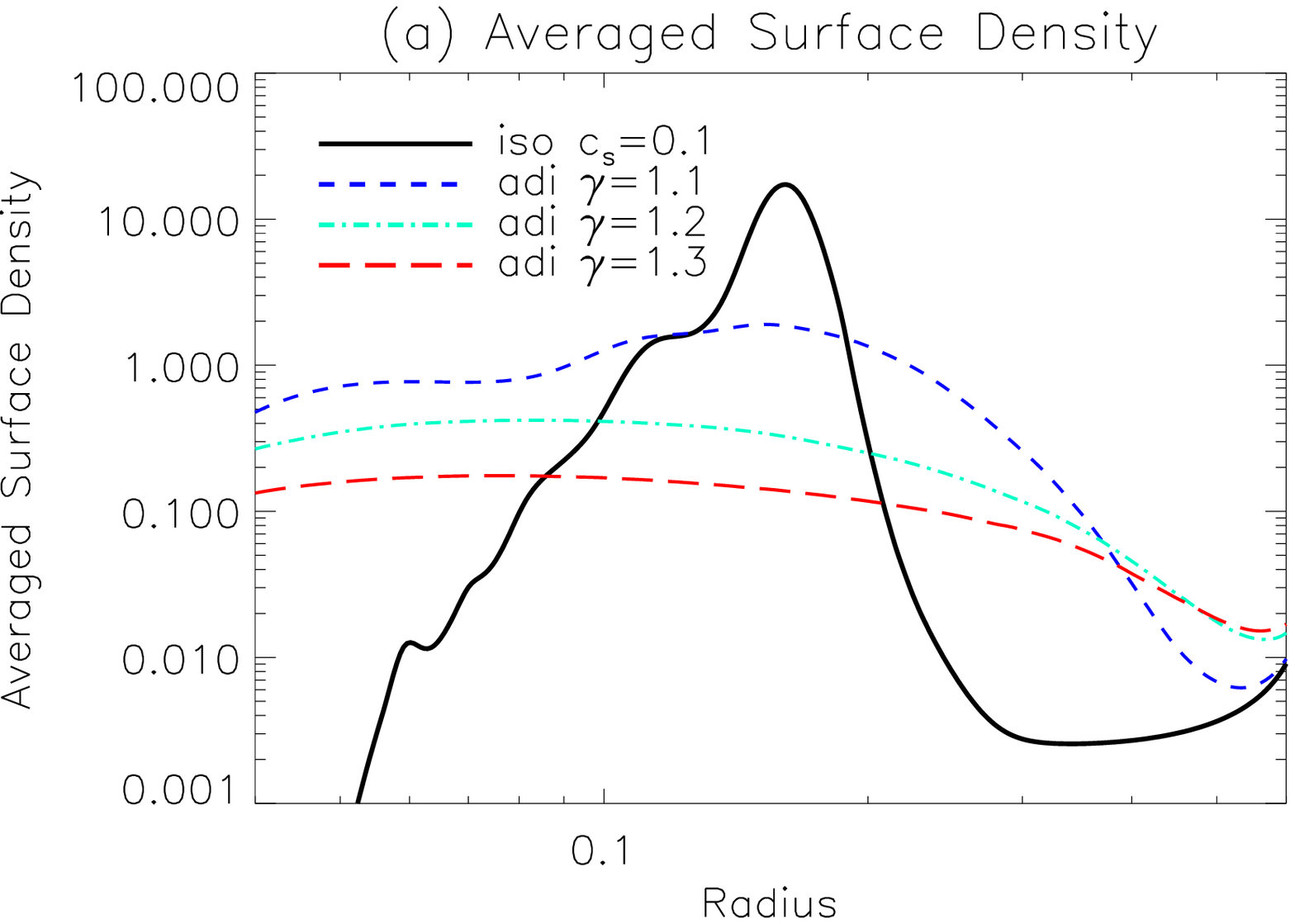}
\includegraphics[width=0.49\textwidth]{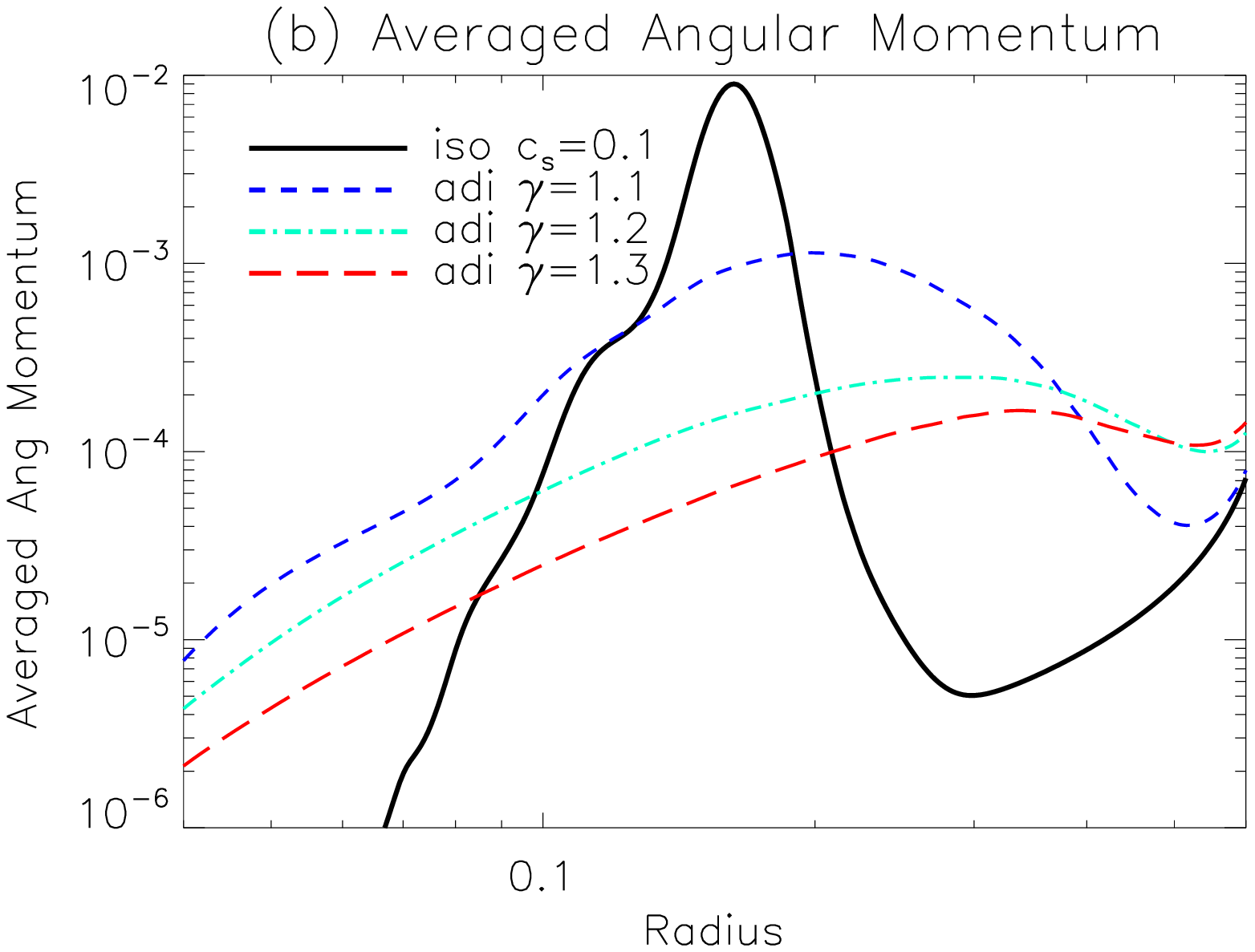}
\includegraphics[width=0.49\textwidth]{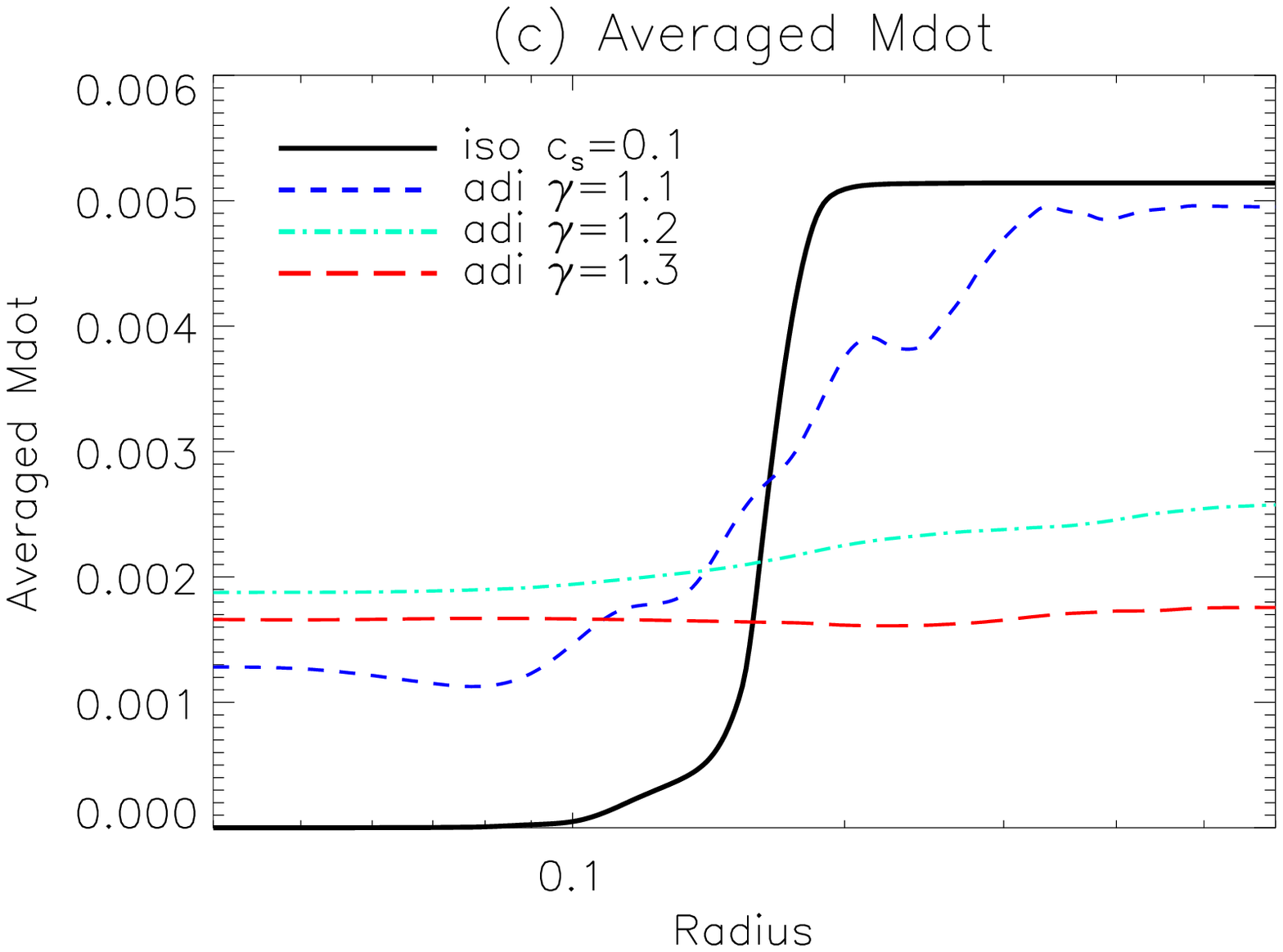}
\includegraphics[width=0.49\textwidth]{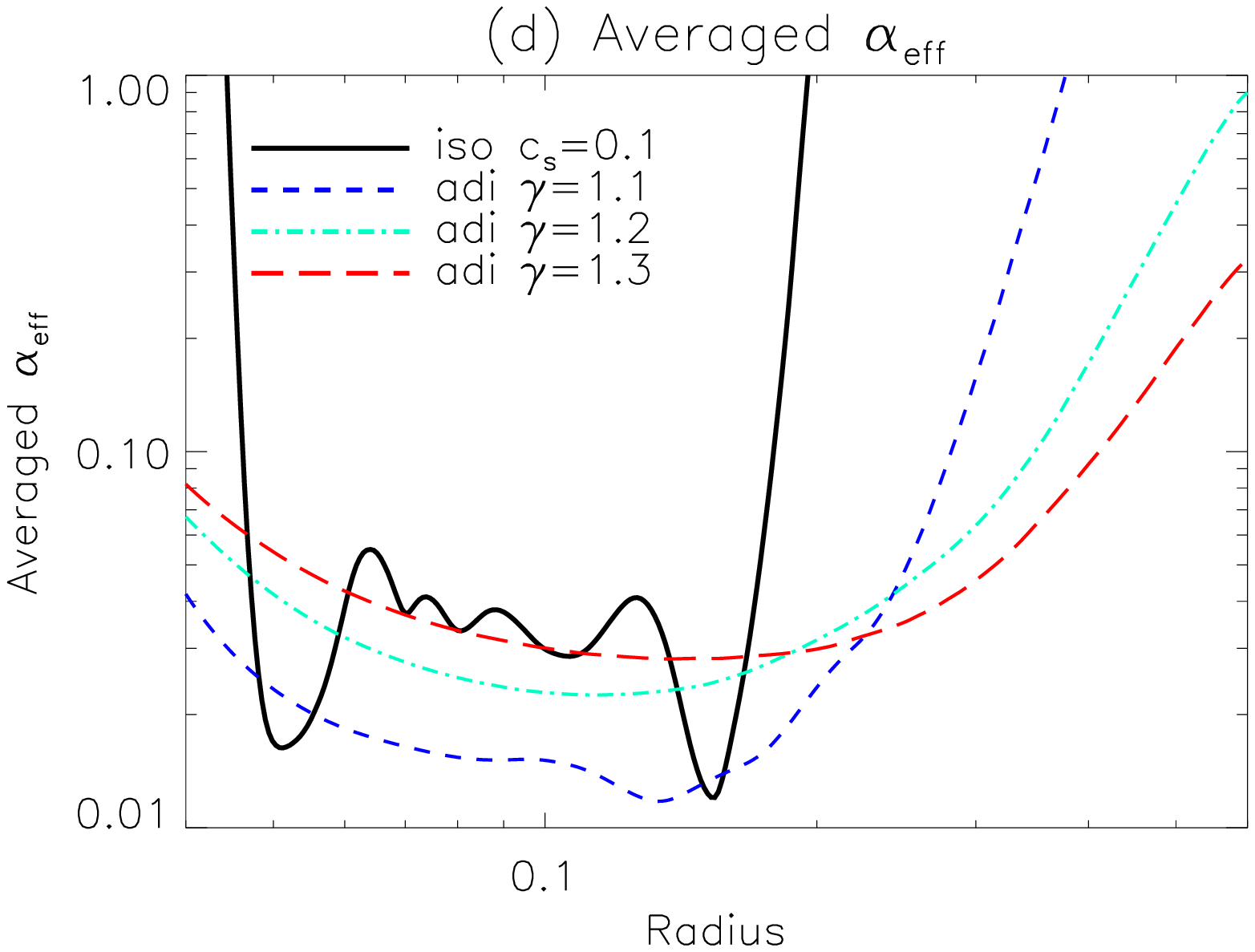}
\includegraphics[width=0.49\textwidth]{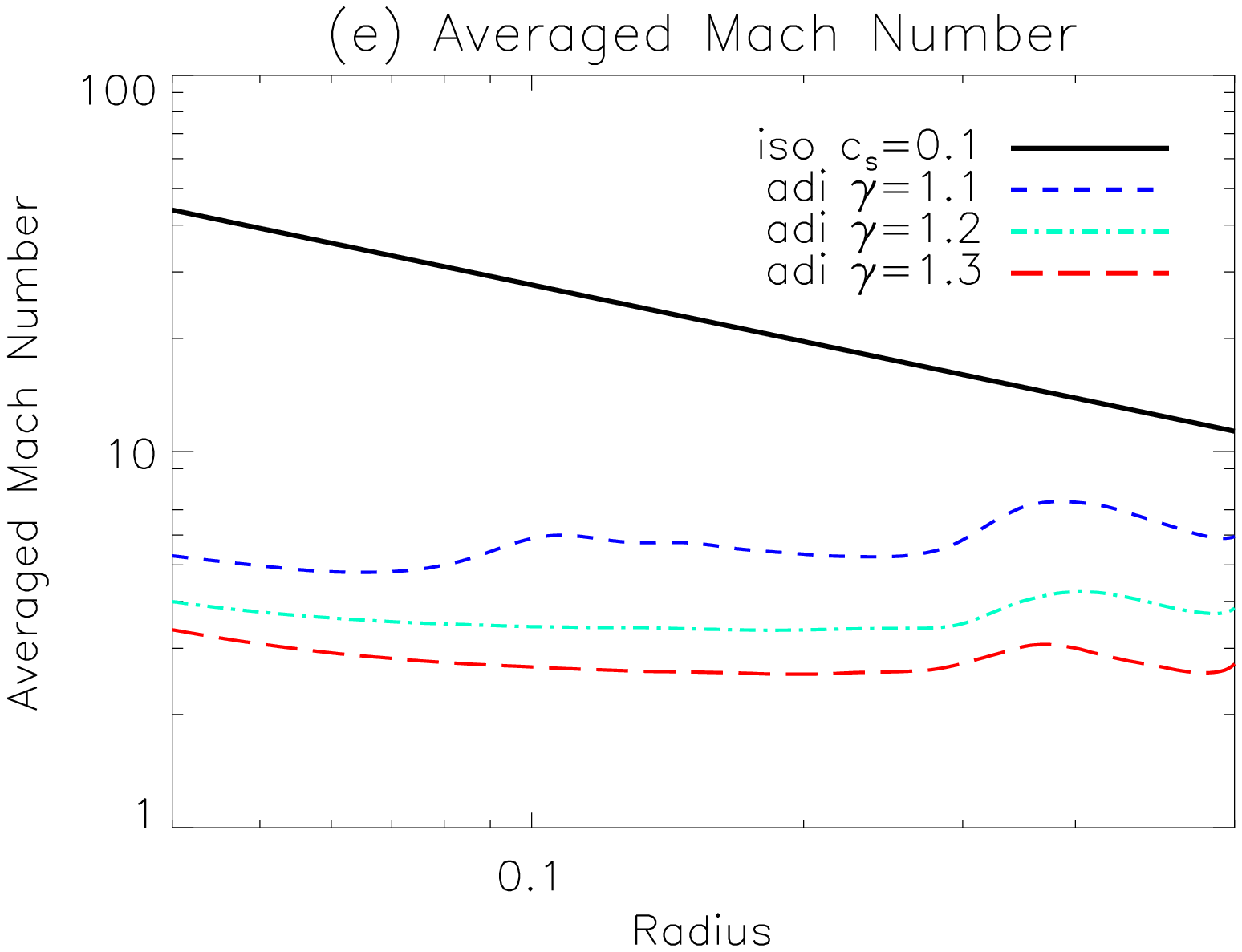}
\caption{Time- and azimuthal- averaged profiles over 25 time units ($\sim$ 4 binary orbits) of surface density(panel a), angular momentum(panel b), mass accretion rate($\dot{M}$, panel c), effective $\alpha$(panel d) for the with-inflow hydro models. Both $\dot{M}$ and $\alpha_{eff}$ have flat radial profiles in the whole disk with $\alpha_{eff} \sim 0.02 - 0.05$ which is consistent with $\alpha$ from DNe observations during quiescence states. }
\label{fig:inflow-average}
\end{figure*}

In Fig. \ref{fig:noinflow-average} and Fig. \ref{fig:inflow-average}, we show the time- and azimuthal- averaged radial profiles over 25 time units ($\sim$ 4 binary orbits) of the following quantities for the no-inflow models and with-inflow-models respectively:  surface density (\ref{fig:noinflow-average}a, \ref{fig:inflow-average}a), angular momentum (\ref{fig:noinflow-average}b, \ref{fig:inflow-average}b), mass accretion rate(\ref{fig:noinflow-average}c, \ref{fig:inflow-average}c), $\alpha_{eff}$ (\ref{fig:noinflow-average}d, \ref{fig:inflow-average}d), and Mach number $\mathcal{M}$ (\ref{fig:noinflow-average}e, \ref{fig:inflow-average}e). The values of $\alpha_{eff}$ are calculated using Eq. \ref{eq:effalpha}.

For the no-inflow models, the $\alpha_{eff}$ is in the range of [0.003, 0.03], varying with radius and EOS. In the isothermal disk with sound speed $c_s=0.1$, mass accretion mostly happens at outer edge of the disk where $\alpha_{eff} \sim 0.02$, but $\alpha_{eff}$ decreases to $\sim 0.003$ at $R<0.1$. This is because, spiral arms have larger pitch angles and stronger shock dissipation at outer part of the disk, but as the spiral waves propagated to smaller radii, the amplitudes of the waves become weaker, the opening angles become smaller, and thus shock dissipation becomes much weaker. Since the projection of  azimuthal velocity onto the normal of wave front surface is much smaller as the waves are more oblique. This $\alpha_{eff}$ profile is consistent with previous work such as \citet{2000NewA....5...53B}. As a result of this $\alpha_{eff}$ profile, most of mass is accumulated at outer edge of the disk (Fig. \ref{fig:noinflow-average}a) due to low accretion efficiency at inner part of the disk.

The $\alpha_{eff}$ has different profiles in the adiabatic disks. $\alpha_{eff} \sim 0.01$ in the majority of the disk and increases towards smaller radii. Shocks are stronger at the inner part of the disk rather than at the outer edge as in the isothermal case (see Fig. \ref{fig:noinflow-dispersion} for comparison). Shock dissipation heats up the inner part of the disk and makes the opening angle of spirals large at small radii. The large opening angles help sustain strong spiral arms and shock dissipation. The $\dot{M}$ reaches a constant non-zero value throughout the whole disk (Fig. \ref{fig:noinflow-average}c) meaning the disk reaches quasi-steady state. Therefore, $\alpha_{eff}$ increases towards smaller radii as the surface density decreases. At the outer edge of the adiabatic disks, there is significant outflow for the disks with $\gamma=1.2$ and $\gamma=1.3$. This is because as $\gamma$ increases, the disk temperature increases and some gas may be blown out due to radial pressure gradient. Efficient angular momentum transport also contribute to mass escape at the outer edge of the disk.

In the with-inflow models (Fig.\ref{fig:inflow-average}), $\alpha_{eff} \sim 0.02 - 0.05$ throughout the whole disk, larger than the no-inflow models. In the isothermal model with $c_s=0.1$, the disk has $\alpha_{eff} \sim 0.03$ down to very small radii, meaning the spiral shocks are still strong at inner part of the disk, and accreted material could reach surface of the WD. This indicates that the L1 accretion stream could enhance angular momentum transport to some extent. The values of $\alpha_{eff}$ are order-of-magnitude consistent with those from DNe observations during quiescence states.

The averaged radial profiles of $\mathcal{M}$ for the no-inflow models and the with-inflow models are shown in Fig. \ref{fig:noinflow-average}e and \ref{fig:inflow-average}e respectively. As the ratio of specific heats increases from $\gamma=1$ (the isothermal model) to $\gamma=1.3$, $\mathcal{M}$ decreases from $\sim 10 - 50$ to $\sim 3-7$. For most of the models (all no-inflow models and the adiabatic with-inflow models), as $\mathcal{M}$ decreases (disk temperature increases), $\alpha_{eff}$ increases, especially at the inner part of the disk where the majority of the disk resides. This trend has several causes: 1) As discussed in \S \ref{sec:waveexication}, lower $\mathcal{M}$ leads to stronger tidal response of the disk to the perturbation of the companion star; 2) Disk with lower $\mathcal{M}$ is usually more radially extended, which also strengthens the tidal coupling of the disk and the companion star; 3) As $\mathcal{M}$ decreases, the spiral arms are more open which helps form stronger shocks since the projected velocity gradient pertpenticular to the shock surface is larger. 

As we mentioned earlier in \S \ref{subsec:units-scaling} and \S \ref{sec:result-mhd}, the Mach numbers $\mathcal{M}$ in our models ($\mathcal{M} \sim 3-50$) are lower than real CV disks according to eclipse mapping observations ($\mathcal{M} \sim 50 - 200$ during DNe outbursts). Extrapolating the trend we observed above, we may infer that $\alpha_{eff}$ in a real CV disk is smaller than the range of $\sim 0.02 - 0.05$ observed in our simulations. However, there are several points worth noting. Firstly, the $\mathcal{M}$ values from eclipse mapping observations are derived from blackbody brightness temperature which is smaller than the midplane temperature that we refer to in our models. Therefore, the midplane $\mathcal{M}$ in CV disks is supposed to be smaller than quoted in eclipse mapping observations \citep[e.g.][]{1998MNRAS.298.1079B}. Secondly, one exception for the trend of decreasing $\alpha_{eff}$ with increasing $\mathcal{M}$ is the isothermal with-inflow model (see solid black line Fig. \ref{fig:inflow-average}d): although this model has the highest $\mathcal{M}$, its $\alpha_{eff}$ is comparable with that in adiabatic models but much larger than that in the isothermal no-inflow model. This may be secon-order effects due to constant injection of gas from L1 point.

%%%%%%%%%%%%%%%%%%%%%
\subsection{Energy Budget}
\label{sec:energybudget}
%Thermodynamics of CV disks is of potential interest because the spiral structures as well as strength of shock dissipation and thus mass accretion rate are all sensitive to disk temperature. Since we do not include extra cooling source in our adiabatic simulations (both hydro and MHD), the factors that affect the thermal evolution of the disks in our simulations are: 1) heating from mass accretion during which gravitational energy is released; 2) heating from work done by the external forces from companion star; 3) cooling from radial advection of energy. We observe that the temperature profiles of our adiabatic simulations either have reached steady state (e.g. the MHD model, the with-inflow hydro model with $\gamma=1.3$) or increase very slowly compared with energy flow within the same time period (e.g. the with-inflow hydro models with $\gamma=1.1$ and $\gamma=1.2$). Therefore, the heating and cooling processes must nearly balance each other in CV disks in our simulations. Since radial advection is the only possible cooling process in our cases, we preclude that radial advection may play an important role in CV disks. Due to complexity of energy diagnostics, we do not explore the detailed energy budget terms in the conservation equation as we did for angular momentum in above sections. But we hope to explore the roles of radial advection and radiative cooling in CV disk thermodynamics in our future work.

%[Energy released by radial advection since we have no cooling] 
Thermodynamics plays an essential role in angular momentum transport and mass accretion in CV disks since the pattern of spiral density waves and strength of shock dissipation are both sensitive to disk temperature. In an accretion disk, local heating by turbulence and shock dissipation is balanced by radial energy advection and radiative cooling. In our adiabatic simulations without local radiative cooling, we observe that the temperature profiles either have reached steady state (e.g. the MHD model and the with-inflow hydro model with $\gamma= 1.3$) or increase very slowly compared to the radial advection time (e.g. the with-inflow hydro models with $\gamma= 1.1$ and $\gamma = 1.2$). This implies that the local heating by shock dissipation is balanced by the radial energy advection in these simulations. This is in stark contrast to standard thin-disk theory, where local dissipation is everywhere balanced by radiative cooling \citep{1973A&A....24..337S, 1999ApJ...521..650B}. 

In future papers, we will report results from MHD simulations with radiative cooling that treat disk thermodynamics more realistically. Such models are the only way to investigate whether traditional thin viscous disk theory accurately predicts the temperature structure in CV disks when the radial advection of energy or work done by the binary torque is included. Such questions are beyond the scope of the present paper.

%%%%%%%%%%%%%%%%%%%%%
\subsection{Convergence Study}
\label{sec:convergence}

In order to assure that our results are not affected by numerical resolution, we perform a convergence study using the with-inflow hydro model with adiabatic EOS $\gamma=1.1$. As listed in Table \ref{tab:parameter}, our fiducial resolution is $384(R) \times 704(\phi)$. For comparison, we conduct a second model with half the fiducial resolution $192(R) \times 352(\phi)$ and a third model with double the fiducial resolution $768(R) \times 1408(\phi)$. All three models are using the same parameters and logarithmic spacing in radial direction. In Fig.\ref{fig:inflow-average}, we show the snapshots of the spiral patterns at $t=80$ and compare the volume- and time-averaged effective $\alpha$ for the three resolutions. The effective $\alpha$ are all averaged over the time period of $70-80$.

\begin{figure*}
\centering
\includegraphics[width=0.49\textwidth]{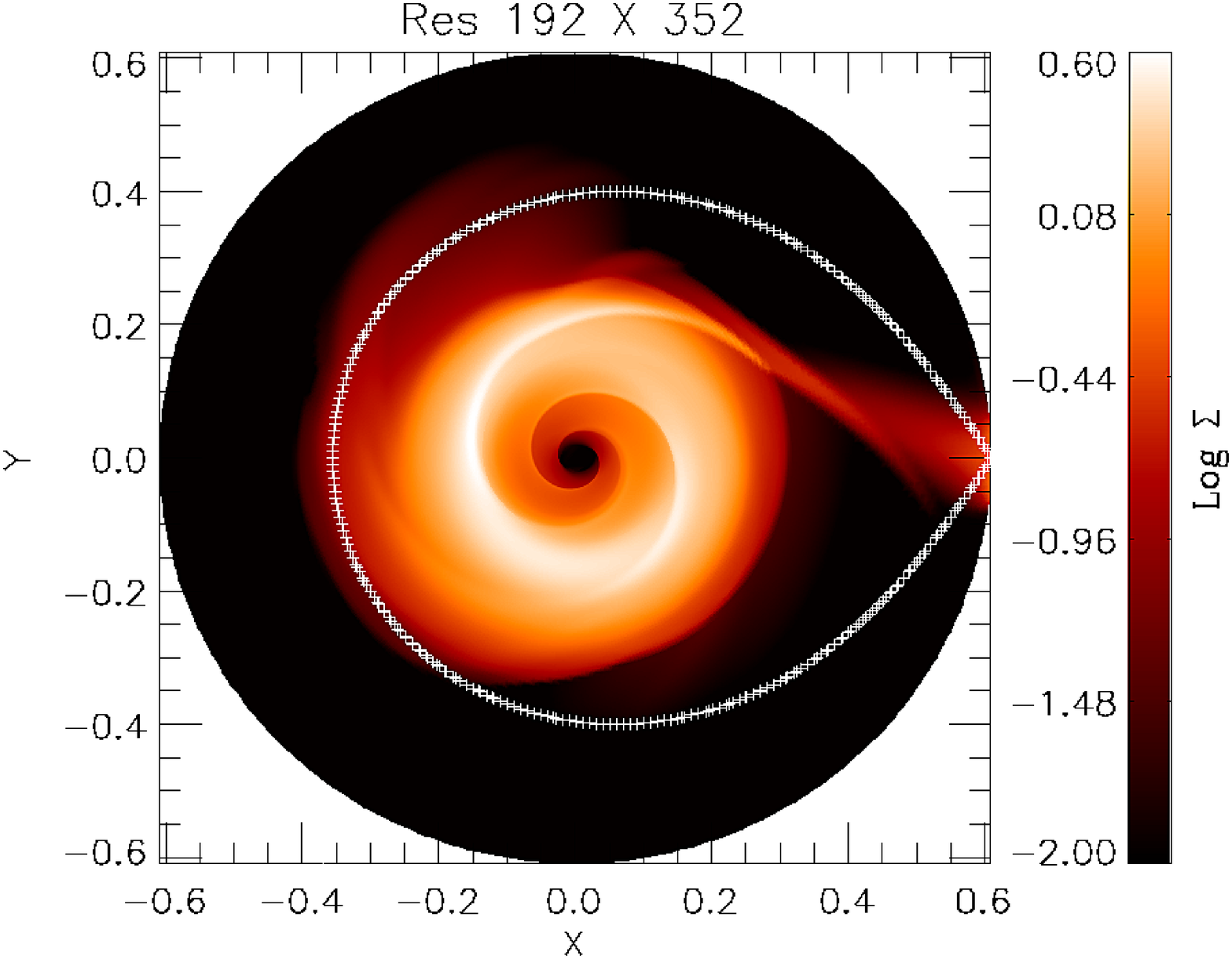}
\includegraphics[width=0.49\textwidth]{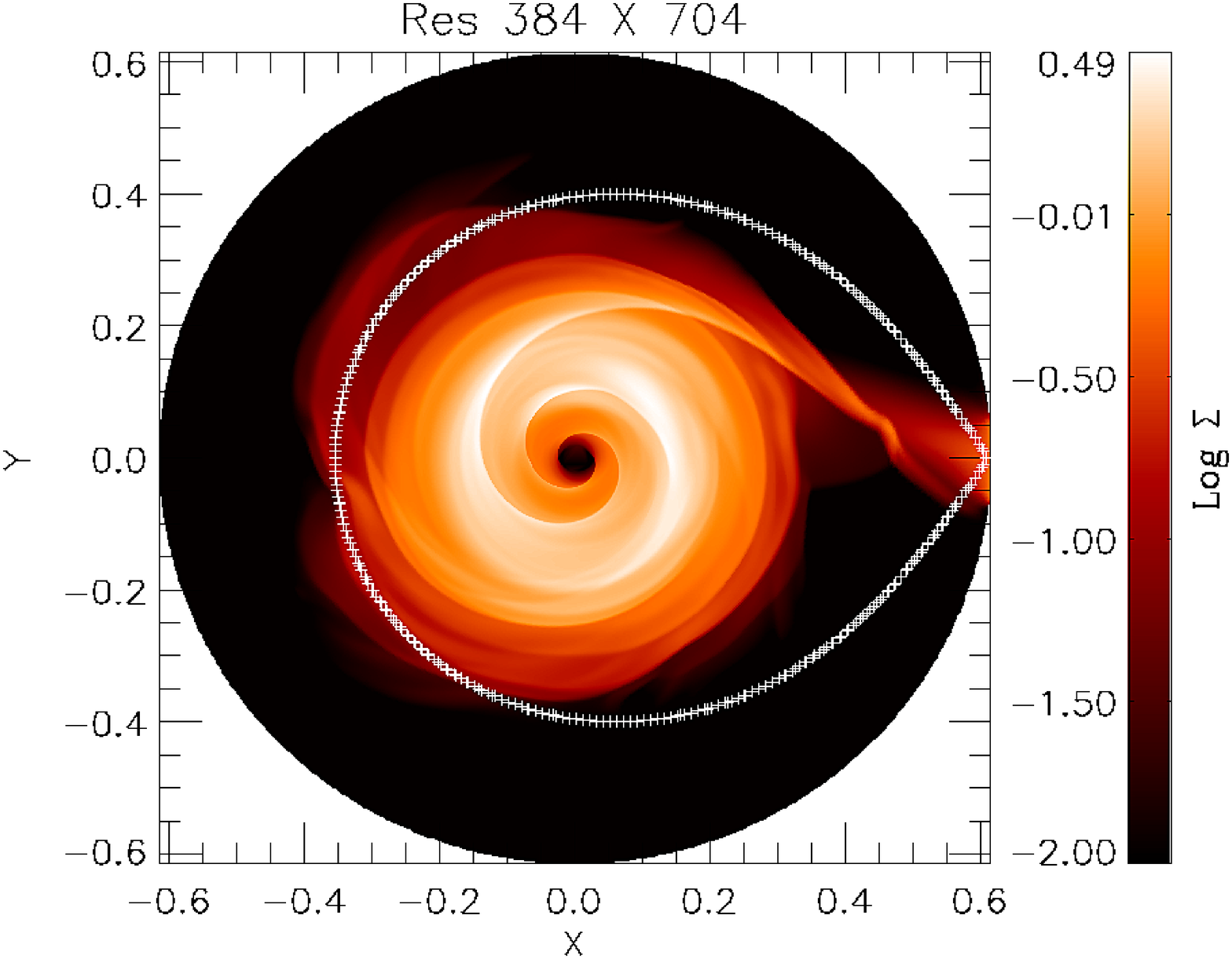}
\includegraphics[width=0.49\textwidth]{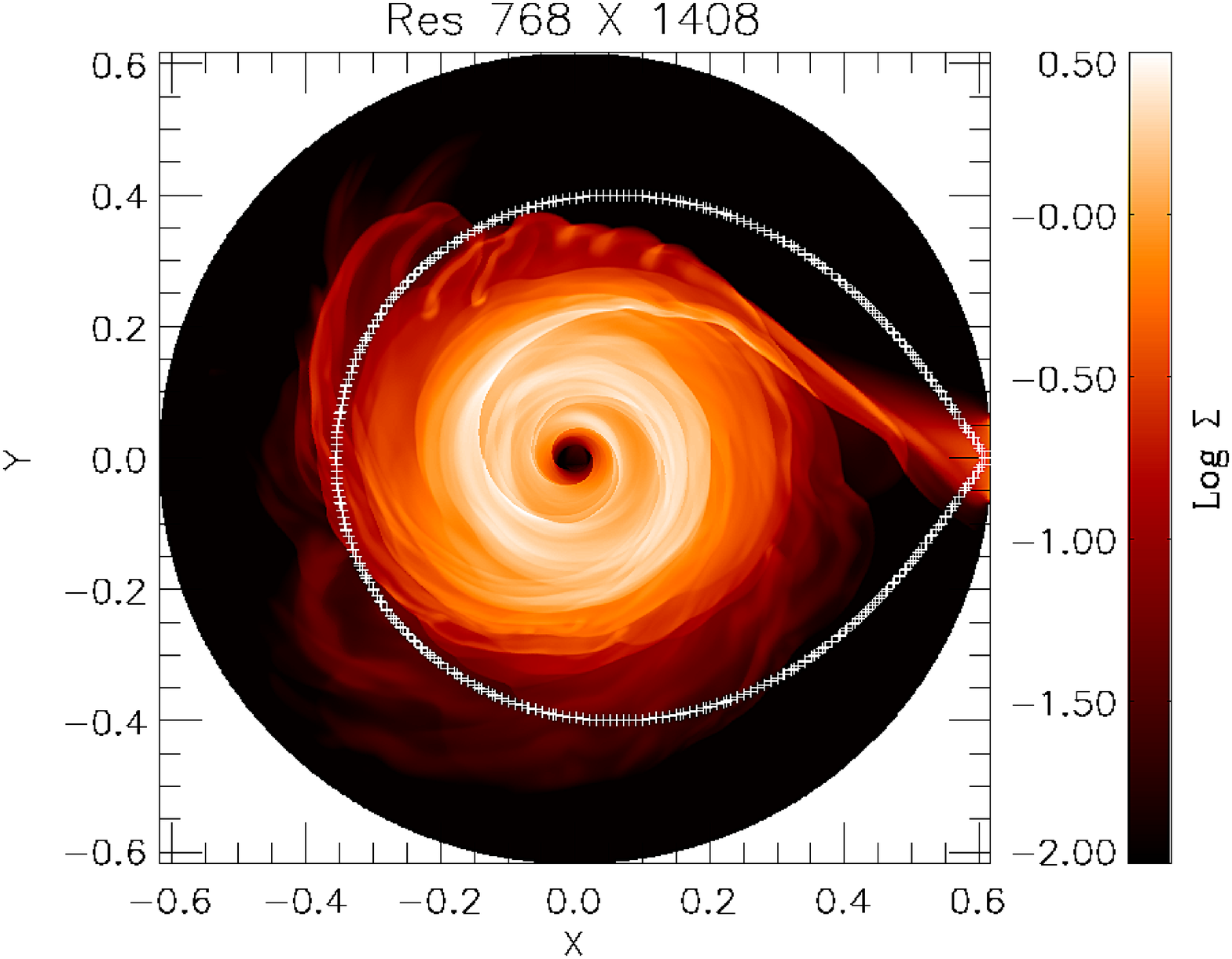}
\includegraphics[width=0.49\textwidth]{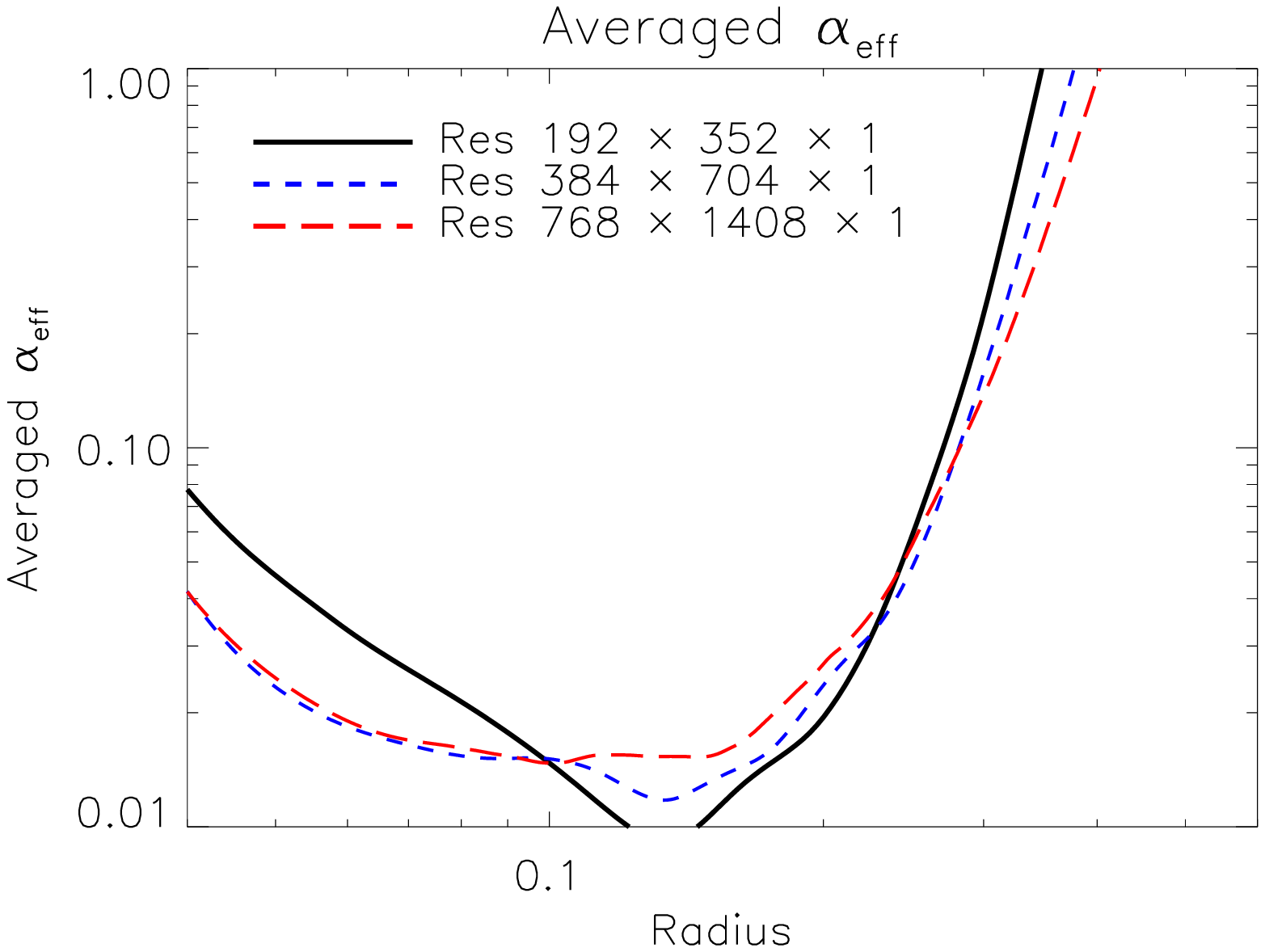}
\caption{Convergence study using the with-inflow hydro model with adiabatic EOS $\gamma=1.1$. The first three plots show the snapshots of surface density at $t=80$ for resolution $192(R) \times 352(\phi)$, $384(R) \times 704(\phi)$ and $768(R) \times 1408(\phi)$ respectively. The last plot shows the effective $\alpha$ averaged over time $70-80$ for the three resolutions where our fiducial resolution ($384(R) \times 704(\phi)$) and the higher resolution ($768(R) \times 1408(\phi)$) are converged.}
\label{fig:convergence}
\end{figure*}

As seen in Fig.\ref{fig:convergence}, the spiral patterns have the same shape in general and the disks have similar sizes. This indicates that the general evolution of the spiral patterns converges. The only difference we notice is that as the resolution increases, the disk appears more turbulent. This is because higher resolution resolves more fine scale structures produced by instabilities \citep[e.g.][]{1983ApJ...274..152V}. Regarding the parameter of the most interest $\alpha_{eff}$, the fiducial resolution ($384(R) \times 704(\phi)$) and the higher resolution ($768(R) \times 1408(\phi)$) models are consistent, with $\alpha_{eff} \sim 0.02$ in most of the disk. This indicates that our fiducial resolution has reached convergence in terms of $\alpha_{eff}$. We did similar convergence tests for other hydro models, especially for the isothermal models with $c_s=0.1$ which have the smallest spiral pitch angles and thus are the most difficulty cases to resolve, and found our fiducial resolution are also converged in the measured value of $\alpha_{eff}$. 

%%%%%%%%%%%%%%%%%%%%%%%%%%%%%%%%%%%%%%%%%%%%%%%%%%%%%%%%%%%%%%%%%
%%%%%%%%%%%%%%%%%%%%%%%%%%%%%% Conclusion %%%%%%%%%%%%%%%%%%%%%%%%%%%%%
%%%%%%%%%%%%%%%%%%%%%%%%%%%%%%%%%%%%%%%%%%%%%%%%%%%%%%%%%%%%%%%%%
\section{Discussion}
\label{sec:discussion}

The results of the work presented in this paper point to a complex interplay between angular momentum transport due to MRI turbulence and spiral shocks. On the one hand, the MRI can enhance spiral shocks, and hence the angular momentum transport efficiency, for the following reasons. Outward transport of angular momentum by MRI turbulence tends to expand the disk, which makes the outer region of the disk closer to the Lindblad resonance. This increase of disk size can significantly strengthen the spiral waves. MRI turbulence can also heat up the disk, which lowers the Mach number. With lower Mach numbers, the opening angles of spiral waves are larger, which helps to sustain stronger shocks due to a larger velocity jump projected perpendicular to the wave surface. At the same time, lower Mach numbers also strengthen the spiral arms due to an increase of the wavelengths of perturbations excited by Lindblad resonance. This is confirmed by the MHD model we show in \S \ref{sec:result-mhd}, where the Reynolds stress in the MHD model with MRI turbulence is indeed larger than that in the corresponding hydro model (see Fig. \ref{fig:mhd-stress}). 

However, MRI turbulence can also potentially damp the spiral shocks in some circumstances and inhibit angular momentum transport by spiral shocks. If the disk is cold, the spiral waves may be very weak. In this case, the inward propagation of the spiral waves may be significantly damped by the turbulent motions induced by MRI essentially due to an effective turbulent viscosity. Therefore, it is important to investigate the relative importance of MRI turbulence and spiral shocks in various physical circumstances including different field geometries and different Mach numbers, which will be the focus of the second paper in this series.

Another factor of potential impact on the interplay between MRI turbulence and the spiral shocks is vertical stratification. All simulations in this work focus on unstratified disks where vertical structures of density and temperature are neglected. If vertical stratification is included, vertical resonant forcing would become important, and waves that propagate radially can channel and tilt along the vertical direction due to continuous change of vertical profile \citep{1998ApJ...504..983L, 2002MNRAS.332..575B, 2015ApJ...814...72L}. Along with the vertical propagation of waves, wave amplitude changes and may steepen into shocks \citep{2015ApJ...813...88Z}. The 3D structure of spiral waves in a stratified CV disk and the interplay between the 3D waves with magnetic field can only be known with non-linear global MHD simulations of stratified disks, which is the goal of our future work and is beyond the scope of the current paper.

Finally, it is interesting to examine whether the disk in our MHD models show growth in eccentricity driven by the elliptical instability \citep{1991ApJ...381..259L, 2008A&A...487..671K}. This instability requires visocosity, and therefore we do not expect it in an inviscid hydro simulation. However it is often suggested that MRI turbulence acts as a ``viscosity". Therefore, will it drive the elliptical instability? We measure the eccentricity of the disk using the same method as in \citet[see their Eq.12]{1991ApJ...381..259L} and \citet[see their Eq.2]{2009MNRAS.393.1423C}, which utilizes the perturbation of the velocity vectors. In each ring at radius $R$, the averaged eccentricity is 
\begin{equation}
e(R) = \frac{ | \int d\phi \Sigma(R, \phi) v_R \exp(i \phi) |  }{\int d\phi \Sigma(R,\phi) v_\phi}.
\end{equation}
Then the averaged eccentricity of the whole disk is the mass-weighted average of $e(R)$ as \citet{2008A&A...487..671K} did. The mass average is to exclude the effects in the outer part of our computational domain where there is barely gas but the velocity perturbations are big due to gravitational force from the companion.

\begin{figure}[ht]
\centering
\includegraphics[width=0.49\textwidth]{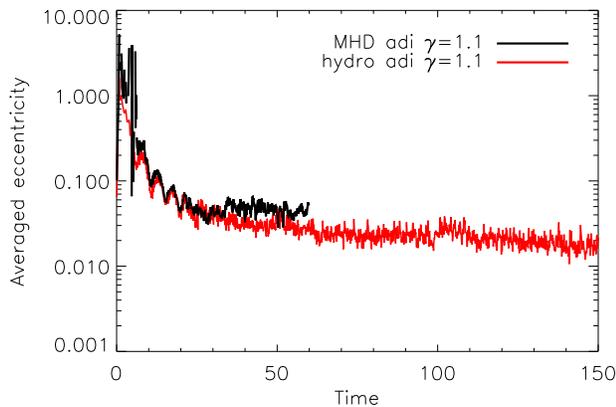}
\caption{Time evolution of eccentricity in the MHD model (black line) and the with-inflow hydro model using adiabatic EOS with $\gamma=1.1$. No elliptical instability is observed in either model within the time of our simulations.}
\label{fig:ecc}
\end{figure}

In Fig.\ref{fig:ecc} we show the time evolution of the mass averaged eccentricity of the disk in the MHD model and the with-inflow hydro model using adiabatic EOS with $\gamma=1.1$. In both of the models, we do not observe significant growth in eccentricity within the time of our simulations ($t \sim 300$ in the hydro model and $\sim 60$ in the MHD model), and the eccentricity eventually stays stable at $e \sim 0.02 - 0.04$.
%In the no-inflow hydro cases, the $c_s=0.1$ isothermal model keeps constant eccentricity at 0.04, while the $cs=0.3$ isothermal model and the $\gamma=1.1$ adiabatic model have decrease in eccentricity. For the $\gamma=1.2$ and $\gamma=1.3$ adiabatic models, they have fast accretion due to strong and open spiral arms whereas no mass supply at outer boundary, so they depletes mass very early. Before their depletion of gas, their eccentricity both stays below 0.1. For the with-inflow hydro cases, the eccentricity all becomes stable at a constant value which increases as the disk temperature increases. In contrast to the high eccentricity observed in \citet{2008A&A...487..671K}, we do not observe elliptical instability in our models. The reason for this discrepancy is that we include no artificial viscosity in our simulations although we do have angular momentum transport by spiral shocks. As \citet{2008A&A...487..671K} reports, the eccentricity of the disk drops to below 0.01 as they decrease the kinetic viscosity to $3.3 \times 10^{-9}$ (see their Fig. 8). This indicates that the spiral shocks are fundabmentally different from $\alpha$ viscosity in terms of angular momentum transport.
Adopting $\alpha_{eff} \sim 0.01$ from Maxwell stress (see Fig. \ref{fig:mhd-stress}) and $\mathcal{M} \sim 10$ in our MHD model, the corresponding dimensionless viscosity is $\nu = \alpha / \mathcal{M}^2 \sim 10^{-4}$. According to the $\nu = 10^{-4}$ model in \citet{2008A&A...487..671K}, the disk eccentricity is expected to start growing from $e \sim 0.03$ at $\sim 10$ binary orbits ($t \sim 60$ in our time units) to $e \sim 0.5$ at $\sim 100$ binary orbits ($t \sim 600$ in our time units). However, our 3D global MHD models are expensive so that we only run to $t=60$. At this time we do not observe eccentricity growth yet (see black line in Fig.\ref{fig:ecc}).  We hope to investigate this problem further with longer MHD simulations.

%%%%%%%%%%%%%%%%%%%%%%%%%%%%%%%%%%%%%%%%%%%%
%%%%%%%%%%%%%%%%%%%%%%%%%%%%%%%%%%%%%%%%%%%%
\section{Conclusion}
\label{sec:conclusion}

In this work, we conducted the first global MHD simulations of unstratified CV disks with particular focus on the dynamics of spiral waves observed in our simulations. Our major conclusions are as follows: 
\begin{enumerate}
\item The strength of spiral waves are very sensitive to the size and Mach number of the disk because the size controls how closely the outer edge of the disk overlaps with the Lindblad resonance of the companion, while the Mach number controls the net amplitude of perturbations (see \S \ref{sec:waveexication}). Spiral waves are stronger if the disk size is larger or the Mach number is lower. 

\item The pitch angles of spiral waves follow the linear wave dispersion relation with $m=2$ (see \S \ref{sec:wavepropagation}). As the local Mach number is decreased, the spiral arms are wound more openly. The self-similar shock model by \citet{1987A&A...184..173S} is not a good fit to the spiral pattern, instead linear wave theory works well.

\item By analyzing the angular momentum budget of the disk, we conclude that the local shock dissipation leads to angular momentum loss of the disk gas and mass accretion(see \S \ref{sec:shockdissipation}). The effective $\alpha_{eff}$ is $\sim 0.003 - 0.03$ in the no-inflow cases, and $\alpha_{eff} \sim 0.02 - 0.05$ in the with-inflow cases, similar to observed values of DNe in quiescence.
\end{enumerate}

In the second paper of this series, we focus on global MHD simulations of unstratified CV disks to investigate the relative importance of MRI and spiral shocks in angular momentum transport and how their relative importance depends on Mach number of the disk and geometry and strength of the injected seed magnetic field.

In the future, we plan global MHD simulations of stratified CV disks including radiative cooling to further study the interplay between MRI turbulence and spiral shocks with realistic thermodynamics. In addition, it would also be interesting to connect our simulations to recent studies of WD boundary layers \citep{2013ApJ...770...68B} since our current inner boundary is only a few WD radii.

%%%%%%%%%%
\acknowledgements We thank the anonymous referee for substantially improving this work. We thank J. Cannizzo, J. Goodman, D. Lin, J. Papaloizou, R. Rafikov, S. Tremaine for insightful discussions and suggestions. This project is supported by NSF grant AST-1312203. Numerical simulations in this work are carried out using computational resources supported by the Princeton Institute of Computational Science and Engineering, and the Texas Advanced Computing Center (TACC) at The University of Texas at Austin through XSEDE grant TG-AST130002. Z.Z. acknowledges support by NASA through Hubble Fellowship grant HST HF-51333.01-A awarded by the Space Telescope Science Institute, which is operated by the Association of
Universities for Research in Astronomy, Inc., for NASA, under contract NAS 5-26555.

%%%%%References

%\begin{thebibliography}{115}
%\expandafter\ifx\csname natexlab\endcsname\relax\def\natexlab#1{#1}\fi

%\end{thebibliography}

\bibliographystyle{apj}
\bibliography{reference}

\end{document}